\documentclass[12pt,preprint]{aastex}

\usepackage{graphics}
\usepackage{mathptmx}
\usepackage{natbib}

\newcommand{\teff}{$T_{\!\mbox{\scriptsize\em eff}}$}

\newcommand{\teffq}{$T_{\!\mbox{\scriptsize \em eff}}^4$}

\slugcomment{}

\shorttitle{A supergiants in NGC 300}
\shortauthors{Kudritzki et al.}

\begin{document}


\title{Quantitative Spectroscopy of 24 A supergiants in
the Sculptor galaxy NGC 300$^1$: Flux weighted gravity luminosity
relationship, metallicity and metallicity gradient\\[2mm]}\footnotetext[1]{Based
on VLT observations for ESO Large Programme 171.D-0004.}

\author{Rolf-Peter Kudritzki} \affil{Institute for Astronomy, 2680 Woodlawn
Drive, Honolulu, HI 96822; kud@ifa.hawaii.edu}

\author{Miguel A.~Urbaneja} \affil{Institute for Astronomy, 2680 Woodlawn
Drive, Honolulu, HI 96822; urbaneja@ifa.hawaii.edu}

\author{Fabio Bresolin} \affil{Institute for Astronomy, 2680 Woodlawn
Drive, Honolulu, HI 96822; bresolin@ifa.hawaii.edu}

\author{Norbert Przybilla} \affil{Dr. Remeis-Sternwarte Bamberg,
Sternwartstr. 7, D-96049 Bamberg, Germany;
przybilla@sternwarte.uni-erlangen.de}

\author{Wolfgang Gieren} \affil{Universidad de Concepci\'on, Departamento de F\'isica, Casilla 160-C, Concepci\'on, Chile; wgieren@astro-udec.udec.cl}

\and

\author{Grzegorz Pietrzy\'nski} \affil{Universidad de Concepci\'on, Departamento de F\'isica, Casilla 160-C, Concepci\'on, Chile;  pietrzyn@hubble.cfm.udec.cl}

\begin{abstract}
A quantitative spectral analysis of 24 A supergiants in the Sculptor Group 
spiral galaxy NGC 300 at a distance of 1.9 Mpc is presented. A new method 
is introduced to analyze low resolution ($\sim 5 \AA $) spectra, which 
yields metallicities accurate to 0.2 dex including the uncertainties 
arising from the errors in \teff\/ (5\%) and $log~g$ (0.2 dex).
For the first time the stellar metallicity 
gradient based on elements such as titanium and iron in a galaxy beyond the 
Local Group is investigated. Solar metallicity is measured in the center and 0.3 solar in the
outskirts and a logarithmic gradient of $-0.08$ dex/kpc. An average reddening 
of E(B-V) $\sim$ 0.12 mag is obtained, however with
a large variation from 0.07 to 0.24 mag. We also determine stellar radii, luminosities 
and masses and discuss the evolutionary status. Finally, 
the observed relationship between absolute bolometric magnitudes M$_{bol}$ and flux 
weighted gravities $g_{F} = g/$\teffq~is investigated. At high temperatures the 
strengths of the Balmer lines depends solely on the flux-weighted gravity,
which allows a precise direct determination of $log~g_{F}$ with an accuracy of 
0.05 to 0.1 dex. We find a tight relationship between M$_{bol}$ and $log~g_{F}$ 
in agreement with stellar 
evolution theory. Combining these new results with previous 
work on Local Group galaxies we obtain a new flux weighted gravity luminosity 
relationship (FGLR), which is very well defined and appears to be an excellent 
alternative tool to determine distances to galaxies.  

\end{abstract}

\keywords{galaxies: --- distances galaxies: abundances --- galaxies: stellar content --- galaxies:
  individual (NGC300) --- stars: early-type}
 

\section{Introduction}\label{intro}

It has long been the dream of stellar astronomers to study individual stellar 
objects in distant galaxies to obtain detailed spectroscopic information about 
the star formation history and chemodynamical evolution of galaxies and to
determine accurate distances based on the determination of stellar 
parameters and interstellar reddening and extinction. The most promising objects 
for such studies are massive 
stars in a mass range between 15 to 40 $M_{\odot}$ in the short-lived evolutionary
phase, when they leave the hydrogen main-sequence and cross the HRD in a few 
thousand years as blue supergiants of late B and early A spectral type. Because of 
the strongly reduced absolute value of bolometric correction when evolving towards
smaller temperature these objects increase their brightness in visual light and 
become the optically brightest stars in the universe with absolute visual 
magnitudes up to $M_{V} \cong -9.5$ rivaling with the integrated light brightness 
of globular clusters and dwarf spheroidal galaxies.

There has been a long history of quantitative spectroscopic studies of these 
extreme objects. In a pioneering and comprehensive paper on Deneb 
\citet{groth61} was the first to obtain stellar parameters and detailed 
chemical composition. This work was continued by \citet{wolf71}, \citet{wolf72},
 \citet{wolf73} in studies of A supergiants in the Milky Way and the Magellanic 
Clouds. \citet{kud73} using newly developed NLTE model atmospheres found 
that at the low gravities and the correspondingly low electron densities of these
objects effects of departures from LTE can become extremely important. With strongly
improved model atmospheres \citet{venn95a}, \citet{venn95b} and \citet{aufdenb02}
continued these studies in the Milky Way. Most recently, \citet{przybilla06} used
very detailed NLTE line formation calculations including ten thousands of lines in NLTE
(see also \citealt{przybilla00}, \citealt{przybilla01a}, \citealt{przybilla01b},
\citealt{przybilla01}, \citealt{przybilla02}) to determine stellar parameters and
abundances with hitherto unkown precision (\teff\/ to $\le 2$\%, $log~g$ to $\sim 0.05$ dex,
individual metal abundances to $\sim 0.05$ dex).
At the same time, utelizing the power of the new 8m to 10m class telescopes, high resolution 
studies of A supergiants in many Local Group galaxies  were carried out by \citet{venn99} (SMC),
\citet{mccarthy95} (M33), \citet{mccarthy97} (M31), \citet{venn00} (M31),
\citet{venn01} (NGC 6822), \citet{venn03} (WLM), and \citet{kaufer04} (Sextans A) 
yielding invaluable information about the stellar chemical composition in these galaxies.

The concept to go beyond the Local Group and to study A supergiants by means of 
quantitative spectroscopy in galaxies out to the Virgo cluster has been first
presented by \citet{kud95} and \citet{kud98}. Following-up on this idea,
\citet{bresolin01} and \citet{bresolin02} used the VLT and FORS at 5 $\AA$ resolution
for a first investigation of blue supergiants in NGC 3621 (6.7 Mpc) and NGC 300 (1.9 Mpc). They were able
to demonstrate that for these distances and at this resolution spectra of sufficient S/N 
can be obtained allowing for the quantitative determination of stellar parameters 
and metallicities. \citet{kud03} extended this work and showed that stellar 
gravities and temperatures determined from the spectral analysis can be used to 
determine distances to galaxies by using the correlation between absolute bolometric
magnitude and flux weighted gravity $g_{F} = g/$\teffq\/ (FGLR).
However, while these were encouraging steps towards the use of A supergiants as quantitative
diagnostic tools of galaxies beyond the Local Group, the work presented in these papers
had still a fundamental deficiency. At the low resolution of 5 $\AA$ it is not possible
to use ionization equilibria for the determination of \teff\/ 
in the same way as in the high resolution 
work mentioned in the previous paragraph. Instead, spectral types were determined and
a simple spectral type - temperature relationship as obtained for the Milky Way was 
used to determine effective temperatures and then gravities and metallicities. Since
the spectral type - \teff\/ relationship must depend on metallicity (and also gravities),
the method becomes inconsistent as soon as the metallicity is significantly different from
solar (or the gravities are larger than for luminosity class Ia) and may lead to inaccurate
stellar parameters. As shown by  \citet{evans03}, the uncertainties introduced in this 
way could be significant and would make it impossible to use the FGLR for distance 
determinations. In addition, the metallicities derived might be unreliable. This posed
a serious problem for the the low resolution study of A supergiants in distant galaxies.

In this paper we present a solution to this problem and provide the first self-consistent
determination of stellar parameters and metallicities for A supergiants in galaxies 
beyond the Local Group based on the quantitative model atmosphere analysis of low 
resolution spectra. We apply our new method on 24 supergiants of spectral type
B8 to A5 in the Scultor Group spiral galaxy NGC 300 and obtain temperatures, gravities,
metallicities, radii, luminosities and masses. Note that while we have included a number 
of late type B supergiants in this work, we refer to the whole sample as ``A supergiants'' 
in the following, for reasons of simplicity. We will compare some of our results with the 
work by \citet{urbaneja05}, who performed a quantitative study of six early B-type 
supergiants (B0.5 to B3). We will refer to those objects as ``B supergiants''. The reason 
for this simplifying destinction is the difference in the analysis method. The temperatures 
of the B supergiants (in this nomenclature) are determined from the ionization equilibria 
of helium and silicon and metallicities result from the fit of individual spectral lines 
of $\alpha$-elements such as oxygen, magnesium, silicon and carbon, whereas the 
temperatures of the A supergiants are obtained from the Balmer jump and the metallicities 
are determined from the spectral synthesis of many overlapping lines of different species 
including heavy elements such as iron, chromium and titanium.

The paper is organized as follows. After a brief description of the observations in 
section 2 we provide a detailed discussion of the analysis method, i.e. the 
determination of \teff, $log~g$, metallicity, E(B-V), radius, luminosity and mass 
from the low resolution spectra and HST photometry, in section 3. Section 4 presents
the results of the analysis together with a discussion of their uncertainties. The 
metallicity of NGC 300 obtained from this work is discussed in section 5 and compared
with previous work. Section 6 investigates stellar properties and evolutionary time scales.
The new FGLR is presented in section 7 together with a detailed discussion of the
uncertainties. Conclusions and outlook towards future work are given in section 8.

\section{Observations}\label{sec_obs}

The spectroscopic observations were obtained with FORS1 at the ESO VLT in multiobject 
spectroscopy mode in two consecutive nights of excellent seeing conditions (typical 0.7 
arcsec, but long spells of 0.4 to 0.5 arcsec) on 2000 September 25 and 26 as part of the Auracaria Project (\citet{gieren05b}). Four 
fields (A, B, C, D) were observed in five exposures, each lasting 45 minutes. The spectral
resolution is approximately 5~\AA~and the spectral coverage extends from shortward of 
the Balmer jump to H${\beta}$ or beyond for most of our targets. The observational 
data set has been described by \citet{bresolin02}. The paper also contains finding 
charts, coordinates, photometry (ESO/MPI 2.2m WFI) and spectral types. Of the 62 confirmed
supergiants found with spectral types ranging from late O to F, 29 were of spectral type
B8 to A5 - the spectral type of interest for this study - and were not severely contaminated by HII emission or composite 
(indicating a contribution of a star of different spectral type). 24 of those had a
S/N per pixel of about 40 or better, sufficient for our quantitative analysis. This 
is the sample listed in Table 1 and selected for the work presented in the following. 
The target designation is the same as in \citet{bresolin02}. Note that a quantitative spectral
analysis of the early type B supergiants (spectral type B0 to B3) has already been published by 
\citet{urbaneja05}.

The analysis method presented in the next section makes use of the strength of the 
observed Balmer jump $D_{B}$. (Fortunately, FORS was developed with high UV throughput 
as a design goal, which turns out to be an enormous advantage for our study).
To measure $D_{B}$ we selected one 45 minute exposure for each 
of the observed fields (usually the exposure with the best seeing) and used standard 
stars to obtain a flux calibration of our spectra, where special care was given to the 
calibration in the near UV. While the single 
exposures are somewhat noisy, they
do allow for a rather accurate measurement of $D_{B}$ (see next section).
Slit losses due to the 1 arcsec slitlets can be corrected for by renormalizing the 
observed flux calibrated spectrum to the HST B-band photometry. They are unimportant 
to first order, since $D_{B}$ is a relative measurement of fluxes shortward and longward
of 3647~\AA. However, we have to worry about second order effects caused by the
wavelength dependence of seeing and resulting wavelength dependent slit losses. 
Fortunately, FORS is equipped with an atmospheric dispersion compensator, which was used
for our observations, thus, already correcting for slit losses caused by atmospheric dispersion.
The effects of wavelength dependent seeing were addressed in the following way. We 
measured the FWHM of our spectra on the detector perpendicular to the direction of dispersion
as a function of wavelength. This provided us with an estimate of the seeing as a function 
of wavelength. Assuming a Moffat-function (with exponent $\beta = 2$) for the PSF we were then 
able to calculate slit-losses at every wavelength from the measured seeing FWHM using
the simple formulae

%
%

\begin{equation}
\begin{array}{l}
\displaystyle PSF(\phi) \propto 1/\{1+(\phi/\alpha)^{2}\}^{\beta} \\
\displaystyle F_{\lambda}^{new} = F_{\lambda}^{old} (1 + (\alpha/w)^{2})^{0.5} \\
\displaystyle \alpha = 0.7769 FWHM(\lambda)
\end{array} 
\end{equation}

w is the slit half-width corresponding to 0.5 arcsec.

The relative flux corrections obtained in this way are small over the wavelength range, 
which we use for measuring or fitting the Balmer jump $D_{B}$ (3585~\AA\/ to 3876~\AA).
The logarithm of ratio of the correction factors at these two wavelength are 0.016,
0.070, 0.024 and 0.004 for fields A, B, C, D respectively, smaller or comparable to 
the stochastic errors caused by the photon noise. Assuming a Moffat-function with $\beta = 3$
yields very similar results.

The original photometry for our targets was obtained with the ESO 2.2m and the 
Wide Field Imager as a result of a multiepoch survey for Cepheids (\citealt{bresolin02}). 
\citet{bresolin04} have used these data for a study of blue supergiant
photometric variability. No significant variability has been found for our targets.

Almost all of our targets were observed with HST/ACS in an investigation of the effects
of blending on the Cepheid distance to NGC 300 (\citealt{bresolin05}), which were found 
to be negligible. With A supergiants being significantly brighter than Cepheids it is 
obvious that blending and multiplicity should not be an issue for this study, as is also
confirmed by a comparison of the ground-based and HST/ACS V and I photometry.
\citet{bresolin05} detected an off-set of 0.1 mag between ESO/WFI and HST/ACS
B-band photometry, which they attributed to a calibration uncertainty of the ground-based
data. We will use the HST/ACS photometry for our targets, where available. For three
objects, we do not have HST data. In these cases, we use the ESO/WFI B, V, I data but 
apply a 0.1 mag correction to the B-band.

\section{Analysis Method}\label{sec_analysis}

\begin{figure}[!]
 \begin{center}
  \includegraphics[width=0.90\textwidth]{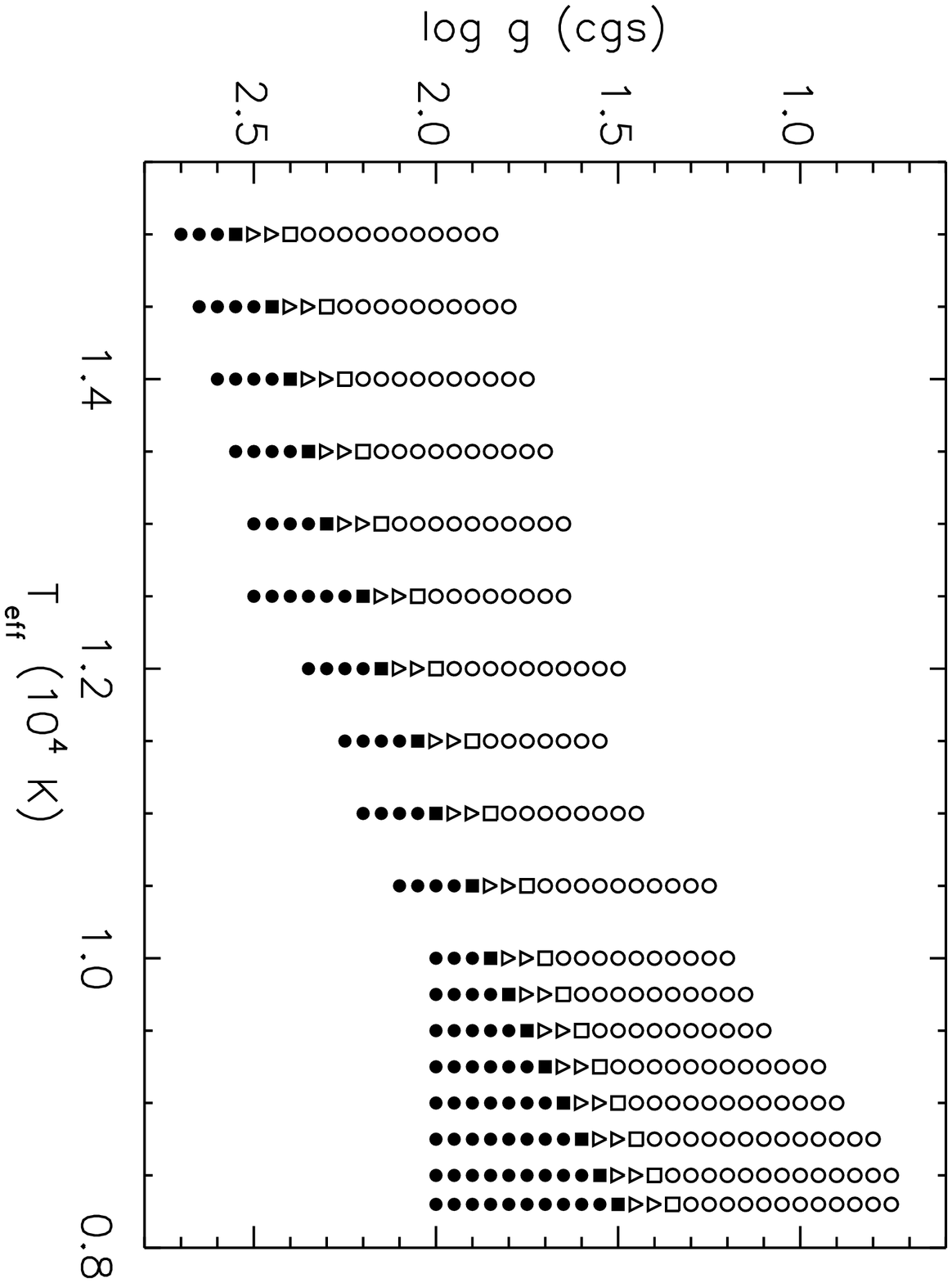}
  \caption[]{Parameter space of the model atmosphere grid used in this study. Different symbols refer
to the microturbulence values used for the models. Open circles: $v_{t}$ = 8 km/s; open squares
7 km/sec; open triangles 6 km/sec; filled squares 5 km/sec; filled circles 4 km/sec. At each grid
point metallicities from between [Z] = 0.3 to -1.30 are available (see text). \label{modelgrid} }
 \end{center}
\end{figure}

For the quantitative analysis of the spectra we will use the same combination
of line blanketed model atmospheres and very detailed NLTE line formation calculations
as \citet{przybilla06} in their high signal-to-noise and high spectral resolution
study of galactic A-supergiants. The model atmosphere assumptions and the individual atomic 
models used for the NLTE calculations have been described in this paper, which also 
contains a variety of tests with regard to the accuracy  and consistency that can be 
obtained. In the spectral range to be used in our work (3500~\AA\/ to 5500~\AA) these 
calculations reproduce the observed normalized spectra and the spectral energy 
distribution, including the Balmer jump, extremely well. For the four objects studied, 
\citet{przybilla06} calculated a relatively small grid of models centered around the 
expected parameters for each star followed by more line formation calculations to obtain 
element abundances, once the stellar parameters were determined. The large sample of 24 
objects in our study requires a different approach. We calculate an extensive, 
comprehensive and dense grid of model atmospheres and NLTE line formation covering the 
potential full parameter range of all the objects in gravity, effective temperature and 
metallicity. We then use the synthetic spectra and energy distributions obtained in this 
multi-dimensional parameter space to constrain the stellar parameters. The gravities and 
temperatures of the grid are displayed in Fig.~\ref{modelgrid}. 

At each grid point we have a set of models with line formation calculations for the 
following metallicities [Z] = log Z/Z$_{\odot}$: -1.30, -1.00, -0.85, -0.70, -0.60, -0.50,
 -0.40, -0.30, -0.15, 0.00, 0.15, 0.30. Z/Z$_{\odot}$ is the metallicity 
relative to the sun in a sense 
that the abundance for each element is scaled by the same factor relative to its solar 
abundance. Solar abundances were taken from \citet{grevesse98}. We note that by 
scaling abundances in this way we are not able to detect 
deviations from the solar abundance pattern but only general changes in 
metallicity. We are confident that it will be possible to extent this work to at least a 
study of the relative ratio of $\alpha$- to iron-group elements as a function of 
galactocentric distance. But at this stage we restrict our work to a discussion of stellar 
metallicities and the metallicity gradient and postpone a more detailed study to 
follow-up work.

A determination of metallicities always requires an estimate of the photospheric 
mictroturbulence velocity v$_{t}$ normally obtained from the constraint that 
weak and strong lines of each ion should yield the same element abundance. At our 
spectral resolution it is difficult to constrain mictroturbulence in this way, thus, 
we rely on the results of the high resolution work done for A supergiants in the Milky Way 
and Local Group galaxies, which, fortunately, shows a very consistent pattern of v$_{t}$ 
as a function of stellar gravities (\citealt{przybilla02}, \citealt{przybilla06}, 
\citealt{venn95a}, \citealt{venn95b}, \citealt{venn00}, \citealt{venn01}, 
\citealt{venn03}, \citealt{kaufer04}). At low 
gravities the values of v$_{t}$ are about 8 km/sec and decrease to 
lower values of about 4 km/sec at higher gravities. The values of v$_{t}$ adopted for this 
study, accordingly,  are also displayed in Fig.~\ref{modelgrid}.

The analysis of the each of the 24 targets proceeds in three steps. First, the stellar 
parameters (\teff\/ and $log~g$) are determined together with interstellar reddening 
and extinction, then the metallicity is determined  and finally, assuming a distance to 
NGC 300, we obtain stellar radii, luminosities and masses. For the first step,
a well established method to obtain the stellar parameters of supergiants of late B to 
early A spectral type is to use ionization equilibria of weak metal lines (OI/II; MgI/II;
NI/II etc.) for the 
determination of effective temperature \teff\/ and the Balmer lines for the 
gravities $log~g$. The most recent high resolution work cited in section~\ref{intro} has
very convincingly shown that very accurate results can be obtained in this way, when
detailed non-LTE line formation calculations are used. However, at the low resolution
of 5 \AA~the weak spectral lines of the neutral species disappear in the noise of the spectra
and an alternative technique is required to obtain temperature information.

One simple way
is to use the information from stronger lines in the spectrum which define the 
spectral type and then to apply a spectral type - temperature calibration based on the high
resolution quantitative spectroscopic work. This technique was used by \citet{bresolin01},
\citet{bresolin02} and \citet{kud03}. The disadvanteage of this method is obvious. As 
soon as the metallicity of the objects investigated with this low resolution method is 
significantly different from the metallicity of the objects which are used to calibrate 
the spectral type effective temperature relationship, substantial errors can be introduced.
For instance, objects at a lower metallicity will have weaker  characteristic metal lines
and will thus be assigned a spectral type earlier than an object of the same effective 
temperature but higher metallicity. Thus, without a priori knowledge of the metallicity 
this method might indeed introduce large uncertainties. \citet{evans03} have investigated
this effect for A supergiant of luminosity class II in the Magellanic Clouds by studying 
the CaII lines and found significant effects. However, our targets are more luminous 
(luminosity class I) and the spectral type classification used in the papers mentioned above
did not use the CaII lines, but the metal line spectrum in the range from 4400 to 4700 \AA.
Thus, an extended investigation is needed.

\begin{figure}
\begin{center}
\includegraphics[width=0.85\textwidth]{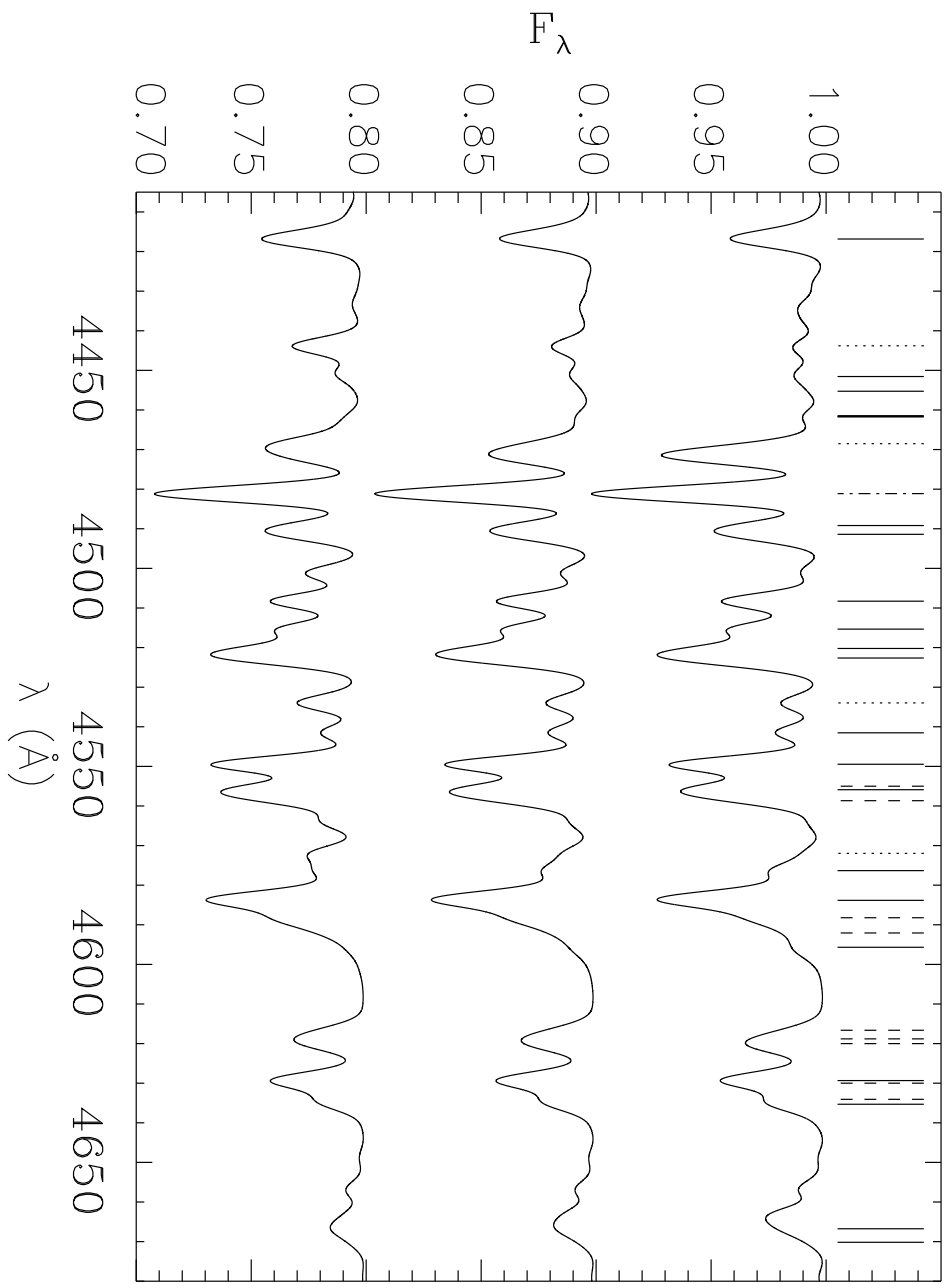}
\caption{The temperature-metallicity degeneracy of spectral types. 
Synthetic spectra of three A supergiant model atmospheres degraded to 5~\AA~resolution
 with (\teff, $log~g, [Z]$) = (10500~K, 1.40, 0.15) (top); (9500~K, 1.20, -0.30) (middle); 
 (8750~K, 1.00, -0.50) (bottom). The latter two are vertically shifted by 0.1 and 0.2. Metal
line identifications are given by the bars at the top (solid: FeII; dashed: CrII; dotted: TiII;
dashed-dotted: MgII). Note the change of the strength of the HeI line at 4471~\AA~from the hottest 
to the cooler models.} 
\label{degen}
\end{center} 
\end{figure}

Fortunately, the comprehensive model atmosphere and line formation grid described above 
makes it easy to investigate the degeneracy of the metal line spectrum with temperature and 
metallicity. All we have to do is to follow the iscontours of constant Balmer line 
equivalent widths in the $(log~g, \log$~\teff) plane (see Fig.~\ref{isohdbgt}) (along 
which the Balmer lines as a gravity indicators have identical profiles at 5 \AA 
~resolution) and to compare synthetic metal line spectra of models with different \teff\/ 
and [Z]. A rather dramatic example obtained in this way is shown in Fig.~\ref{degen} with 
metal line spectra of three models at different pairs of \teff\/ and $log~g$, where the 
Balmer line profiles are exactly the same. The models have different metallicity [Z].
The hottest model at \teff\/ = 10500~K with a metallicity slightly higher than solar
([Z] = 0.15) has an almost identical metal line spectrum as the coolest model at 
\teff\/ = 8750~K with a metallicity significantly lower than solar  ([Z] = -0.50). 

Thus, while all three models have exactly the same Balmer line profiles and would fit 
the observed Balmer lines of a hypothetical star to be analyzed, we would have a hard 
time to independently determine temperature and metallicity from the rest of the spectrum, which looks 
very similar for all three models. Fortunately, the strength of HeI $\lambda$ 4471 allows, in principle, to 
distinguish between the hottest model and the two cooler ones, and this will be one of the 
alternative techniques that we will use in the following. However, it seems impossible
to extract the temperature difference between the two cooler models from the spectra 
displayed. We note the slightly enhanced strengths of the TiII lines at low temperature, 
but using these lines as an independent temperature indicator while simultaneously fitting the Balmer lines would be based on an
assumption on the abundance ratio of titanium to iron, and, thus, the ratio of $\alpha$
 -elements to iron-group elements. While the investigation of the possible variation of 
the ratio as a function of galaxy evolution, is beyond the scope of this paper (see above),
 it will 
certainly become important in future work. We, thus, need a temperature 
indicator independent from any a priori assumptions about general metallicity and 
relative abundance ratios of metals. This is in particular important, as the spiral 
galaxies, which we intend to investigate with our spectroscopic technique, are assumed 
to have significant metallicity gradients and possibly also gradients, for instance, in 
their $\alpha$ to iron-group element ratios.

\begin{figure}
\begin{center}
\includegraphics[width=0.90\textwidth]{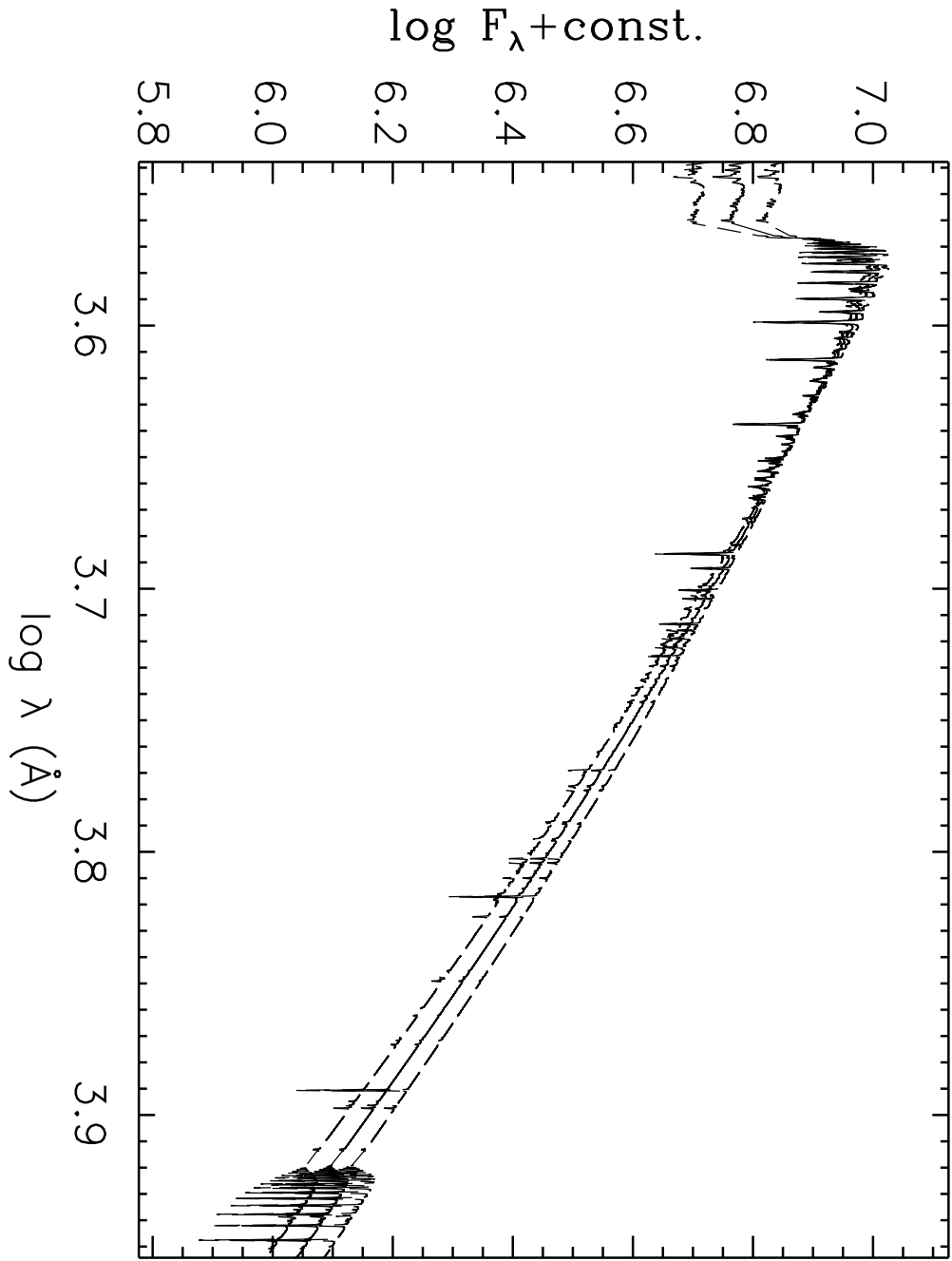}
\caption{Energy distributions of the models of Fig.~\ref{degen}. The fluxes are normalized at
4380 \AA. The hottest model has the steepest SED.} 
\label{seddegen}
\end{center}
\end{figure}

\begin{figure}
\begin{center}
\includegraphics[width=0.90\textwidth]{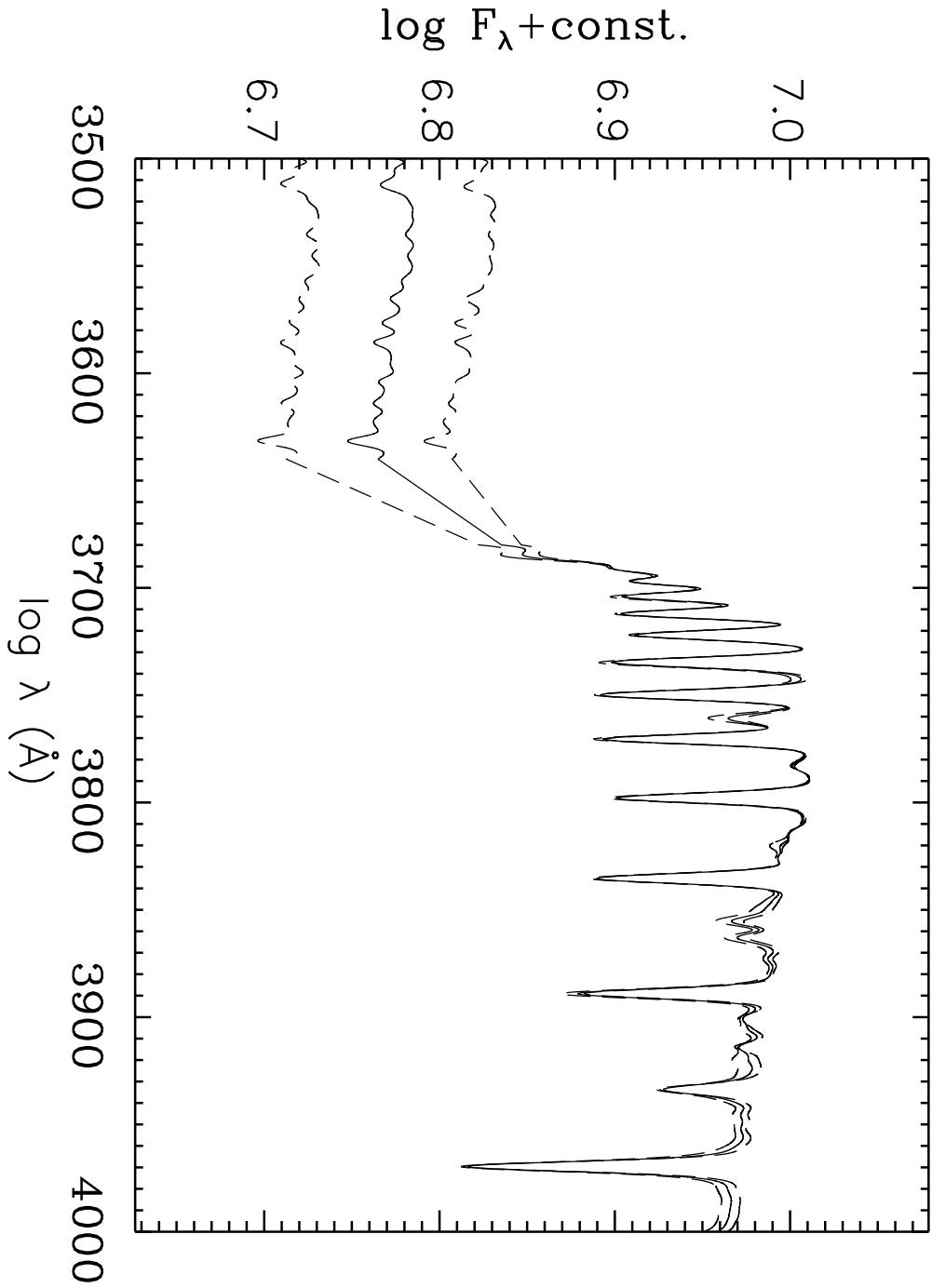}
\caption{Balmer jump fluxes of the models of Fig.~\ref{degen}. The fluxes are normalized at
3790 \AA. The hottest model has the smallest Balmer jump.}
\label{dbdegen}
\end{center}
\end{figure}

A very obvious way out of this dilemma is the use of the spectral energy distributions
(SEDs) and here, in particular of the Balmer jump $D_B$. Fig.~\ref{seddegen} and ~\ref{dbdegen}
show, how the degeneracy of the temperature vs. metallicity diagnostic is broken. Very 
obviously, the SEDs and $D_B$'s of the three models with almost identical metal line spectra are
significantly different. While the observed photometry from B-band to I-band will be 
used to constrain the interstellar reddening, we can use $D_B$ as a temperature diagnostic.
For stars of spectral type A this idea is certainly not new and goes back to the classical 
work by Chalonge and Divan in the 1950's. For A supergiants it has been successfully 
investigated and applied by \citet{groth61}, \citet{aydin72}, and \citet{kud73}.

The basis for our  diagnostic technique is given in 
Fig.~\ref{isohdbgt}. It shows isocontours of constant H${\delta}$ equivalent width and of 
constant values of the Balmer jump $D_{B}$. The Balmer jump is defined as below

   \begin{equation}
   \begin{array}{l}
\displaystyle D_{B} = \langle log (F_{\lambda}^{long}) \rangle - \langle log (F_{\lambda}^{short}) \rangle \\
\displaystyle \langle log (F_{\lambda}^{long}) \rangle = \lbrace log F_{3782\AA} + log F_{3814\AA} + log F_{3847\AA} + log F_{3876\AA} \rbrace /4 \\
\displaystyle \langle log (F_{\lambda}^{short}) \rangle = \lbrace \sum_{i=1}^{N} log (F_{\lambda_{i}}) \rbrace /N, ~3585\AA \le \lambda_{i} \le 3627\AA
\end{array} 
   \end{equation}

The wavelengths for the individual fluxes on the longward side are selected to avoid 
the wings of the Balmer lines and stronger metal line features. In this definition, 
the dependence of $D_B$ on [Z] is very weak and can be neglected, as an inspection of
our model grid shows. The same is true for the strengths of the Balmer lines and, thus,
the diagnostic diagram of Fig.~\ref{isohdbgt} is metallicity independent. While the two 
sets of isocontours are not perfectly orthogonal, they certainly lead to well defined 
intersection points except very close to the Eddington-limit.
 
\subsection{Effective temperature, Gravity and Reddening}

\begin{figure}
\begin{center}
\includegraphics[width=0.90\textwidth]{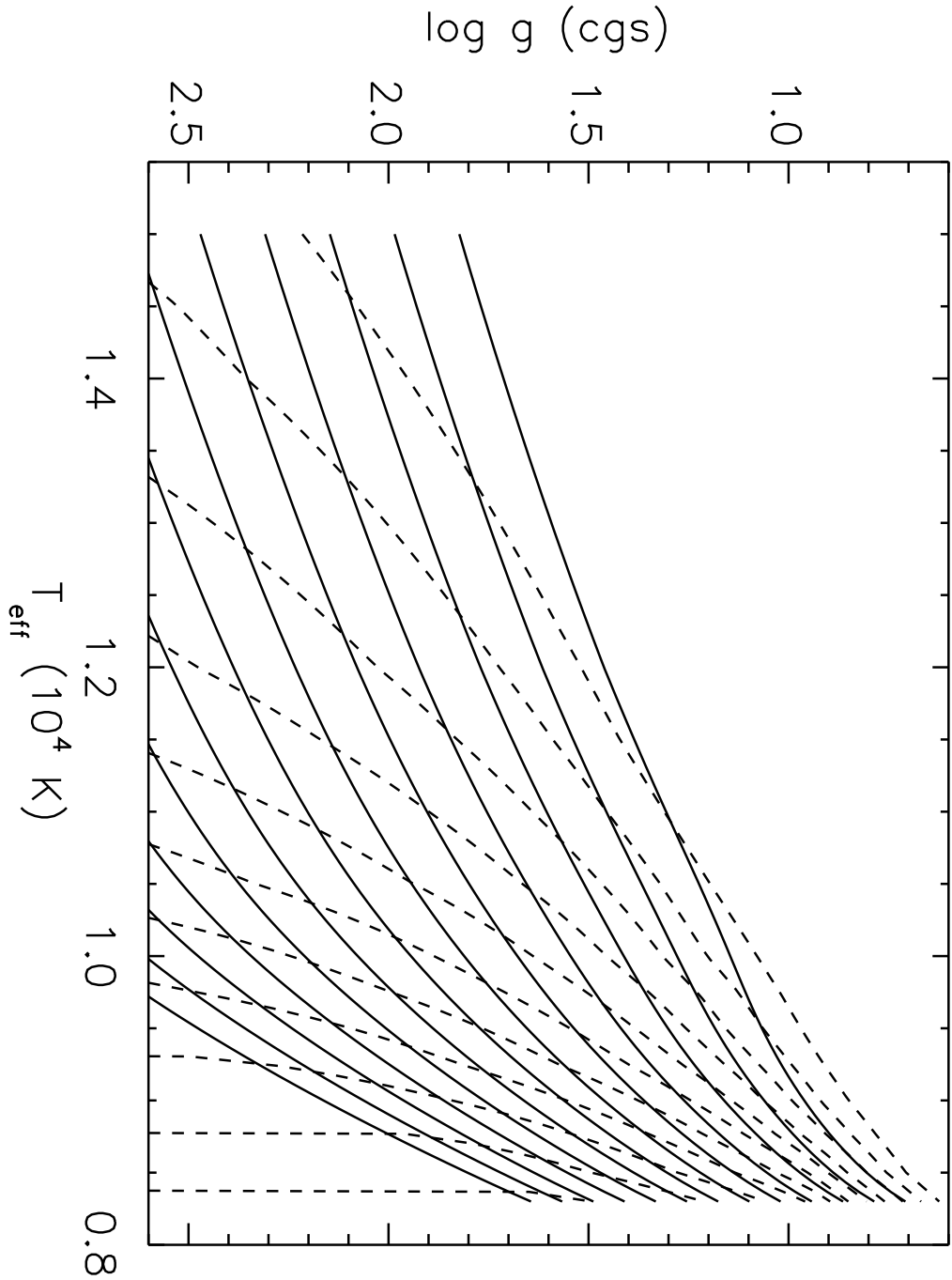}
\caption{Isocontours of H${\delta}$ (solid) and $D_{B}$ (dashed) in the 
$(log~g, log$~\teff) plane. H${\delta}$ isocontours start with 1 \AA~equivalent width and 
increase in steps of 0.5 \AA. $D_{B}$ isocontours start with 0.1 dex and increase by 0.1 dex.
The iscontours are calculated fro the model atmosphere and non-LTE line formation grid
described in section 3 and Fig. 1.} 
\label{isohdbgt}
\end{center}
\end{figure}

Guided by Fig.~\ref{isohdbgt} we apply the following diagnostic technique to determine 
effective temperature and gravity. We use the normalized spectra of our targets to 
interactively fit the Balmer lines profiles H$_{10}$ to H$_{\gamma}$ with our model 
spectra grid in the $(log~g, \log$~\teff) plane (the fit curves obtained in this 
way follow very closely the isocontours of Balmer lines as in the example of 
Fig.~\ref{isohdbgt}). Then, we fit the observed SED and the Balmer jump in the same 
diagnostic plane. Our multi-color photometry from B-band to I-band compared with 
the model fluxes allows to determine reddening and extinction and the calibrated
FORS fluxes around the Balmer jump (see section ~\ref{sec_obs}) fitted with the 
model fluxes yield a second fit curve for $D_{B}$ (again very similar to the 
isocontours in Fig.~\ref{isohdbgt}). The intersection of the two fit curves defines
temperature and gravity.

\begin{figure}
\plotone{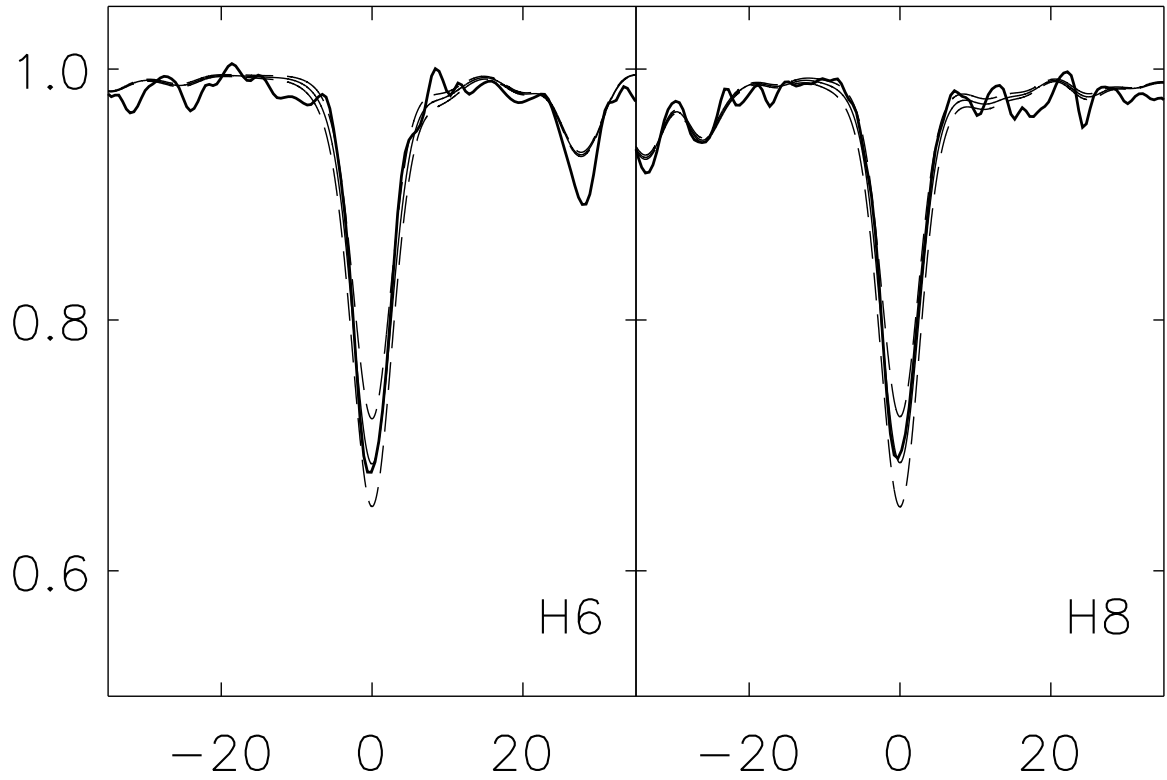}
\caption{Model atmosphere fit of two observed Balmer lines of target No. 21 of Table 1 for 
\teff\/ = 10000~K and $log~g$ = 1.55 (solid). Two additional models with same \teff\/
but $log~g$ = 1.45 and 1.65, respectively, are also shown (dashed). Fits of other Balmer lines 
such as H$_{\gamma,8,9,10}$ are similar. Note that the calculated line profiles were folded 
with a Gaussian of 5~\AA~FWHM to account for instrumental broadening}
\label{fitbalm}
\end{figure}

\begin{figure}
\begin{center}
\includegraphics[width=0.90\textwidth]{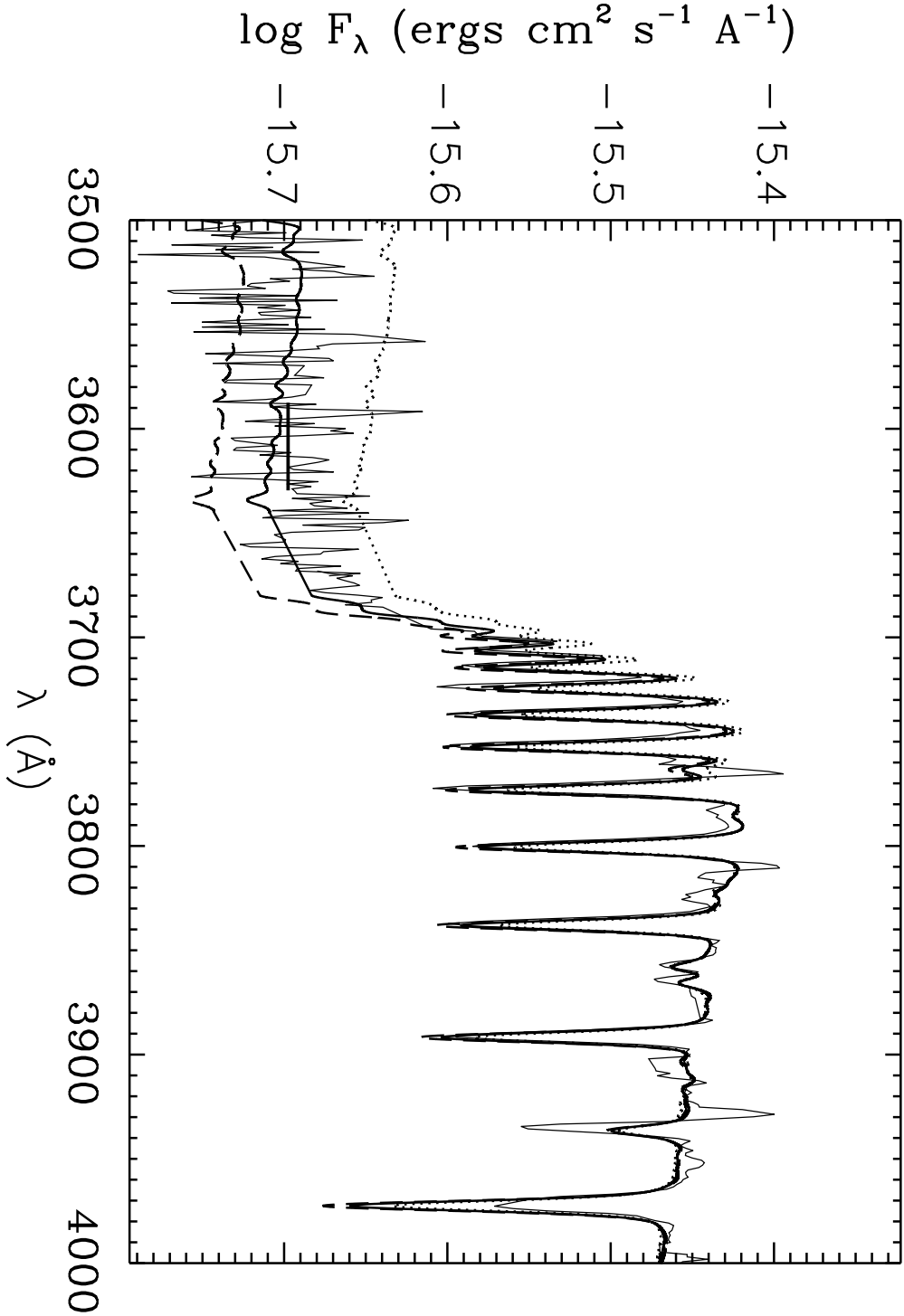}
\caption{Model atmosphere fit of the observed Balmer jump of target No. 21 of Table 1 for 
\teff\/ = 10000~K and $log~g$ = 1.55 (solid). Two additional models with the same $log~g$
but \teff\/ = 9750~K (dashed) and 10500~K (dotted) are also shown. The horizontal bar at 
3600~\AA~represents the average of the flux logarithm over this wavelength interval, which 
is used to measure D${_B}$ (see text).}
\label{fitsed}
\end{center}
\end{figure}

\begin{figure}
\begin{center}
\includegraphics[width=0.90\textwidth]{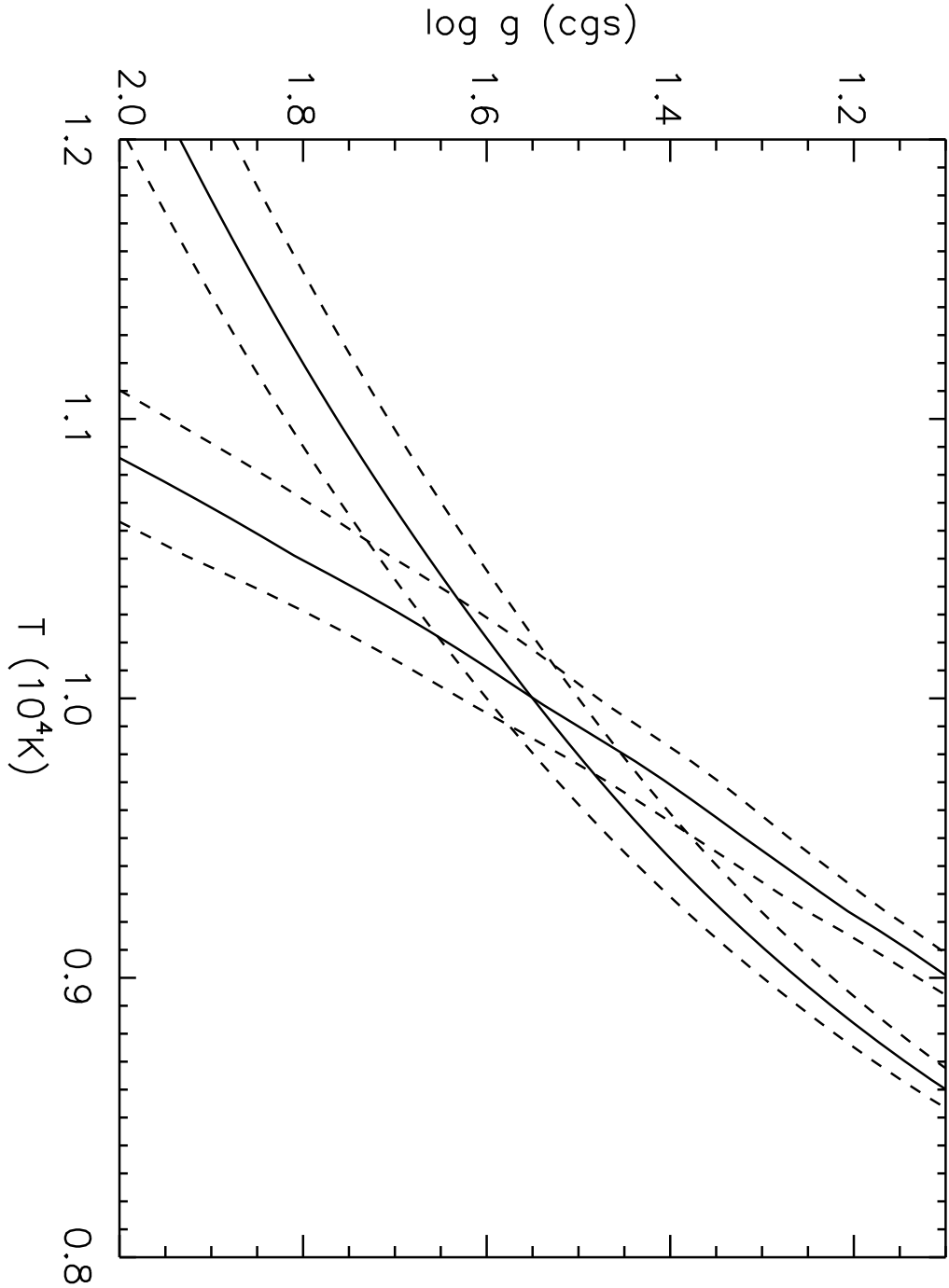}
\caption{($log~g, log$ \teff\/)-fit diagram for target No.21 of Table 1. The solid curves 
represent the best fits of the Balmer jump and the Balmer lines. The dashed curves 
correspond to the maximum fitting errors. }
\label{lgtfit}
\end{center}
\end{figure}

An example is given in Figs.~\ref{fitbalm}, ~\ref{fitsed}, 
and  ~\ref{lgtfit} for object No. 21 of our sample. At a fixed value of effective 
temperature the Balmer lines allow for a determination of $log~g$ within 0.05 dex
uncertainty. The uncertainty of Balmer jump $D_{B}$ measured from the SED fit can
be (conservatively) estimated to be 0.02 dex corresponding to a temperature 
uncertainty $\rm^{+200}_{-150}$K at the fixed gravity of 1.55 at \teff\/ = 10000~K.
(The values of $D_{B}$ obtained from SED fit are given in Table 1).
Applying the same fit procedure at different effective temperatures in the
$(log~g, log$~\teff) plane leads to the fit diagram of Fig.~\ref{lgtfit}. While
the uncertainties at a fixed pair of \teff\/ and $log~g$ are rather small, the fact
that the two sets of fit curves for the Balmer lines and the Balmer jump are not 
orthogonal leads to the significantly larger error box, which is obtained from the 
intersection of the dashed curves. The uncertainty of our low resolution analysis 
obtained in this way are $\rm^{+600}_{-470}$K for \teff\/ and $\rm^{+0.19}_{-0.17}$ 
dex for $log~g$, respectively. This is roughly a factor of two larger than what can 
be obtained with high quality, high resolution spectra and the use of ionization 
equilibria from weak lines (ses \citealt{przybilla06}) but still good enough to 
determine metallicities [Z] with a reasonable accuracy, as we will demonstrate below.
The fit of \teff\/ and $log~g$ also yields the value of interstellar reddening
E(B-V) = 0.13 mag from a comparison of computed model fluxes with fluxes obtained
from HST photometry (see \citealt{bresolin04} for the photometric data). As for
the calculation of the reddening law we use the formulae given by \citet{cardelli89}
and R$_{V}$ = 3.1 for the the ratio of visual extinction A$_{V}$ and reddening E(B-V).

\begin{figure}
\begin{center}
\includegraphics[width=0.90\textwidth]{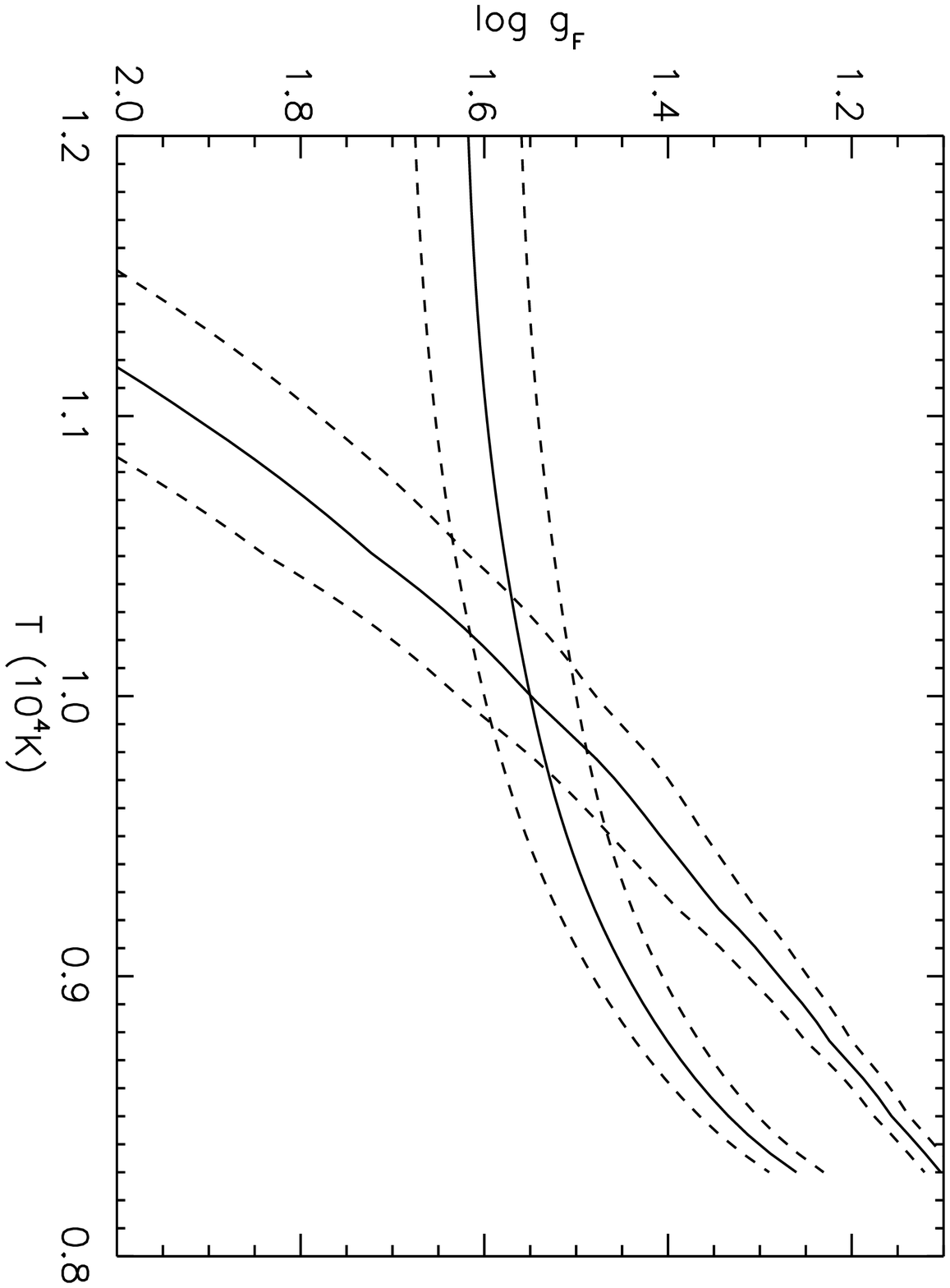}
\caption{Same as Fig.~\ref{lgtfit} but for the flux weighted gravity $log~g_{F}$ 
instead of gravity $log~g$.}
\label{lgftfit}
\end{center}
\end{figure}

The accurate determination of \teff\/ and $log~g$ is crucial for the use of
A supergiants as distance indicators using the relationship between absolute 
bolometric magnitude $M_{bol}$ and flux weighted gravity $log~g_{F}$ defined as

\begin{equation}
        log~g_{F} = log~g - 4log T_{\!\mbox{\scriptsize\em eff,4}}
   \end{equation}

where $T_{\!\mbox{\scriptsize\em eff,4}} =  T_{\!\mbox{\scriptsize\em eff}}$/10000K(see \citealt{kud03}). 
The relatively large uncertainties obtained with our fit 
method may casts doubts whether $log~g_{F}$ can be obtained accurately enough.
Fortunately, the non-orthogonal behaviour of the fit curves in
Fig.~\ref{lgtfit} leads to errors in \teff\/ and $log~g$, which are correlated in a way 
that reduces the uncertainties of $log~g_{F}$. This is demonstrated in
Fig.~\ref{lgftfit}, which shows the corresponding fit curves of the Balmer lines
and D$_{B}$ in the $(log~g, log$~\teff) plane. We obtain much smaller 
uncertainties of $log~g_{F}$ for target No. 21, namely $\rm^{+0.09}_{-0.08}$.
We will discuss the physical reason behind this in section~\ref{sec_fglr}.

We have applied this diagnostic technique on all targets, for which we were able 
to obtain observations of the Balmer jump.
In two cases (target No. 0 and 10), the set-up of the FORS slitlets did not allow to 
obtain spectral information shortward of the Balmer edge. For those, we use the strong
temperature dependence of the observed HeI lines to constrain \teff\/ (see Fig.~\ref{degen}). 
For targets No. 2 and 4 we have used the observed HeI lines as an additional temperature 
criterium, since for both of them the intersections of the $D_{B}$ and Balmer line fit 
curves are not well defined. The errors in \teff\/ obtained in these four cases are 
comparable to those obtained with the Balmer jump for most of the other stars. The HeI 
lines provide a very useful alternative diagnostic tool, however this technique works 
only at high enough temperatures and low gravities. The
results for \teff\/, $log~g$, $log~g_{F}$ and E(B-V) for all targets are 
summarized in Table 1.

\begin{figure}
\plotone{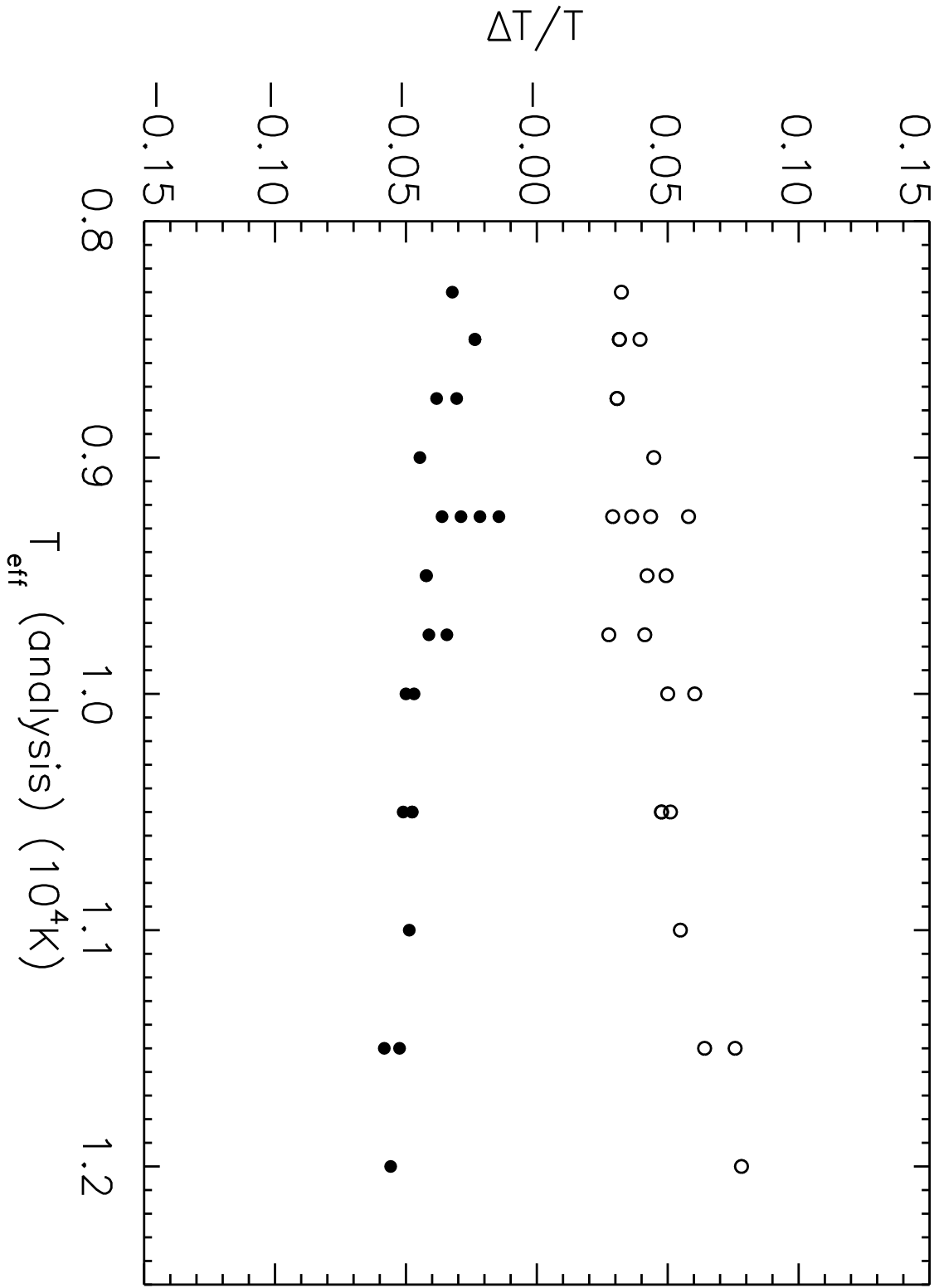}
\caption{Effective temperature uncertainties $\pm \Delta$\teff/\teff\/ as a function of effective 
temperatures (see Table 1). Open and solid symbols refer to positive and negative errors, respectively.}
\label{terror}
\end{figure}

Generally, the uncertainties for \teff\/, $log~g$, and $log~g_{F}$ are similar to 
those obtained for our example target. However, for the effective temperature errors 
$\Delta$\teff/\teff\/ there is a slight trend with temperature as shown by 
Fig.~\ref{terror}, which comes from the fact that the Balmer jump becomes less 
temperature dependent at higher temperatures (see Fig.~\ref{isohdbgt}). On average, 
temperatures are accurate to 5 percent. This is an uncertainty roughly twice as large 
as for the analysis of high resolution and high S/N spectra of similar objects 
(see \citealt{przybilla06}). The uncertainties for gravities and flux weighted 
gravities are 0.2 dex and 0.1 dex on average, again twice as large as for the analysis 
of high quality spectra. This is a result of the larger uncertainty of the temperature 
determination.

\subsection{Metallicity}

\begin{figure}
\plotone{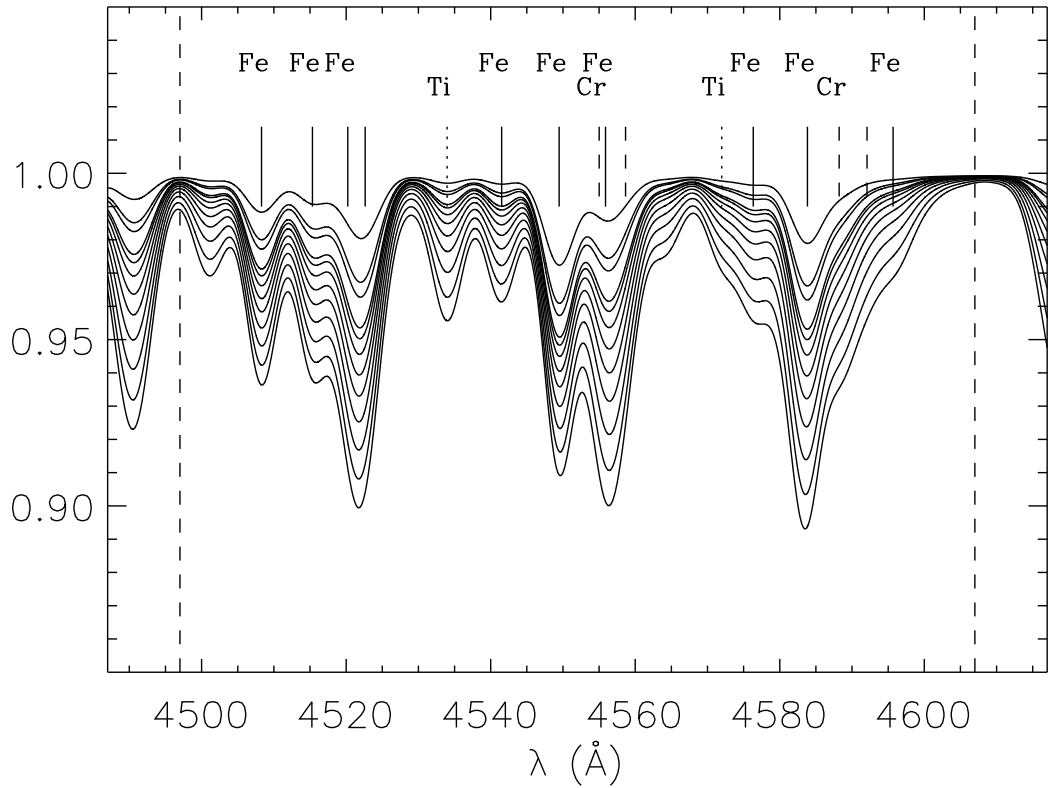}
\caption{Synthetic metal line spectra calculated for the stellar parameters of target 
No.21 as a function of metallicity in the spectral window from 4497~\AA~to 4607~\AA. 
Metallicities range from [Z] = -1.30 to 0.30, as described in the text. The dashed 
vertical lines give the edges of the spectral window as used for a determination of 
metallicity. The synthetic spectra were folded with a Gaussian of 5~\AA~FWHM to 
account for the instrumental broadening. Rotational broadening is negligible 
for A supergiants at this resolution and not taken into account.}
\label{met1}
\end{figure}

\begin{figure}
\plotone{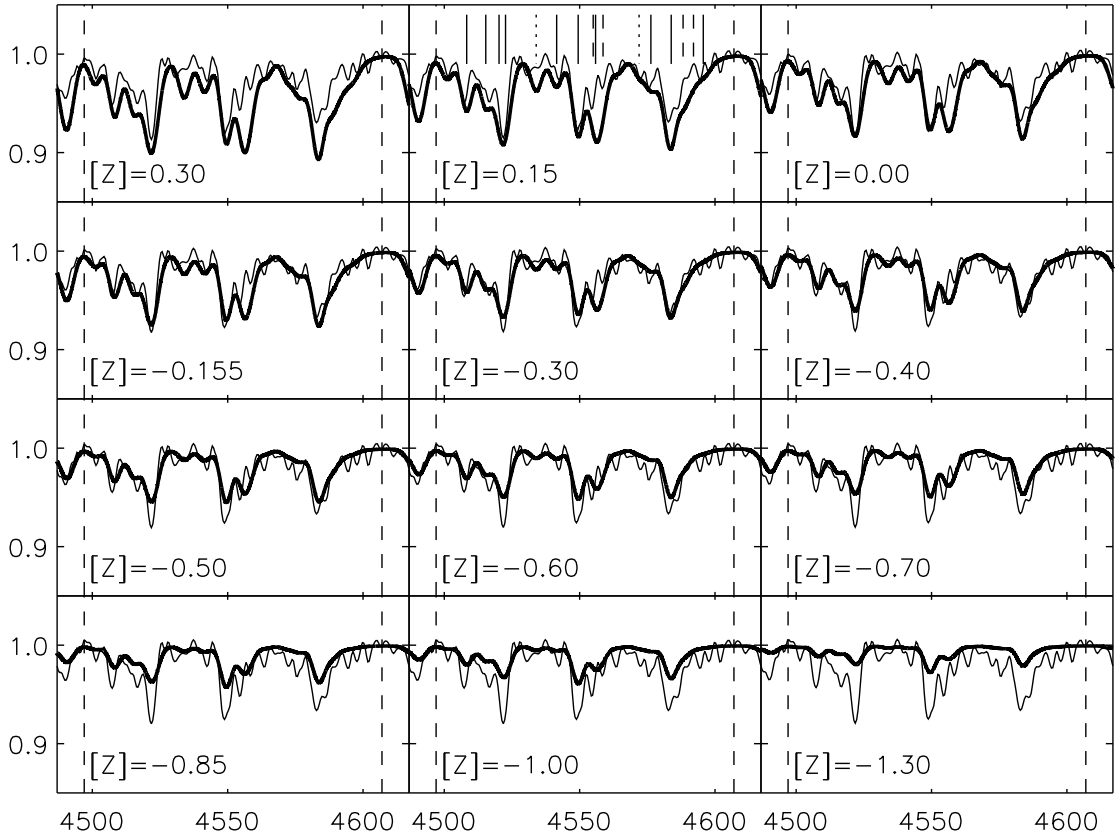}
\caption{Observed spectrum of target No. 21 for the same spectral window as 
Fig.~\ref{met1} overplotted by the same synthetic spectra for each metallicity 
separately. The line identification symbols plotted for [Z] = 0.15 are the same 
as in Fig.~\ref{met1}. For the renormalization of the observed spectrum at each 
different metallicity see discussion in the text. The abscissae are wavelengths in \AA.}
\label{met2}
\end{figure}

Knowing the stellar atmospheric parameters \teff\/ and $log~g$ we are able to 
determine stellar metallicities by fitting the metal lines with our 
comprehensive grid of line formation calculations. The fit procedure proceeds 
in several steps. First, we define spectral windows, for which a good definition 
of the continuum is possible and which are relatively undisturbed by flaws in 
the spectrum (for instance caused by cosmic events) or interstellar emission 
and absorption. A typical spectral window used for all of our targets is the 
wavelength interval 4497~\AA\/  $\le \lambda \le$ 4607~\AA. Fig.~\ref{met1} shows 
the synthetic spectrum calculated for the atmopsheric parameters of target 
No. 21 (the example of the previous subsection) and for all the metallicities 
of our grid ranging from -1.30 $\le$ [Z] $\le$ 0.30. It is very obvious that the 
strengths of the metal line features are a strong function of metallicity. In 
Fig.~\ref{met2} we show the observed spectrum of target No. 21 in this spectral 
region overplotted by the synthetic spectrum for each metallicity. We use separate 
plots for each metallicity, because the optimal relative normalization of the observed and 
calculated spectra is obviously metallicity dependent. We address this problem by 
renormalizing the observed spectrum for each metallicity so that the synthetic 
spectrum always intersects the observations at the same value at the two edges of 
the spectral window (indicated by the dashed vertical lines). The scaling of the 
observed spectrum between the two edges is done by a simple linear interpolation. 
The next step is a pixel-by-pixel comparison of calculated and normalized observed 
fluxes for each metallicity.

\begin{equation}
        (O-C)^{2}([Z])_{i} = \lbrack \sum^{n_{pix}}_{j=1} (F([Z])_{j}^{obs} - F([Z])_{j}^{calc})^{2} \rbrack /n_{pix}
   \end{equation}

\begin{figure}
\plotone{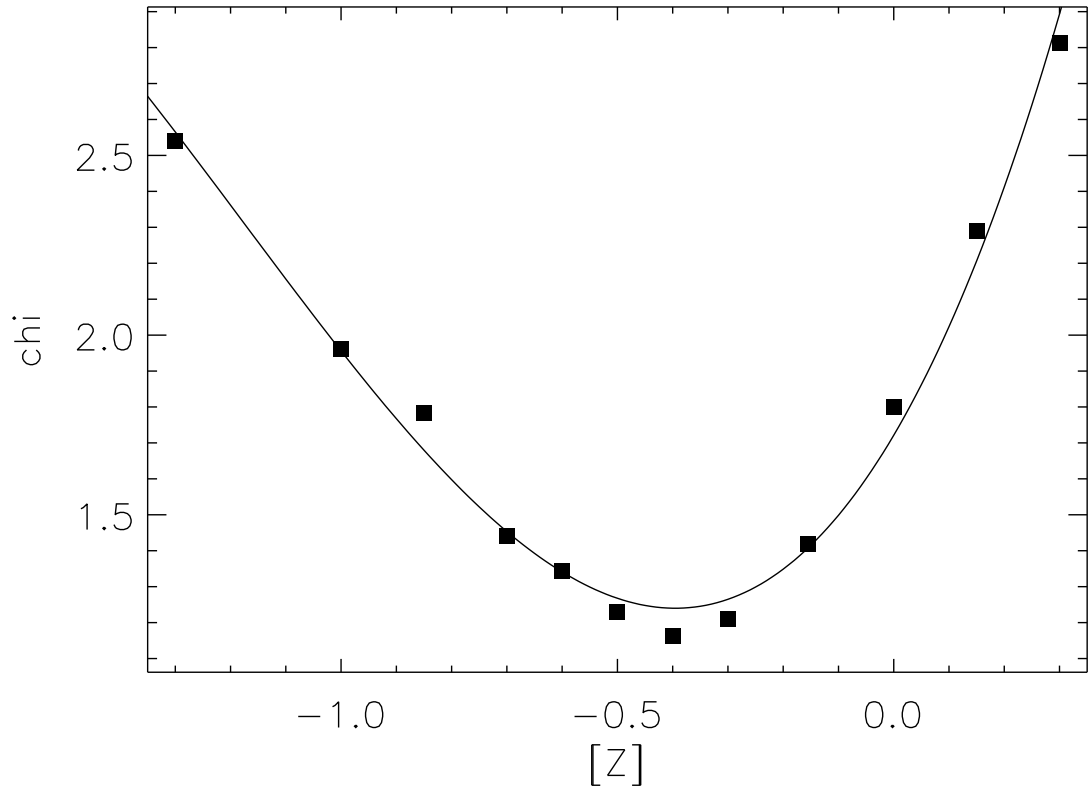}
\caption{$\chi ([Z])_{i}$ as obtained from the comparison of observed and 
calculated spectra in Fig.~\ref{met2}. The solid curve is a third order polynomial 
fit.}
\label{met3}
\end{figure}

\begin{figure}[!]
 \begin{center}
  \includegraphics[width=0.45\textwidth]{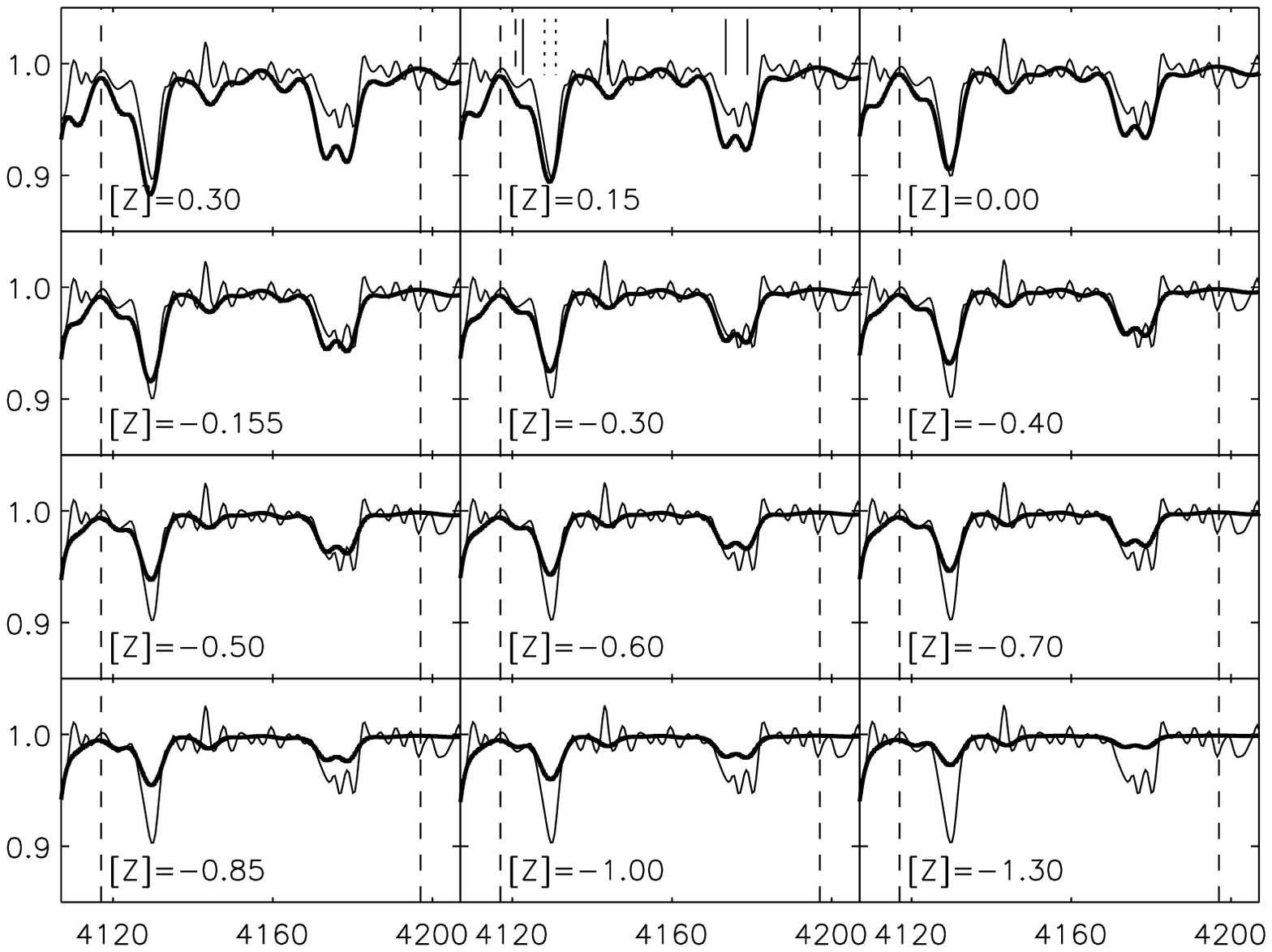}
  \includegraphics[width=0.45\textwidth]{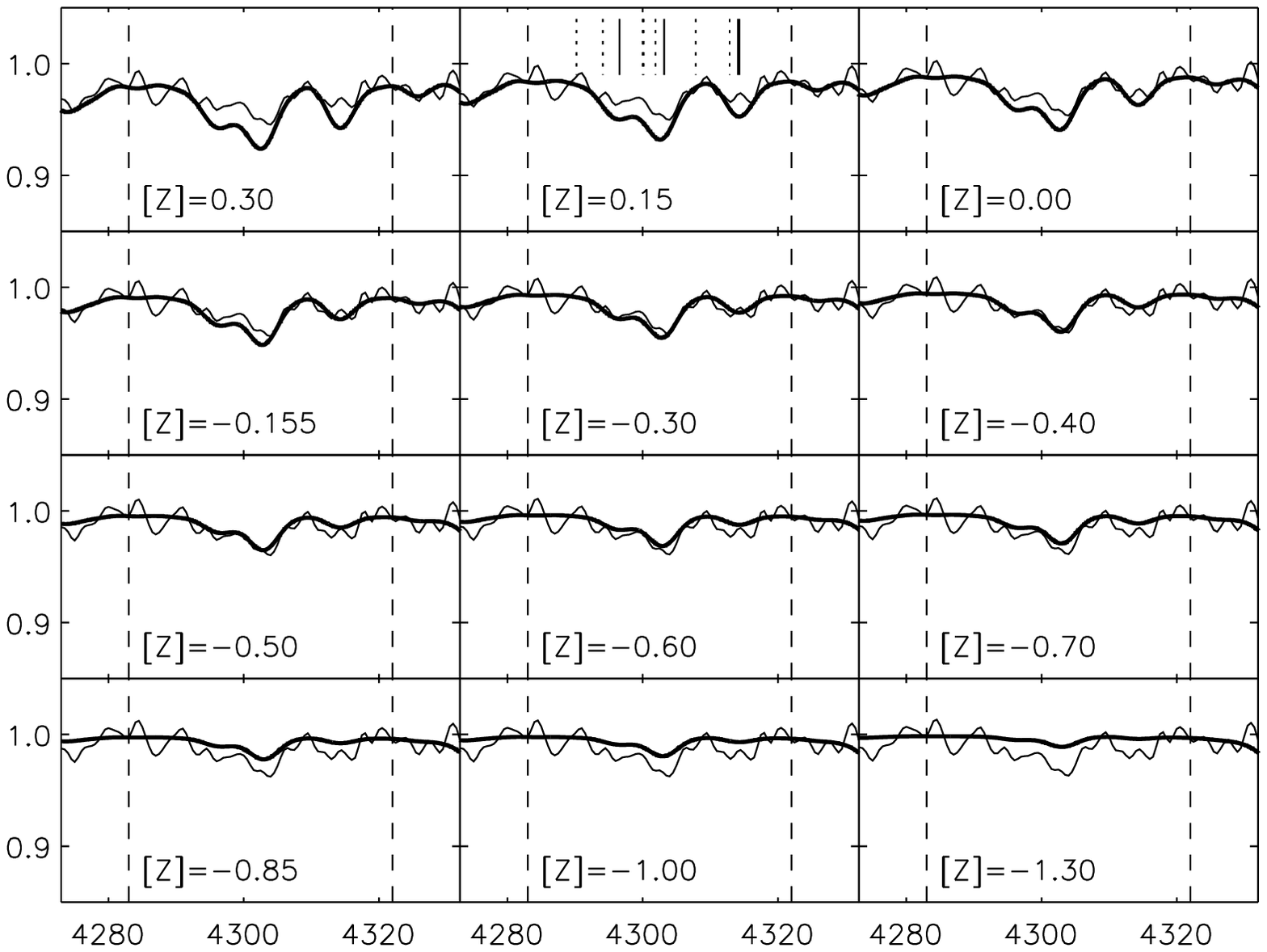}
  \includegraphics[width=0.45\textwidth]{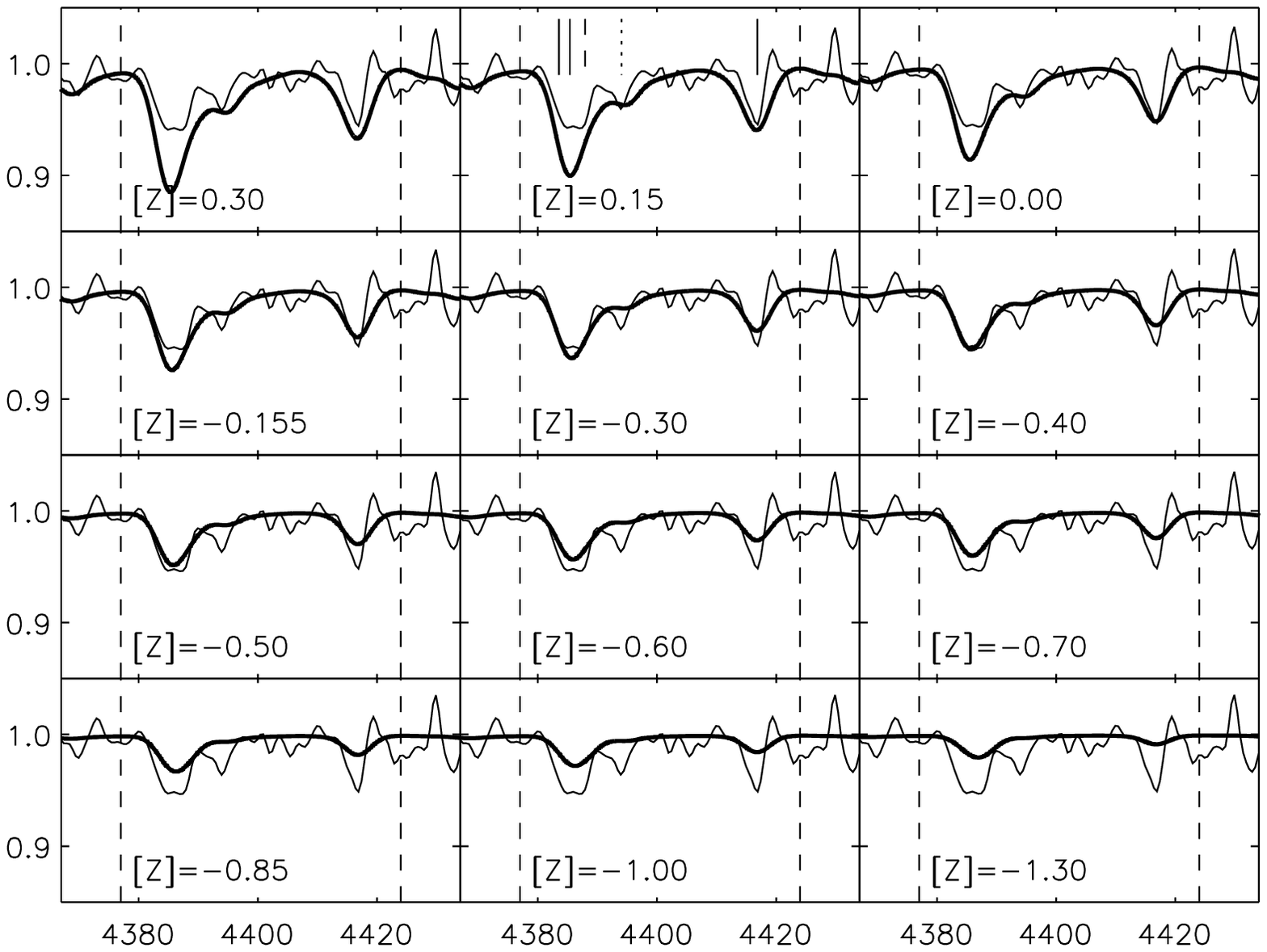}
  \includegraphics[width=0.45\textwidth]{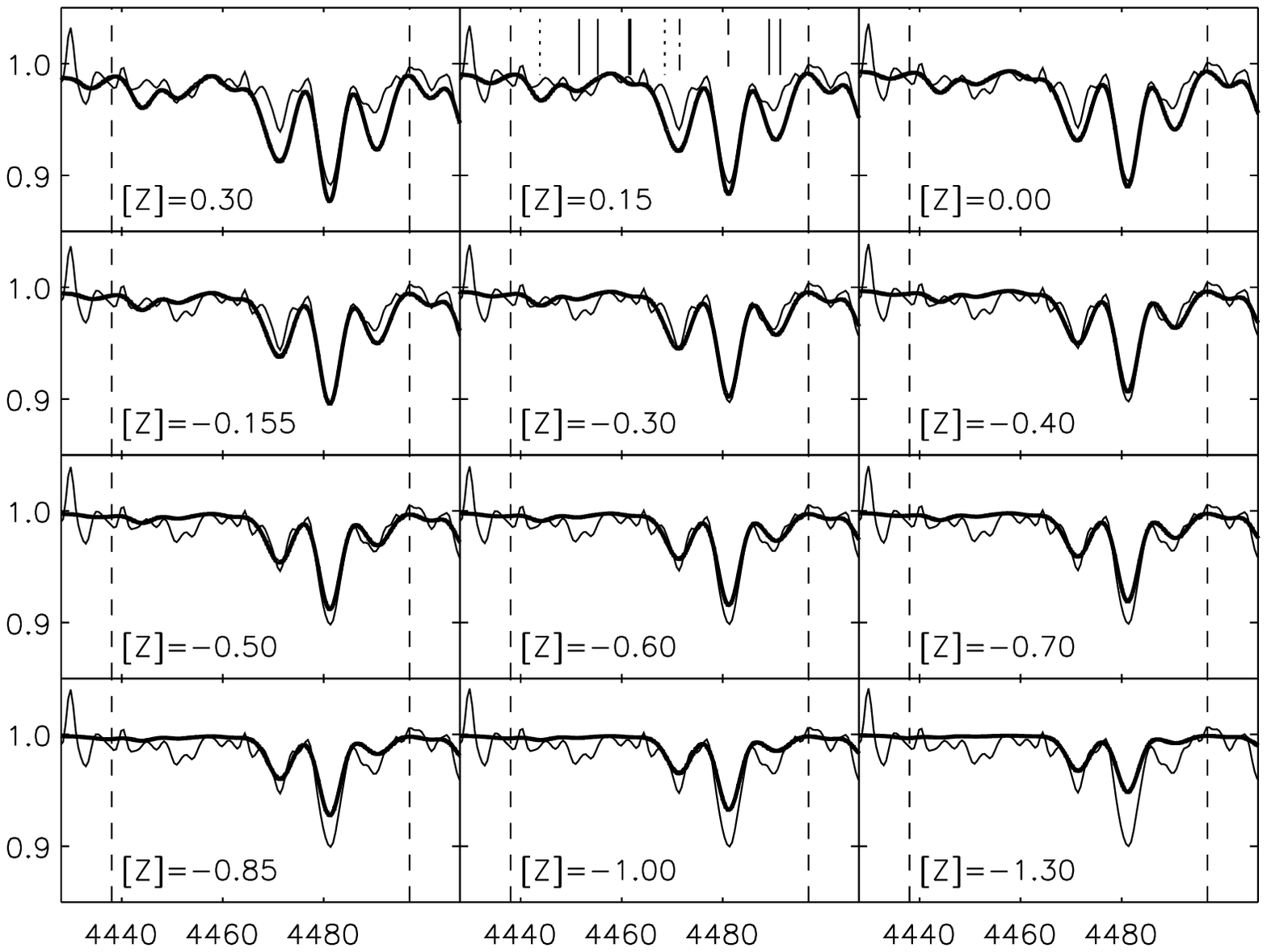}
  \includegraphics[width=0.45\textwidth]{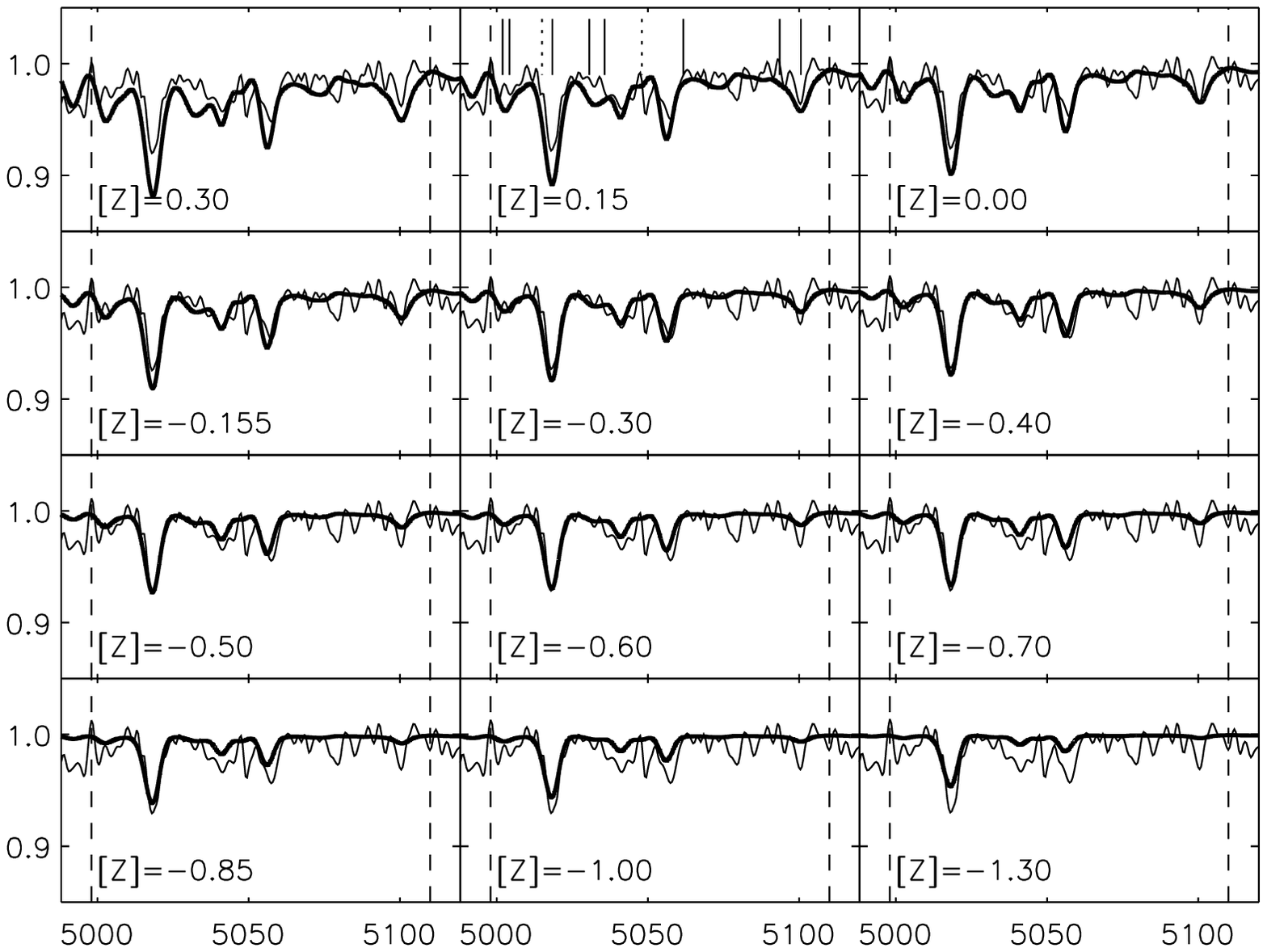}
  \includegraphics[width=0.45\textwidth]{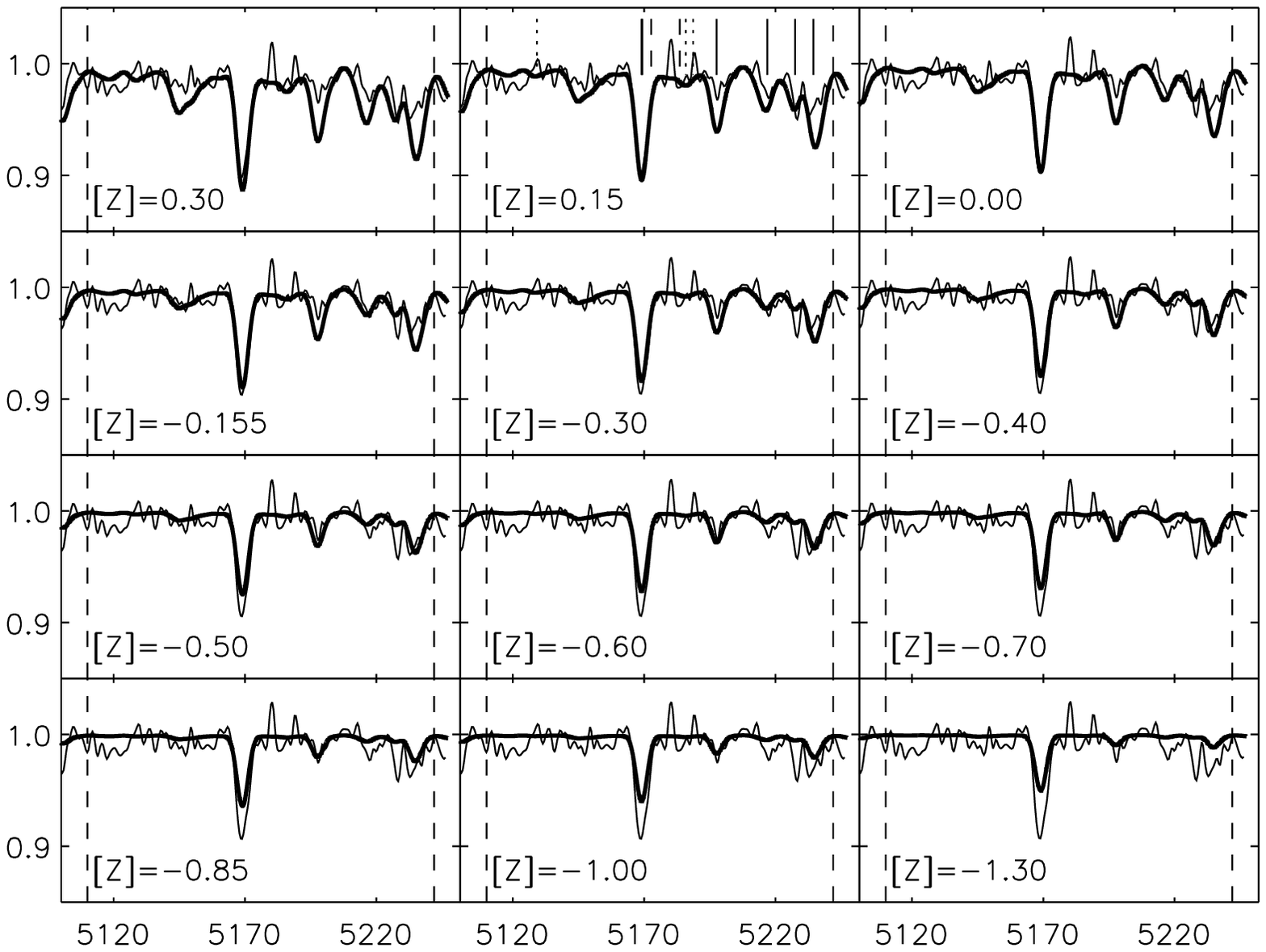}
 \caption[ ]{Determination of metallicity for target No. 21 in six more 
spectral windows. Line identifications are again given for the plots with 
[Z] = 0.15. Solid bars refer to FeII, dotted bars represent TiII in 
windows 2, 3, 4, 6, SiII in window 1 and HeI in window 5. Dashed bars 
indicate HeI in window 1 and 3 and  MgII in window 4 and 6. The 
dashed-dotted bar in window 4 is HeI. The abscissae are wavelengths in \AA.}
{\label{met4}}
 \end{center}
\end{figure}

\begin{figure}
\plotone{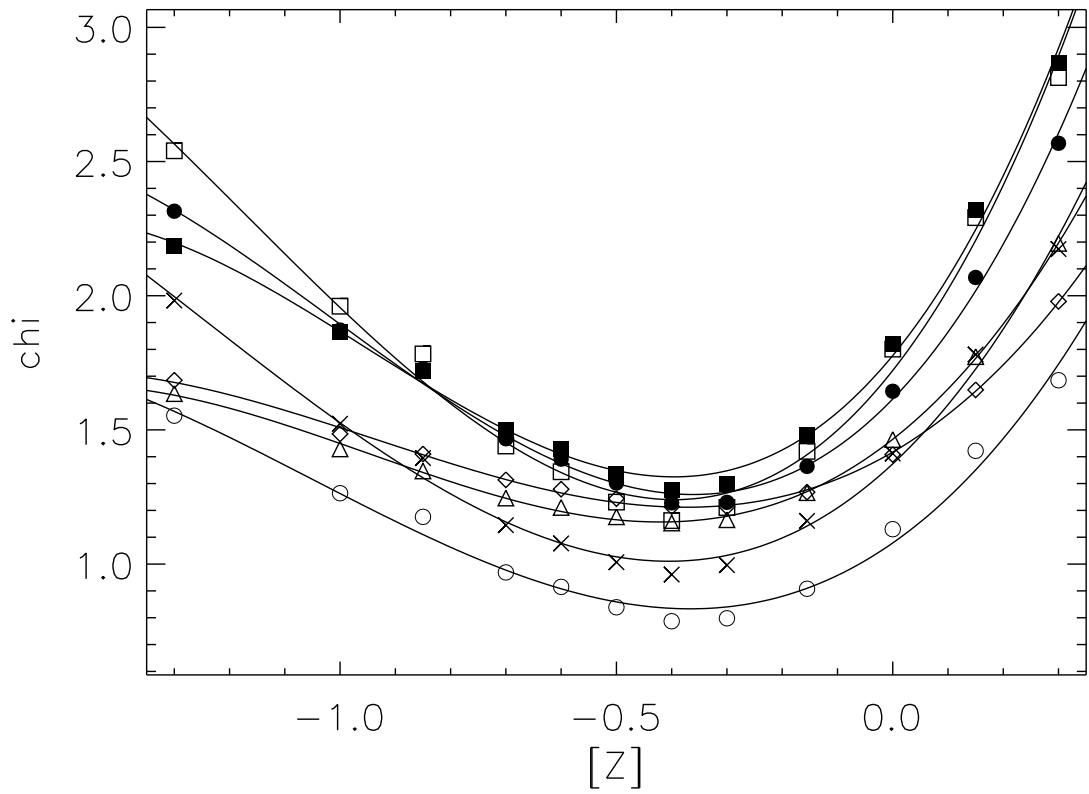}
\caption{$\chi ([Z])_{i}$ for all seven spectral windows of target No.21.}
\label{met5}
\end{figure}

The index i refers to spectral window selected. If we define

 \begin{equation}
        \chi ([Z])_{i} = S/N \ast (O-C)([Z])_{i},
   \end{equation}

where S/N is the signal-to-noise ration per resolution element, we can look for 
the minimum $\chi ([Z])_{i}$ as a function of [Z] to determine metallicity. This 
is done in Fig.~\ref{met3}, which yields [Z] = -0.40. Application of the same method 
on different spectral windows provides additional independent information on 
metallicity as shown in Fig.~\ref{met4} and ~\ref{met5}. The average metallicity 
obtained from all windows is [Z] = -0.39 with a very small dispersion of only 
0.02 dex. 

While it seems that our method yields a very accurate metallicity for a 
fixed pair of stellar parameters \teff\/ and $log~g$, we also need to consider the 
effect of the stellar parameter uncertainties on the metallicity determination. This 
is done by applying the same correlation method for [Z] for models at the extremes 
of the error box for \teff\/ and $log~g$, as given by Fig.~\ref{lgtfit} and 
Table 1. This increases the uncertainty of [Z] to $\pm$ 0.15 dex.

We apply the same method to all the objects of our sample. The results are given in 
Table 1. In some case the uncertainties are larger than for target No. 21, which is 
caused by larger uncertainties of the atmospheric parameters or more noise in the 
spectra or both.

\subsection{Radii, Luminosities and Masses}

The fit of the observed photometric fluxes with the model atmosphere fluxes was 
used to determine interstellar reddening E(B-V) and extinction A$_{V}$ = 3.1 E(B-V). 
Simultaneously, the fit also yields the stellar angular diameter, which provides 
the stellar radius, if a distance is adopted. Alternatively, for the stellar 
parameters (\teff\/, $log~g$, [Z]) determined through the spectral analysis
the model atmospheres also yield bolometric corrections BC, which we use to 
determine bolometric magnitudes. These bolometric magnitudes then also allow us to 
compute radii. The radii determined with these two different methods agree within
a few percent. We have chosen the latter method for the radii given in Table 2. 
\citet{gieren05} in their multi-wavelength study of a large sample of Cepheids in 
NGC 300 including the near-IR have determined a new distance modulus m-M = 26.37 
mag, which corresponds to a distance of 1.88 Mpc. We have adopted these values to 
obtain the radii and absolute magnitudes in Table 2. 

In the temperature range of 8300~K $\le$ 12000~K of our model grid 
(see Fig.~\ref{modelgrid}) the individual BC-values of the models are well 
represented by an analytical fit formula

 \begin{equation}
        BC(T_{\!\mbox{\scriptsize\em eff}}, log~g, [Z]) = f(log~g_{F}) - 4.3~ log~T_{4} (1 + log~T_{4}) + D_{Z}
   \end{equation}

where $T_{4}$ = \teff\//10000~K and
 
\begin{equation}
\begin{array}{c}
\displaystyle D_{Z} = 0.09[Z](1+0.26[Z]) \\
\displaystyle f(log~g_{F}) =f_{max}(1-a~exp(-(log~g_{F}-x_{min})/h))\\
\end{array} 
\end{equation}

and a = 1-$f_{min}$/$f_{max}$, $f_{min}$ = -0.39, $f_{max}$ = -0.265,  
$x_{min}$ = 1.075, h=0.17. The fit is accurate to 0.02 mag over the whole range 
of model parameters. It allows us to estimate the uncertainties of bolometric 
magnitudes arising from the uncertainties of the stellar parameters. The errors for
for M$_{bol}$ given in Table 2 include these uncertainties together with the 
photometric errors.

We note that in most cases fit formulae provided for bolometric corrections do not 
include the effects of gravity. In the case of A supergiants this would lead to 
significant error residuals of the fit of the order of 0.4 mag. Thus, the inclusion 
of the function $f(log~g_{F})$ is essential for the expression of BC.

In our model atmosphere approach to calculate bolometric corrections we convert 
the model flux at the V-band effective wavelength into a V-magnitude in the 
usual way using a constant, which transforms the flux of Vega into the Vega 
V-magnitude. We then use the integral over the total model spectral energy distribution 
and the effective temperature of the Sun (5777K) to calculate the bolometric 
magnitude (assuming $M_{bol}^{\odot}$ = 4.74 mag). The difference between the 
bolometric and the V-magnitudes calculated in this way then yields the bolometric 
correction. This approach provides the correct bolometric correction for Vega,
which is exactly in the effective temperature range of the objects we are dealing 
with. We have, therefore, decided not to re-normalize our BC values to avoid small 
positive values at low effective temperatures, as has been suggested by \citet{buser78} 
(but see also the full discussion on this matter in this paper). As a result, our 
BC-values are about 0.12 mag larger than the ones given by \citet{przybilla06}, 
who have used the approach by \citet{buser78}.

There are two ways to determine stellar masses. We can use the stellar gravities 
together with the radii to directly calculate masses. We call the masses derived 
in this way spectroscopic masses. Alternatively, masses can be estimated by 
comparing the location of our target stars in the HRD with evolutionary tracks (\citealt{meynet05}). 
Masses obtained in such a way are called evolutionary masses. Both values are 
given in Table 2 for each target star. The uncertainties are discussed in section 4.2.

\section{Results}\label{sec_results}

In this section we discuss the main results of our quantitative spectral analysis.

\subsection{Interstellar Reddening}

\begin{figure}
\plotone{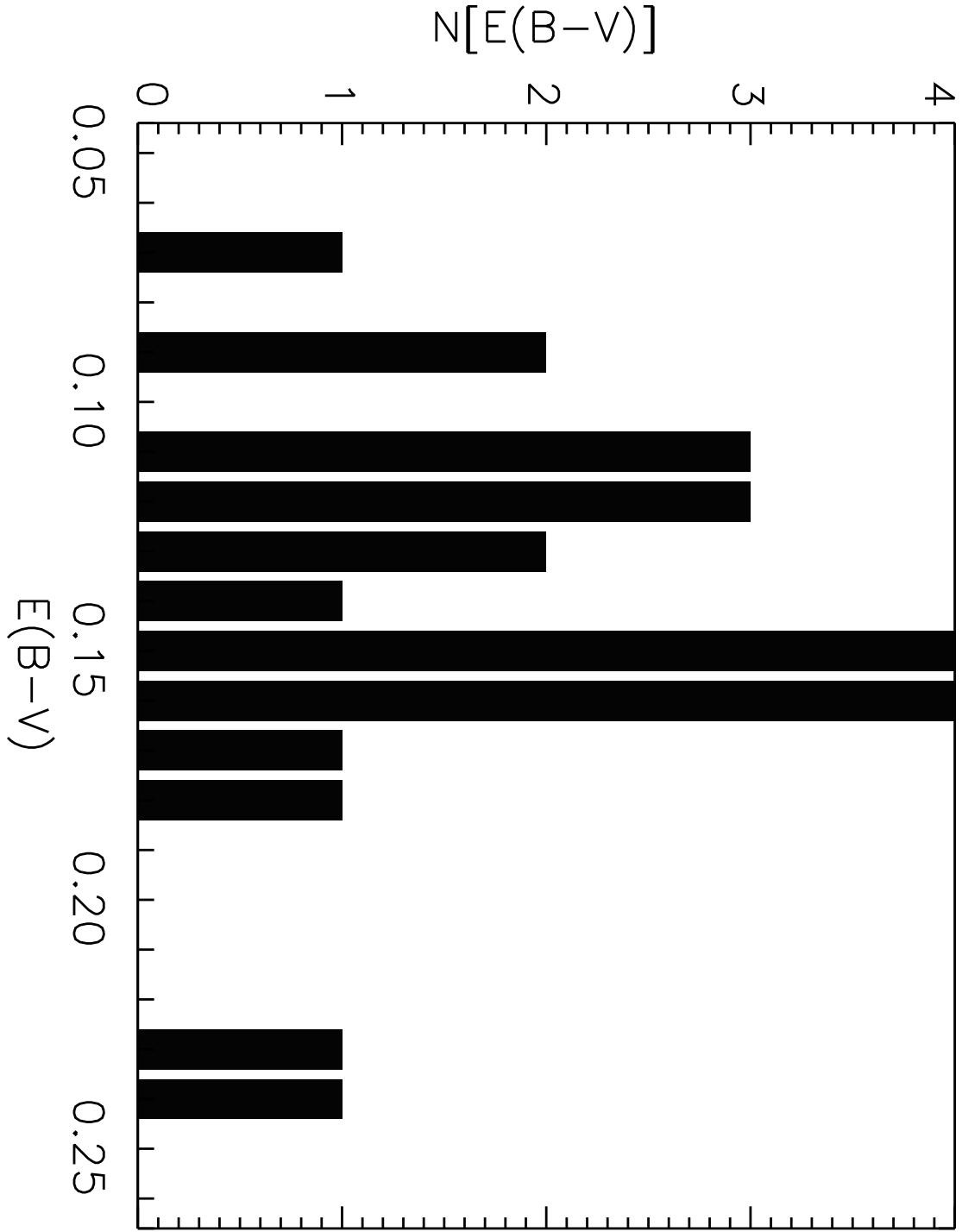}
\caption{The interstellar reddening E(B-V) distribution of our targets.}
\label{nebv}
\end{figure}

Our quantitative spectroscopic method yields interstellar reddening and extinction 
as a by-product of the analysis process. For objects embedded in the dusty disk of 
a star forming spiral galaxy we expect a wide range of interstellar reddening E(B-V).
Indeed, we find a range from E(B-V) = 0.07 mag up to 0.24 mag. Fig.~\ref{nebv} shows
the distribution of interstellar reddening among the 24 stars of our sample. The
individual reddening values are significantly larger than the value of 0.03 mag 
adopted in the HST distance scale key project study of Cepheids by \citet{freedman01}
and demonstrates the need for a reliable reddening
determination for stellar distance indicators, at least as long the study is 
restricted to optical wavelengths. The average over our sample is 
$\langle E(B-V) \rangle $ = 0.12 mag in close agreement with the value of 0.1 mag 
found by \citet{gieren05} in their optical to near-IR study of Cepheids in NGC 300. 
While Cepheids have somewhat lower masses than the A supergiants of our study and 
are consequently somewhat older, they nonetheless belong to the same population and 
are found at similar sites. Thus, we expect them to be affected by interstellar 
reddening in the same way as A supergiants. 

\subsection{Stellar properties}

\begin{figure}
\begin{center}
\includegraphics[width=0.90\textwidth]{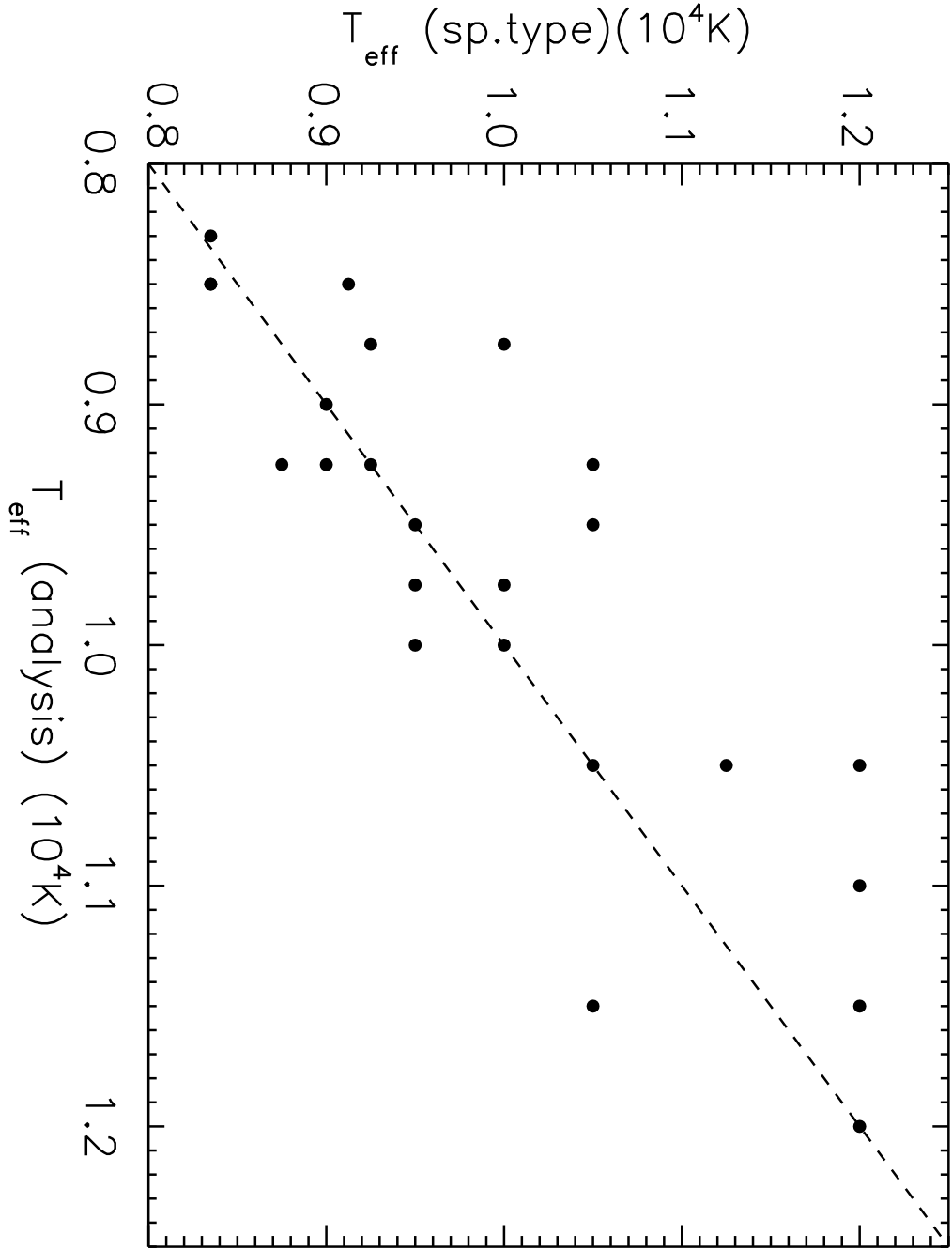}
\caption{Effective temperatures according to spectral types compared with effective 
temperatures obtained from the quantitative spectral analysis. For discussion see text.}
\label{tsvsta}
\end{center}
\end{figure}

\begin{figure}
\begin{center}
\includegraphics[width=0.90\textwidth]{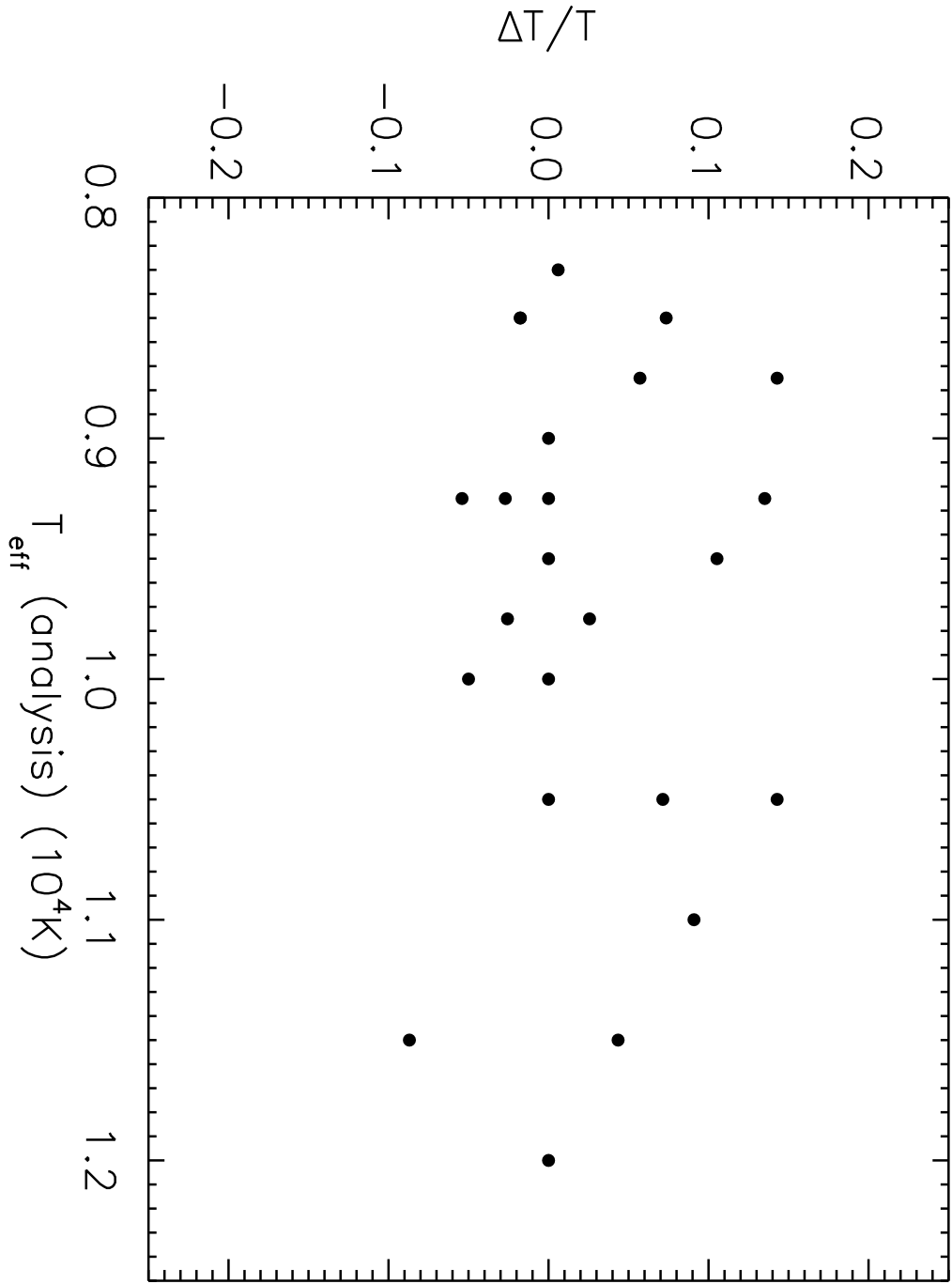}
\caption{Relative difference of spectral type and spectral analysis temperatures as a function 
of effective temperatures obtained from the quantitative spectral analysis.}
\label{dtsvsta}
\end{center}
\end{figure}

\begin{figure}
\begin{center}
\includegraphics[width=0.90\textwidth]{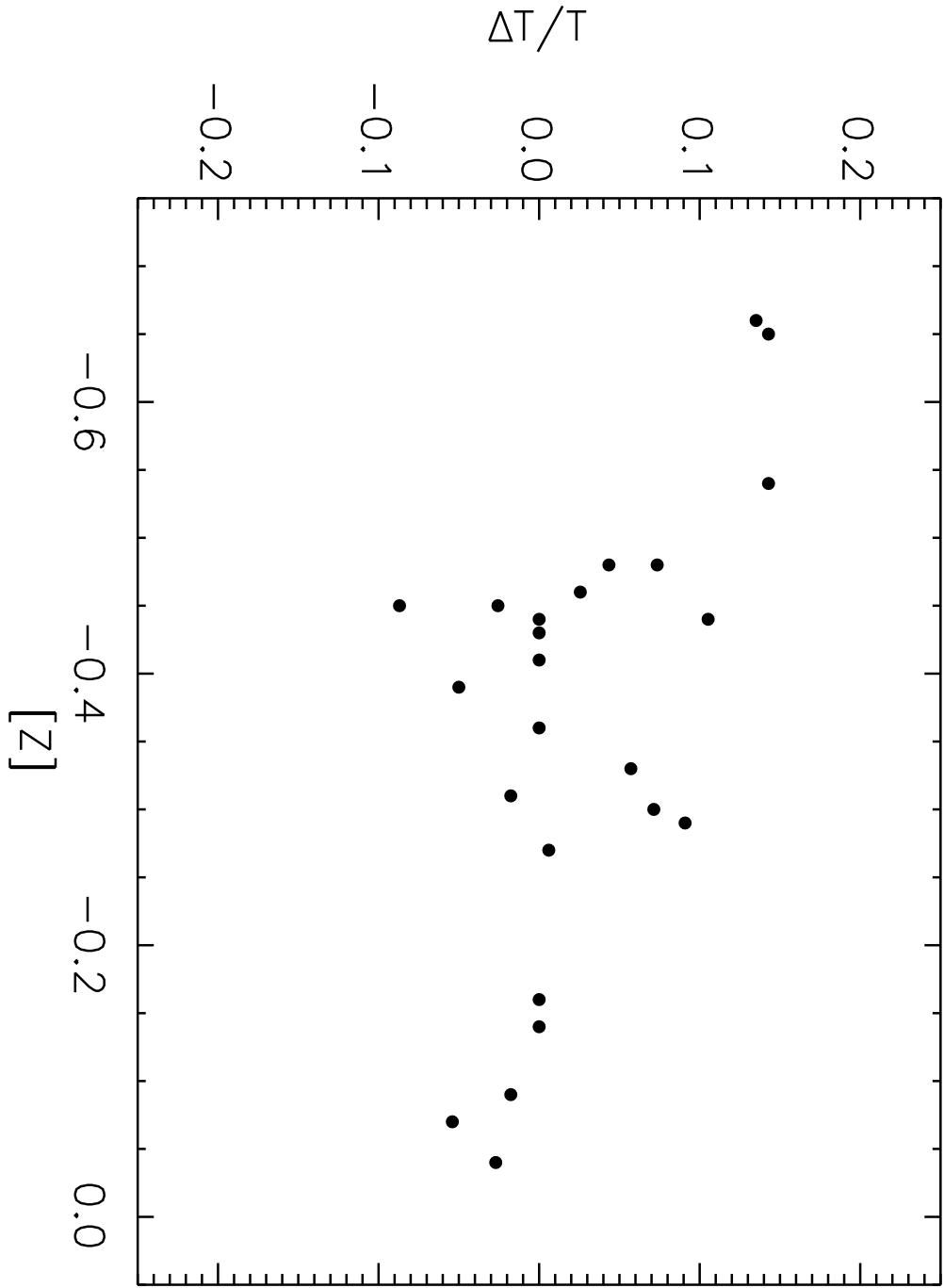}
\caption{Relative difference of spectral type and analysis temperatures as a function 
of metallicity obtained from the quantitative spectral analysis.}
\label{dtsvsz}
\end{center}
\end{figure}

The original motivation for the development of our analysis technique using the Balmer jump 
was the degeneracy between the effective temperature derived from the spectral type and 
metallicity. In Fig.~\ref{tsvsta} and ~\ref{dtsvsta} we compare the temperatures 
obtained with the two different methods. We use the compilation by \citet{kud03} 
for the relationship between spectral type and effective temperature. While there 
is a clear correlation between the two temperatures, there is definitely a shift to 
higher temperatures for \teff\/(spectral type), which is explained by the fact 
that all of our targets have a metallicity smaller than in the solar neighborhood. 
Fig.~\ref{dtsvsz} indicates that the effect becomes larger with decreasing 
metallicity, exactly as expected (for the discussion of metallicities we refer 
to the next section). 

\begin{figure}
\begin{center}
\includegraphics[width=0.90\textwidth]{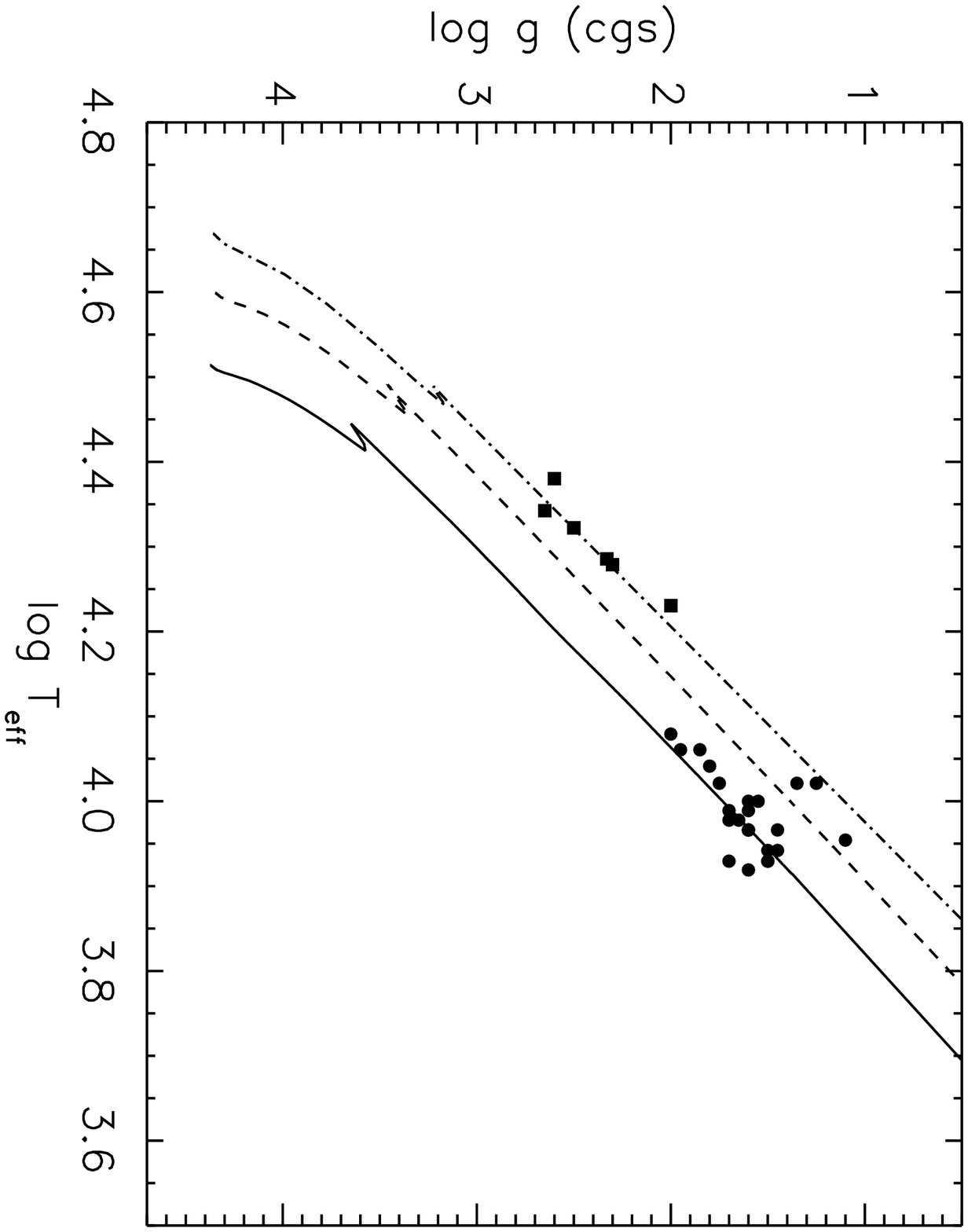}
\caption{NGC 300 A supergiants (circles, this study) and early B supergiants (squares, 
\citealt{urbaneja05}) in the $(log~g, log$~\teff) plane compared with evolutionary 
 tracks by \citet{meynet05} of stars with 15 M$_{\odot}$ (solid), 25 M$_{\odot}$ 
(dashed), and 40 M$_{\odot}$ (dashed-dotted), respectively. The tracks include the 
effects of rotation and are calculated for SMC metallicity (see \citealt{meynet05}).}
\label{lgtngc300}
\end{center}
\end{figure}

Fig.~\ref{lgtngc300} shows the location of our targets in the 
$(log~g, log$~\teff) plane compared to the early B-supergiants studied by
\citet{urbaneja05}. The comparison with evolutionary tracks (\citealt{meynet05}) gives a first 
indication of the stellar masses in a range from 10 M$_{\odot}$ to 40 M$_{\odot}$.
Three targets have obviously higher masses than the rest of the sample and seem to be
on a similar evolutionary track as the objects studied by \citet{urbaneja05}.

\begin{figure}
\begin{center}
\includegraphics[width=0.90\textwidth]{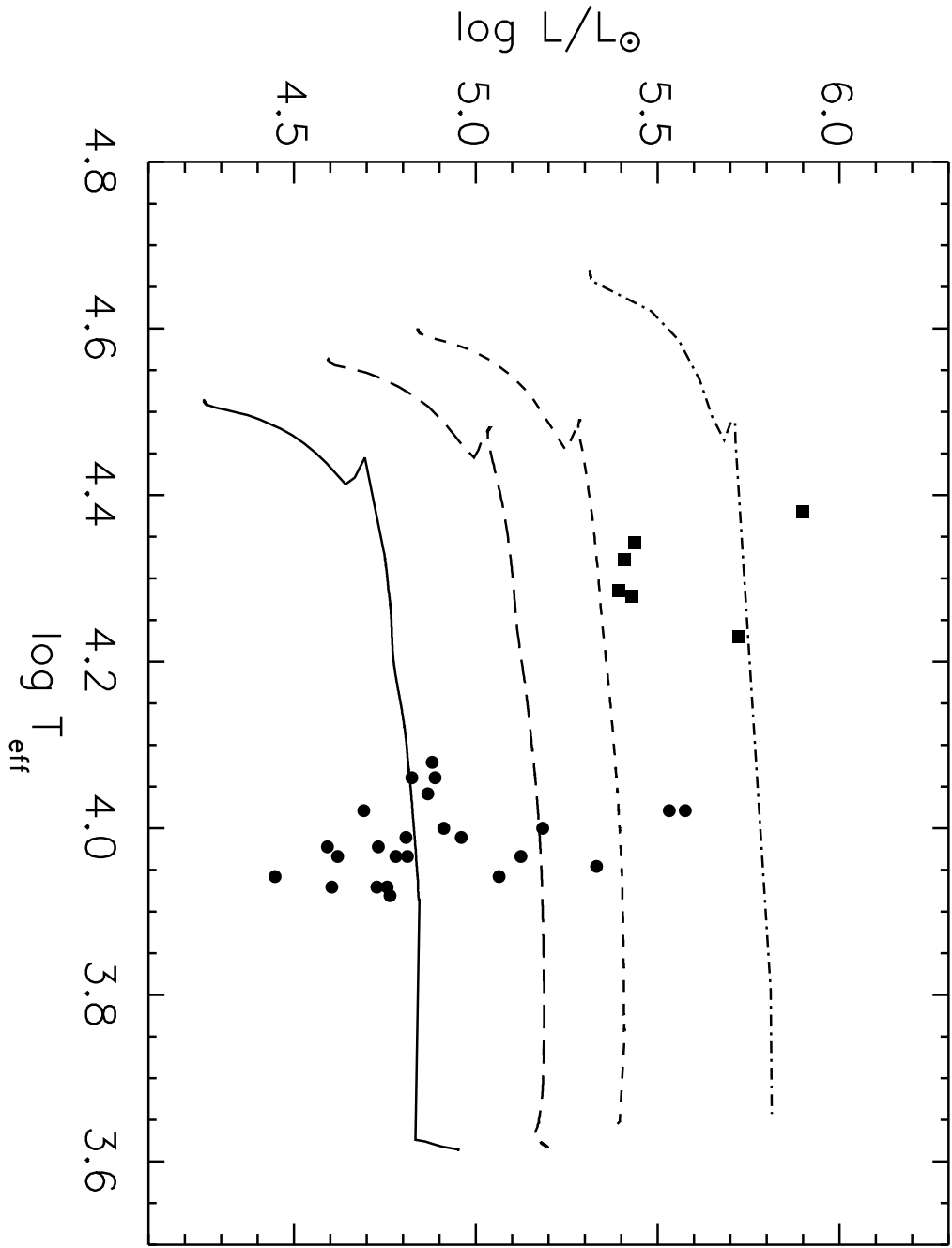}
\caption{
NGC 300 A supergiants (circles, this study) and early B supergiants (squares, 
\citealt{urbaneja05}) in the HRD compared with evolutionary 
 tracks for stars with 15 M$_{\odot}$ (solid), 
20 M$_{\odot}$ (long-dashed), 25 M$_{\odot}$ (short-dashed), 
and 40 M$_{\odot}$ (dashed-dotted), respectively. The tracks include the 
effects of rotation and are calculated for SMC metallicity (see 
\citealt{meynet05}).}
\label{hrdngc300}
\end{center}
\end{figure}

An alternative way to discuss the evolutionary status of our targets is the 
classical HRD. This is done in Fig.~\ref{hrdngc300}. The evolutionary information 
obtained from this diagram compared with Fig.~\ref{lgtngc300} appears to be 
consistent. The B-supergiants seem to be more massive than most of the A 
supergiants. The same three A supergiants apparently more massive than the rest 
in Fig.~\ref{lgtngc300} are also the most luminous objects in Fig.~\ref{hrdngc300}. 
This confirms that quantitative spectroscopy is -at least qualitatively - capable 
to retrieve the information about absolute luminosities. Note that the fact that 
all the B supergiants studied by \citet{urbaneja05} are more massive is simply 
a selection effect of the V magnitude limited spectroscopic survey by 
\citet{bresolin02}. At similar V magnitude as the A supergiants those objects 
have higher bolometric corrections because of their higher effective temperatures 
and are, therefore, more luminous and massive.  

The luminosity uncertainties for our targets range from 0.02 dex to 0.06 dex 
(the latter rather large errors appear only in a few rare cases, see Table 2). 
Correspondingly, Fig.~\ref{hrdngc300} allows to determine stellar masses based on 
evolutionary tracks with a much better resolution than Fig.~\ref{lgtngc300}. We have used 
the evolutionary tracks by \citet{meynet05} (and references therein) to construct 
mass luminosity relationships in the A supergiant stage of the form

 \begin{equation}
       log L/L_{\odot} = (a-bx)x+c,~x = log M/M_{\odot} - log M^{min}/M_{\odot}
  \end{equation}
 
The coefficients for different evolutionary models (with and without rotation, 
Milky Way and SMC metallicity) are given in Table 4. The fit reproduces model 
masses for a given luminosity (in the range $-6.5 \ge M_{bol} \ge -9.7 $) 
to better than 5\% or 0.02 dex. With the bolometric magnitudes in Table 2 we can 
then determine evolutionary masses, which are also given in Table 2. We have chosen 
the mass-luminosity relationship for tracks with rotation and SMC abundance, the 
other relationships give similar results. The reason why we selected Milky Way 
and SMC metallicities as a representative grid of evolutionary tracks is that 
only for those a complete set of tracks is available.

With the small errors in luminosity, we obtain masses which formally have very 
small uncertainties of at most 5\%. However, the masses are, of course, affected 
by the systematic uncertainties of the evolutionary tracks through the effects of 
mass-loss, rotation, mixing and opacity. One particular problem may arise through 
blue loops from the red giant branch (RGB) or crossing to the blue from the RGB. 
Fortunately, for the mass and luminosity range of interest 
($12 M_{\odot} \leq M \leq 30 M_{\odot}$, $-6.5 \leq M_{bol} \leq -9.3$) 
we find that the published tracks (in particular at lower metallicity) do not show 
significant loops or second crossings.

\begin{figure}
\begin{center}
\includegraphics[width=0.90\textwidth]{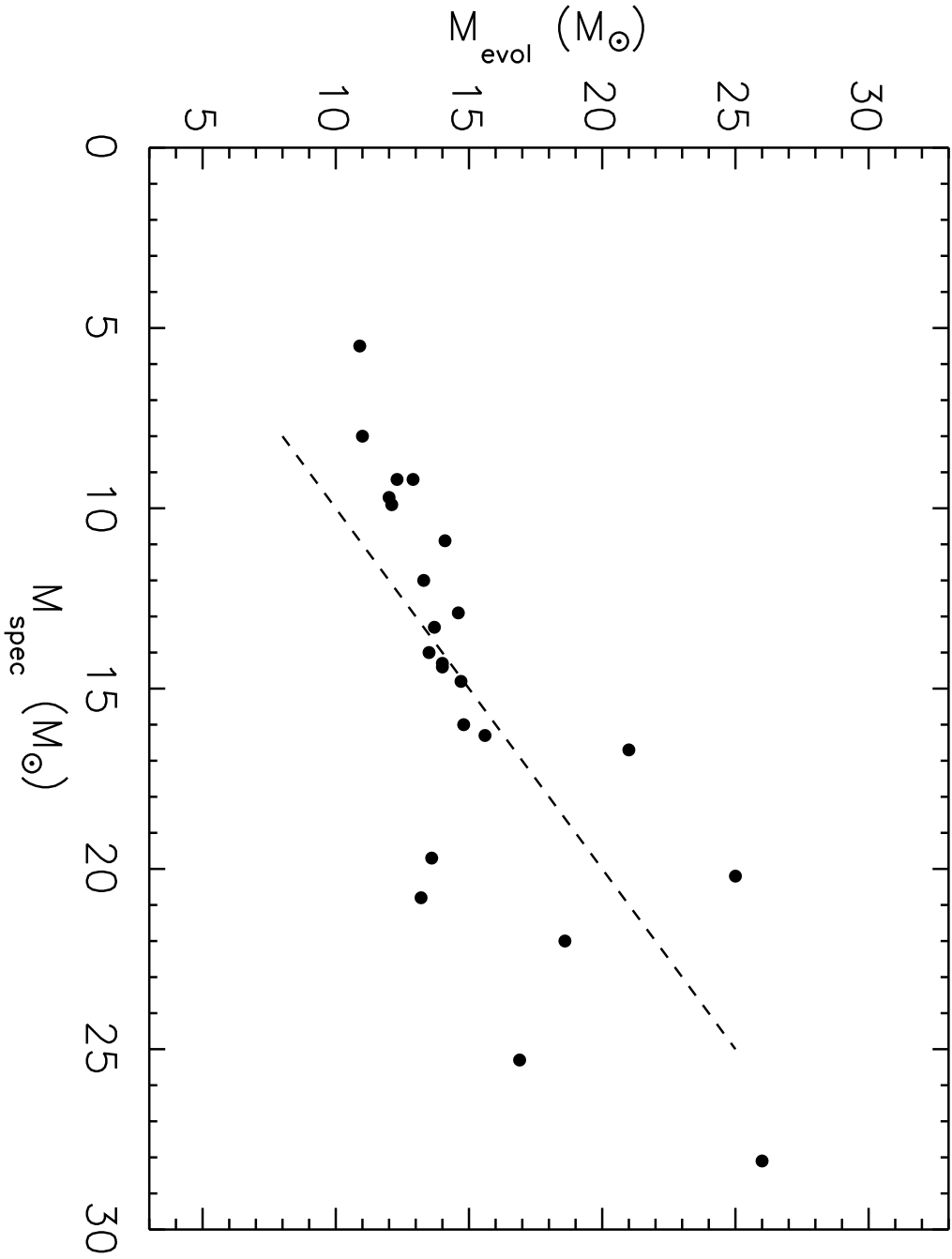}
\caption{Comparison of evolutionary with spectroscopic masses for the NGC 300 
A supergiants of this study}
\label{masses1}
\end{center}
\end{figure}

\begin{figure}
\begin{center}
\includegraphics[width=0.90\textwidth]{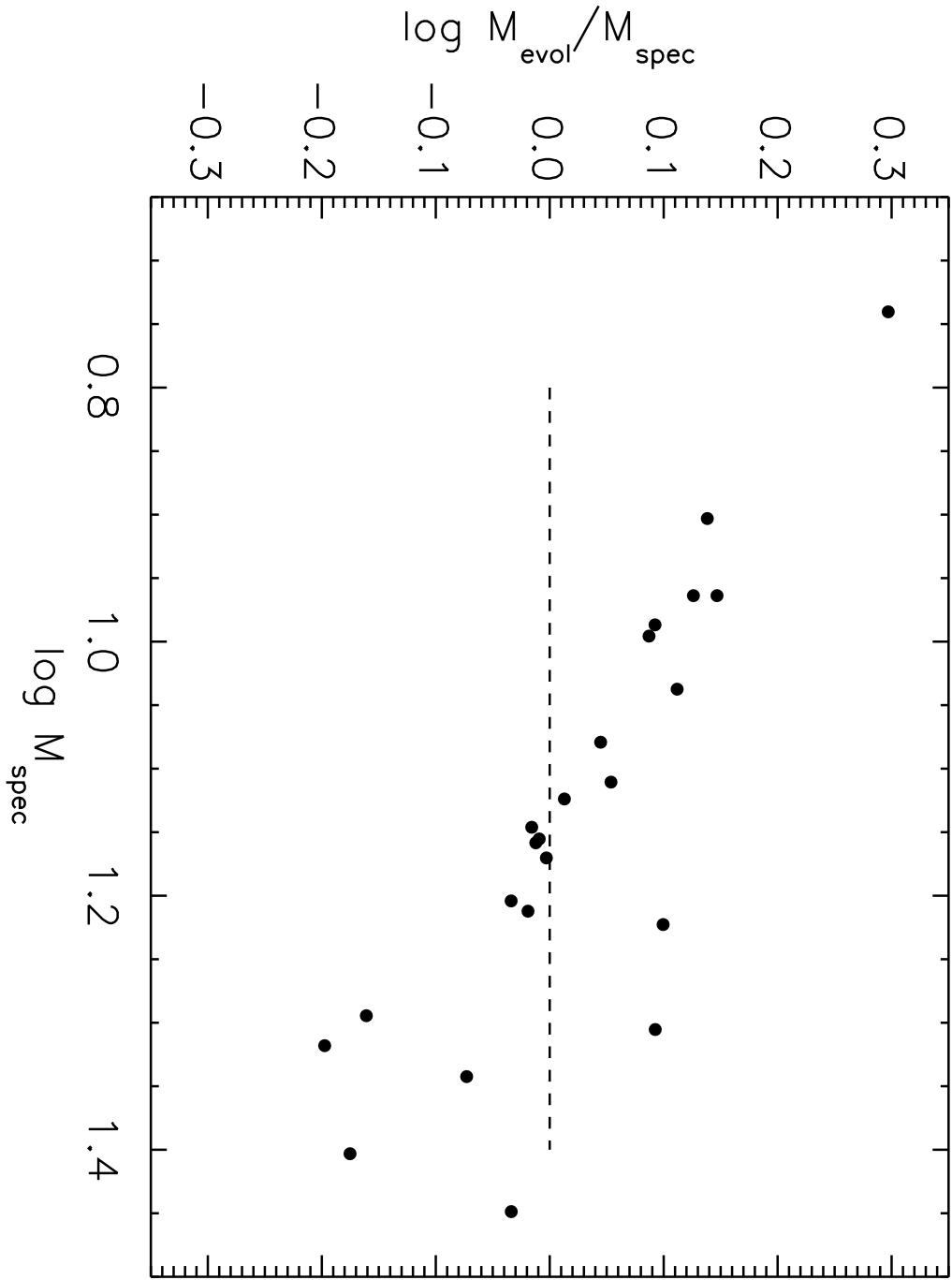}
\caption{Logarithm of the ratio of evolutionary to spectroscopic masses as 
a function of the logarithm of spectroscopic mass.}
\label{masses3}
\end{center}
\end{figure}

An alternative way to determine masses is, of course, to use directly the stellar 
gravities and radii determined through the fitting of stellar spectra and SEDs. 
These spectroscopic masses are also given in Table 2. Note that we have not 
included the effects of centrifugal forces on the effective stellar gravities 
to derive spectroscopic masses. While this is important for O-stars 
(see \citealt{repolust04}), the rotational velocities of A supergiants 
($\sim~$30 km/s) are too small and the radii too large to cause an effect. 
Fig.~\ref{masses1} and
~\ref{masses3} compare the masses derived by the two different 
methods. Given the fact that uncertainties of the spectroscopic masses are roughly 
as large as the gravity uncertainties (0.1 to 0.2 dex) the comparison is generally 
encouraging. At the high mass end, we have agreement within the uncertainties. At 
the low mass end, however, the evolutionary masses seem to be somewhat higher on 
average. There is the indication of a trend of this effect to become stronger with 
decreasing spectroscopic mass. At this point, we have no explanation for this 
result. If it is not an artefact of the spectroscopic technique applied, then it 
may indicate that the effects of blue loops are more important than the evolutionary 
models available to us predict. Blue loops increase the luminosity at a given mass 
( at the low mass end the amount of mass-loss is still small at the RGB), thus, 
their neglect would lead to an overestimate of the evolutionary mass.

The most extreme object in the left part of Figures~\ref{masses1} and 
~\ref{masses3} ist target No. 17 of Tables 1 and 2. It is the object with the 
lowest luminosity in the HRD (Fig.~\ref{hrdngc300}) corresponding to a mass of 
11 M$_{\odot}$. Its position in the $(log~g, log$~\teff) plane of 
Fig.~\ref{lgtngc300} is right on the track with an initial mass of 15 M$_{\odot}$, 
but with the low gravity determined through spectroscopy its radius is too small 
to yield mass high enough in comparison with the evolutionary mass. 

\subsection{Metallicity Gradient}\label{sec_metal}

\begin{figure}
\begin{center}
\includegraphics[width=0.90\textwidth]{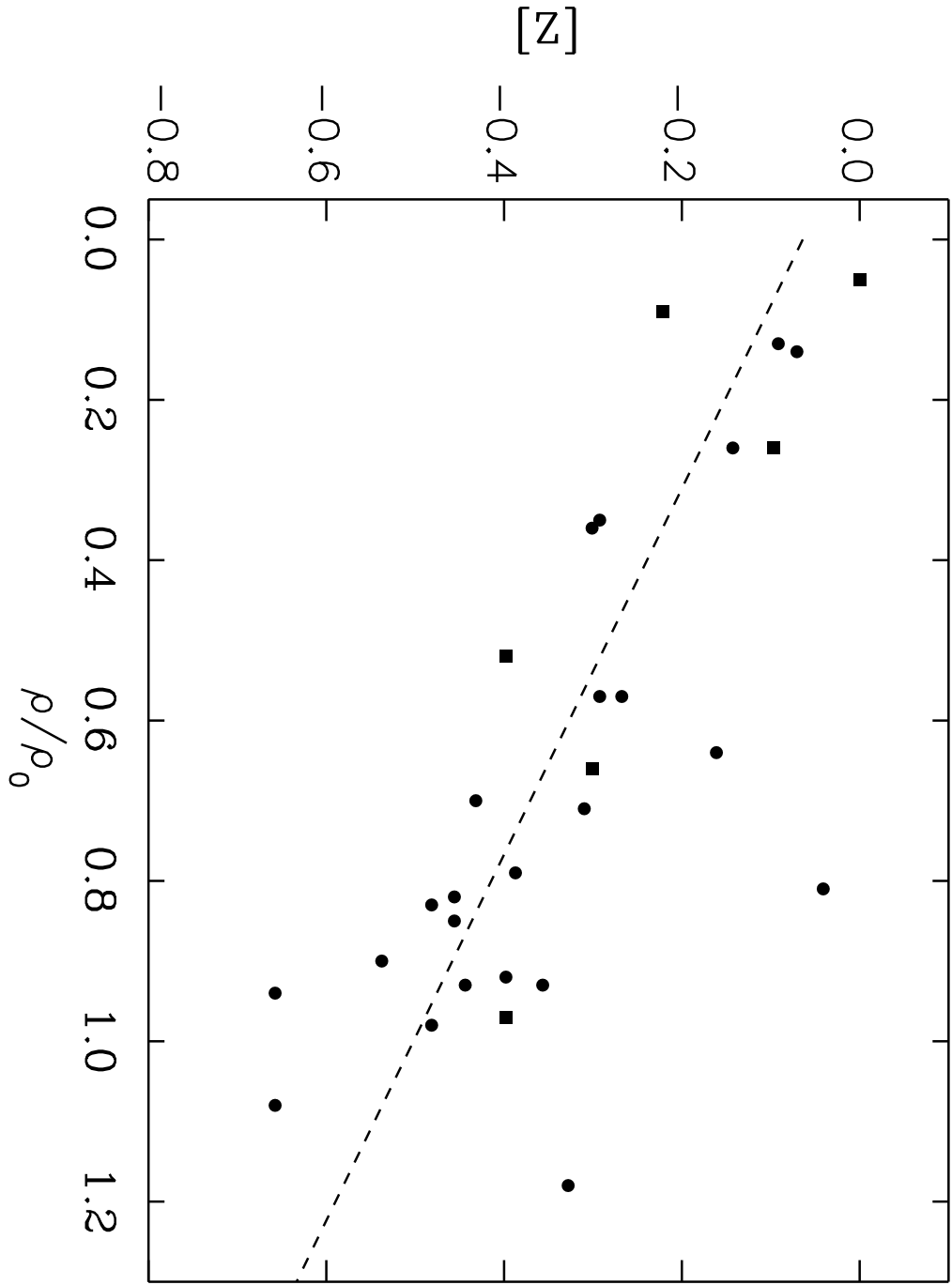}
\caption{Metallicity [Z] as a function of angular galacto-centric distance 
$\rho/\rho_{0}$ (see Table 1) for the A-supergiants of this work (filled circles) and the 
early B-supergiants studied by \citet{urbaneja05} (filled squares). Note 
that for the latter metallicity refers to oxygen only. The dashed curve 
represents the regression discussed in the text. Note that typical uncertainties in [Z] are
$\pm{0.2}$ dex.}
\label{metgrad}
\end{center}
\end{figure}

The metallicities of the individual targets together with their galacto-centric 
distance are given in Table 1 and allow us to discuss 
the stellar metallicity and the metallicity gradient in NGC 300. Fig.~\ref{metgrad} 
shows the stellar metallicities [Z] as function of galactocentric distance, expressed in terms of the isophotal radius, 
$\rho/\rho_{0}$. Despite the scatter caused by the metallicity uncertainties of the 
individual stars the metallicity gradient of the young disk population in NGC 300 
is very clearly visible. A linear regression (excluding target No. 1, the outlier 
at $\rho/\rho_{0}$ = 0.81 and [Z] = -0.04) for the combined A- and B-supergiant 
sample yields 

 \begin{equation}
        [Z] = -0.06\pm{0.09} - 0.44\pm{0.12}~\rho/\rho_{0}
   \end{equation}

for the angular gradient. On a linear scale (d in kpc) we obtain

\begin{equation}
        [Z] = -0.06\pm{0.09} - 0.083\pm{0.022}~d.
   \end{equation}

We obtain a very similar regression, if we restrict the sample to A-supergiants only, 
which indicates that there is no systematic abundance difference between these 
two groups of stars. Our new result confirms the result obtained by 
\citet{urbaneja05}, but with a much larger sample the metallicity gradient and the zero 
point are now much better defined. Note that the metallicities of the B supergiants 
refer to oxygen only with a value of log[N(O)/N(H)] = -3.31 adopted for the sun 
(\citealt{allende01}). On the other hand, the A supergiant metallicities reflect 
the abundances of a variety of heavy elements such as Ti, Fe, Cr, Si, S, and Mg.

With this result we can resume the discussion started by \citet{urbaneja05} 
and compare with oxygen abundances obtained from HII-region emission lines. 
\citet{urbaneja05} used line fluxes published by \citet{deharveng88} and applied 
various different published strong line method calibrations to determine nebular
oxygen abundances, which could then be used to obtain the similar regressions as 
above. The results for zero points and gradients compared to our work are given in 
Table 3 (for a discussion of the individual calibrations we refer to 
\citealt{urbaneja05}).

The different strong line method calibrations lead to significant differences in 
the central metallicity as well as in the abundance gradient. The calibrations 
by \citet{dopita86} and \citet{zaritsky94} predict a metallicity significantly 
supersolar in the center of NGC 300 contrary to the other calibrations. Our work 
yields a central metallicity slightly smaller than solar in good agreement with 
\citet{denicolo02} and marginally agreeing with \citet{kobulnicky99}, 
\citet{pilyugin01}, and \citet{pettini04}. At the isophotal radius, 5.3 kpc 
away from the center of NGC 300, we obtain an average metallicity significantly 
smaller than solar [Z] = -0.50, close to the average metallicity in the SMC. The 
calibrations by \citet{dopita86}, \citet{zaritsky94}, \citet{kobulnicky99} 
do not reach these small values for oxygen in the HII regions either because their 
central metallicity values are too high or the metallicity are gradients too shallow.

In light of the substantial range of metallicies obtained from HII region emission 
lines using different strong line method calibrations it seems to be extremely 
valuable to have an independent method using young stars. It will be very important 
to compare our results with advanced work on HII regions, which will avoid strong 
line methods, and will use direct infomation about nebular electron temperatures 
and densities (Bresolin et al., 2008, in preparation).

As mentioned before, there is a suspicious outlier in Fig.~\ref{metgrad}. We have 
checked the 
stellar parameters and metallicity of this object several times and have not found 
a reason to discard the high metallicity value found, which is 0.38 dex above the 
average value at this galacto-centric distance. A similar example is target No.14 
at the largest galacto-centric distance. While we cannot exclude that 
something went wrong in the analysis, we also want to state very firmly that the 
expectation of homogeneous azimuthal metallicity in patchy star forming spiral 
galaxies seems naive to us. The continuation of this type of work on other galaxies 
will show whether cases like this are common or not. We note that the high 
resolution spectral analysis of A supergiants in WLM by \citet{venn03} has 
detected a similar outlier case.

So far, we have only tried to determine an average metallicity [Z] for each star.
Together with the challenge of an accurate determination of stellar parameters we 
felt that this was a good first reasonable step to obtain quantitative information 
from our low resolution spectra. While [Z] in the way how our method works is mostly 
determined by an overlap of Fe, Ti and Cr lines, we have also observed that 
certain spectral windows are dominated by a majority of lines of specific element species. 
In future work we will try to use this observation to determine relative fractions 
of abundances of $\alpha$- elements such as Mg and Ti to iron-group elements 
such as Fe and Cr. It will be an important goal to investigate whether this 
ratio is constant or not as a function of galactocentric distance.

\section{Stellar Properties and Evolutionary Lifetimes}\label{sec_lifet}

\begin{figure}
\plotone{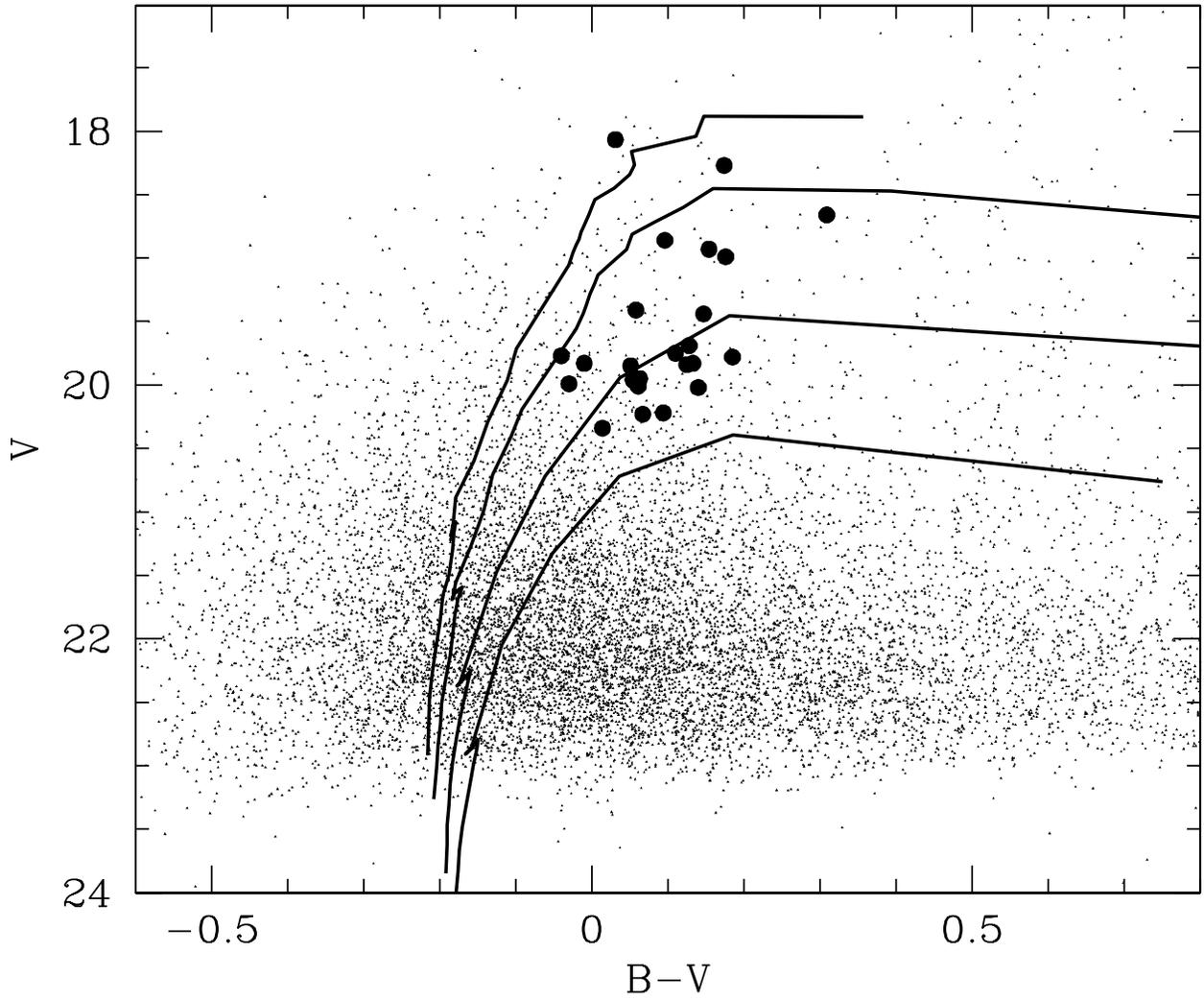}
\caption{HST/ACS color magnitude diagram of the observed fields in NGC 300. 
The position of our targets is indicated by filled circles and evolutionary tracks 
(\citealt{meynet05}) with 12, 15, 20 and 25 M$_{\odot}$ are overplotted.}
\label{cmd}
\end{figure}

\begin{figure}
\plotone{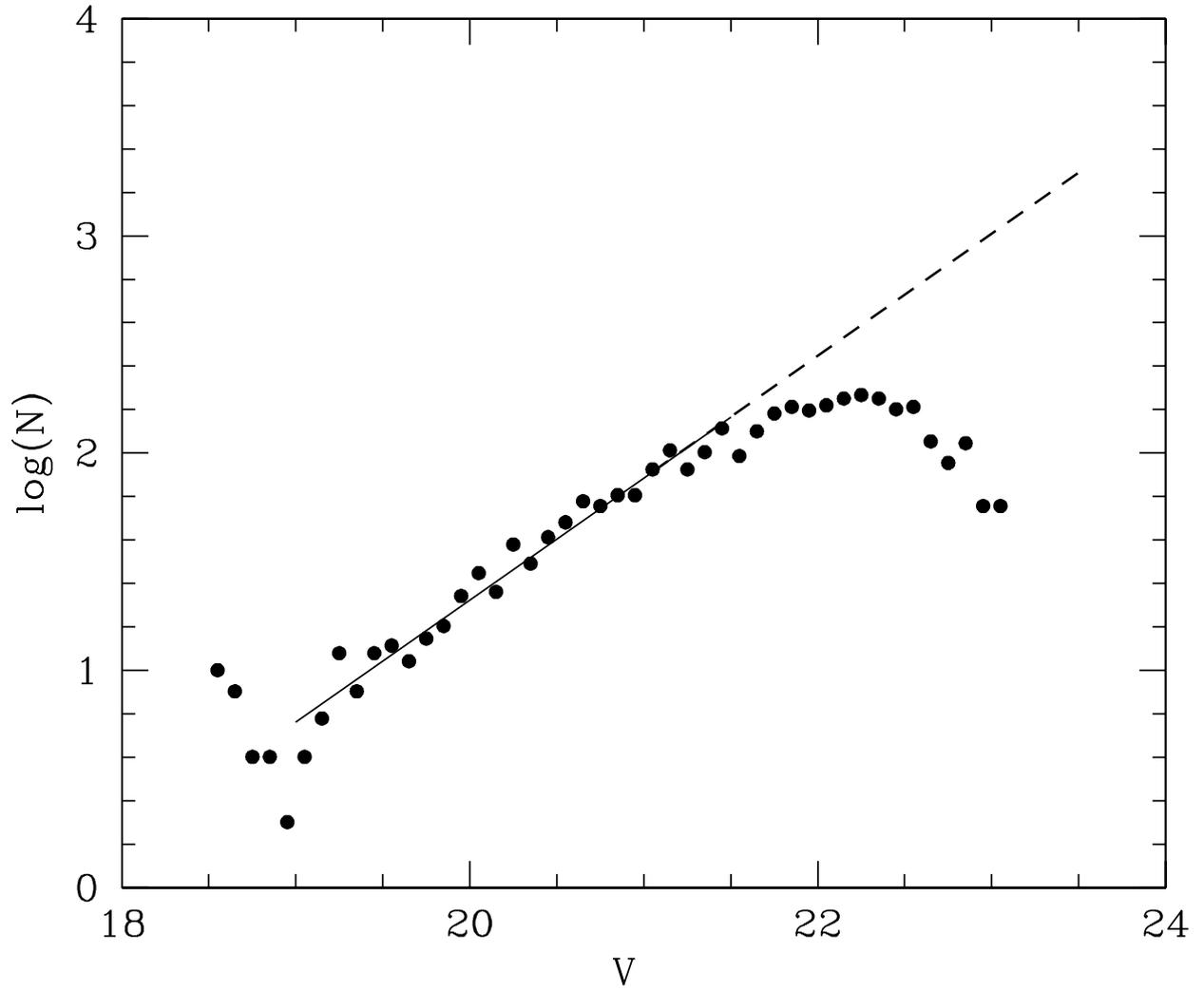}
\caption{The number of stars in Fig.~\ref{cmd} in 0.1 magnitude bins as a 
function of magnitude. The fit discussed in the text is shown as a straight line.}
\label{counts}
\end{figure}

The progenitor stars of the A supergiants of our sample were obviously O-stars 
in a mass range between 15 to 25 M$_{\odot}$. The HST/ACS photometry of 
\citet{bresolin05} allows to estimate how many of such progenitor stars are 
present in exactly the same field. Fig.~\ref{cmd} shows the corresponding combined 
color-magnitude diagram. Adopting our mean reddening of E(B-V) = 0.12 together with 
the corresponding mean visual extinction of A$_{V}$ = 0.37, a distance 
modulus of 26.37 mag, an absolute magnitude of M$_{V}$ = - 2.7 mag for a 15 
M$_{\odot}$ star on the zero age main sequence, we can assume that all stars 
brighter than m$_{V}$ = 24.0 mag are possible progenitors of A supergiants. 
Fig.~\ref{counts} shows the number of star counts in our fields in bins of 
0.1 mag as a function of magnitude. The curve is well fitted by

\begin{equation}
        log~n = a_{\star} m_{V} + b_{\star}
   \end{equation}

and $a{\star}$ = 0.5622, $b_{\star}$ = -9.921 (see also \citealt{bresolin96},
 \citealt{bresolin98}). 
While the number of star counts starts to become incomplete at m$_{V}$ = 21.5 mag 
we can extrapolate the fit to calculate the total cumulative number N of stars 
down to m$_{V}$ = 24.0. We obtain N= $2.9\times10^{4}$ for the present number of 
progenitor stars. Assuming continuous star formation and having 24 confirmed A 
supergiants in the fields the ratio $24/2.9\times10^{4} = 8.3\times10^{-4}$ provides a 
lower limit for the evolutionary lifetime in the A supergiant stage in units of 
the main sequence lifetime. 

The total number of A supergiants in the observed fields in the mass range between 
15 to 25 M$_{\odot}$ can then be estimated from \citet{bresolin02}, who found a total 
of 167 supergiant candidates in the combined color magnitude diagram. They obtained 
spectra for 70 of those and detected 29 A supergiants, i.e. a fraction of 41\%. 
Assuming the same fraction for the original total of 167 candidates leaves us with 
an estimate of 69 A supergiants. The observed ratio of evolutionary lifetime in 
the A supergiant stage to the main sequence lifetime is 
then $69/2.9\times10^{4} = 2.4\times10^{-3}$.

The corresponding lifetime ratio for evolutionary tracks is a strong function of 
metallicity. Comparing calculations including the effects of rotation (\citealt{meynet05}) we obtain 
$2\times10^{-2}$ for SMC metallicity and $2-8\times10^{-4}$ for Milky Way metallicity. 
With metallicities for most of our target stars between these extremes we conclude 
that our empirical lifetime estimate agrees with the prediction of the evolutionary 
models.

\section{Flux Weighted Gravity - Luminosity Relationship (FGLR)}\label{sec_fglr}

As discussed in the previous sections, massive stars with masses in the range from 
12 M$_{\odot}$ to 40 M$_{\odot}$ evolve through the B and A 
supergiant stage at roughly constant luminosity (see Fig.~\ref{hrdngc300}). In addition, 
since the evolutionary timescale is very short when crossing through the B and A 
supergiant domain, the amount of mass lost in this stage is small. This means that 
the evolution proceeds at constant mass and constant luminosity. This has a very 
simple, but very important consequence for the relationship of gravity and 
effective temperature along each evolutionary track. From

\begin{equation}
L \propto R^{2}T^{4}_\mathrm{eff} = \mathrm{const.} ; M = \mathrm{const.}
\end{equation}
follows immediately that
\begin{equation}
M \propto g\;R^{2} \propto L\;(g/T^{4}_\mathrm{eff}) = L~g_{F} = \mathrm{const.}
\end{equation}

Thus, along the evolution through the B and A supergiant domain the 
\emph{``flux-weighted gravity''} $g_{F} = g$/\teffq\/ should remain constant. 
This means each evolutionary track of different luminosity in this domain is characterized 
by a specific value of $g_{F}$. This value is determined by the relationship 
between stellar mass and luminosity, which in a first approximation is a power law
\begin{equation}
L \propto M^{x}\;.
\end{equation}

We note from section 4.2 that the mass-luminosity relationship has higher order 
terms, but for simplicity we use this as a first order approximation to obtain

\begin{equation}
L^{1-x} \propto (g/T^{4}_\mathrm{eff})^{x}\;.
\end{equation}

With the definition of bolometric magnitude 
$M_\mathrm{bol}$\,$\propto$\,$-2.5\log L$ we then derive

\begin{equation}
-M_\mathrm{bol} = a_{FGLR}(\log~g_{F} - 1.5)) + b_{FGLR}\;.
\end{equation}

This is the  \emph{``flux-weighted gravity -- luminosity relationship''} 
(FGLR) of blue supergiants. Note that the proportionality constant
$a_{FGLR}$ is given by the exponent of the mass -- luminosity power law through

\begin{equation}
a_{FGLR} = 2.5 x/(1-x)\;,
\end{equation}

for instance, for $x=3$, we obtain $a_{FGLR}=-3.75$. We have chosen the zero 
point of the relationship at a flux weighted gravity of 1.5, which is in the 
middle of the range encountered for blue supergiant stars.

\begin{figure}
\begin{center}
\includegraphics[width=0.90\textwidth]{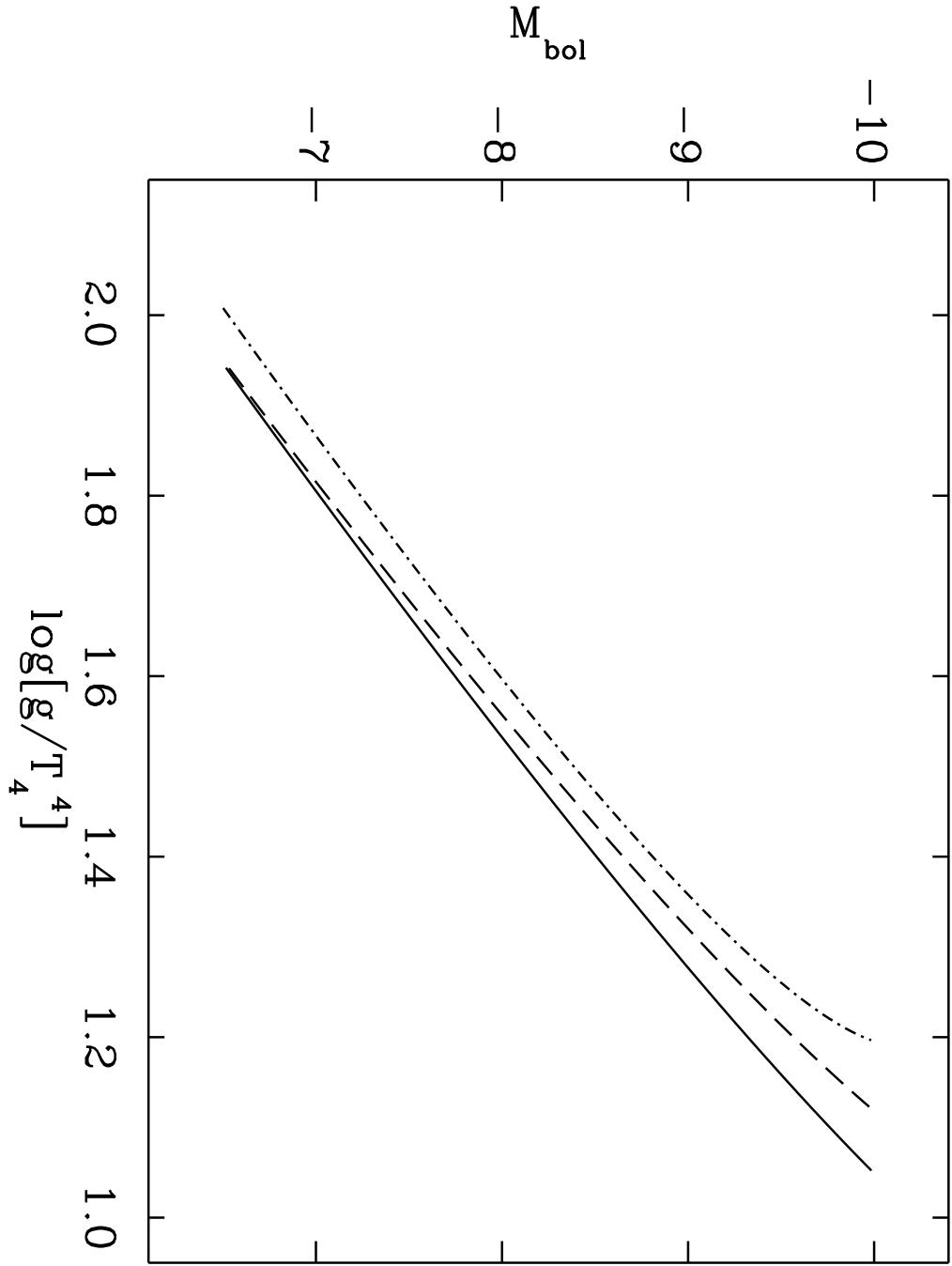}
\caption{The FGLR as predicted by stellar evolution theory (\citealt{meynet05}) using the 
mass-luminosity relationship of eq. 8 and the coefficients of Table 4. The 
solid curve is calculated for solar metallicity and includes the effects of 
stellar rotation. The dashed-dotted curve is for the same metallicity but does 
not include rotation. The long-dashed curves assumes SMC metallicity and accounts 
for rotation (for this metallicity the curve without rotation is very similar).}
\label{fglrev}
\end{center}
\end{figure}

We can use the mass-luminosity relationships of different evolutionary tracks
(with and without rotation, for Milky Way and SMC metallicity) as discussed 
in section 4.2 and characterized 
by the coefficients given in Table 4 to calculate the different stellar 
evolution FGLRs, which are displayed in Fig.~\ref{fglrev}. Very interestingly, 
while different evolutionary model types yield somewhat different FGLRs, 
the differences are rather small. The difference is largest between the two
sets of stellar evolution models for Milky Way metallicity with and without 
the effects of rotation. The models with SMC metallicity and rotation yield 
a FGLR rather close to the case with Milky way metallicity and rotation. Only 
at higher luminosity, where the dependence of mass-loss on metallicity starts 
to become important, is a larger difference from the Milky Way FGLR predicted. 
This is also the parameter domain, where the curvature of the FGLR becomes 
more significant, as to be expected from the non-linear mass-luminosity 
relationship of section 4.2. The set of stellar evolution models at SMC 
metallicity but without the effects of rotation yields a FGLR very close to the 
one with rotation and the same metallicity. 

\citet{kud03} were the first to realize that the FGLR has a very interesting 
potential as a purely spectroscopic distance indicator, as it relates two 
spectroscopically well defined quantitities, effective temperature and gravity, 
to the absolute magnitude. Compiling a large data set of spectroscopic high 
resolution studies of A supergiants in the Local Group and with an approximate 
analysis of low resolution data of a few targets in galaxies beyond the Local 
Group (see discussion in previous chapters) they were able to prove the 
existence of an observational FGLR rather similar to the theoretically 
predicted one.

With our improved analysis technique of low resolution spectra of A supergiants 
and with the much larger sample studied in our new approach we can now resume 
the investigation of the FGLR. We will do this in three steps. We will first 
discuss the determination of flux weighted gravities $log~g_{F}$ from the Balmer 
lines and will then introduce the new FGLR obtained for NGC 300. In this second step we will 
also include the B supergiants analyzed by \citet{urbaneja05}. Finally, we 
will compare the new results obtained for NGC 300 with the previous work on 
Local Group A supergiants to find out whether a universal FGLR exists.

\subsection{Balmer lines and Flux weighted Gravity}

\begin{figure}
\begin{center}
\includegraphics[width=0.90\textwidth]{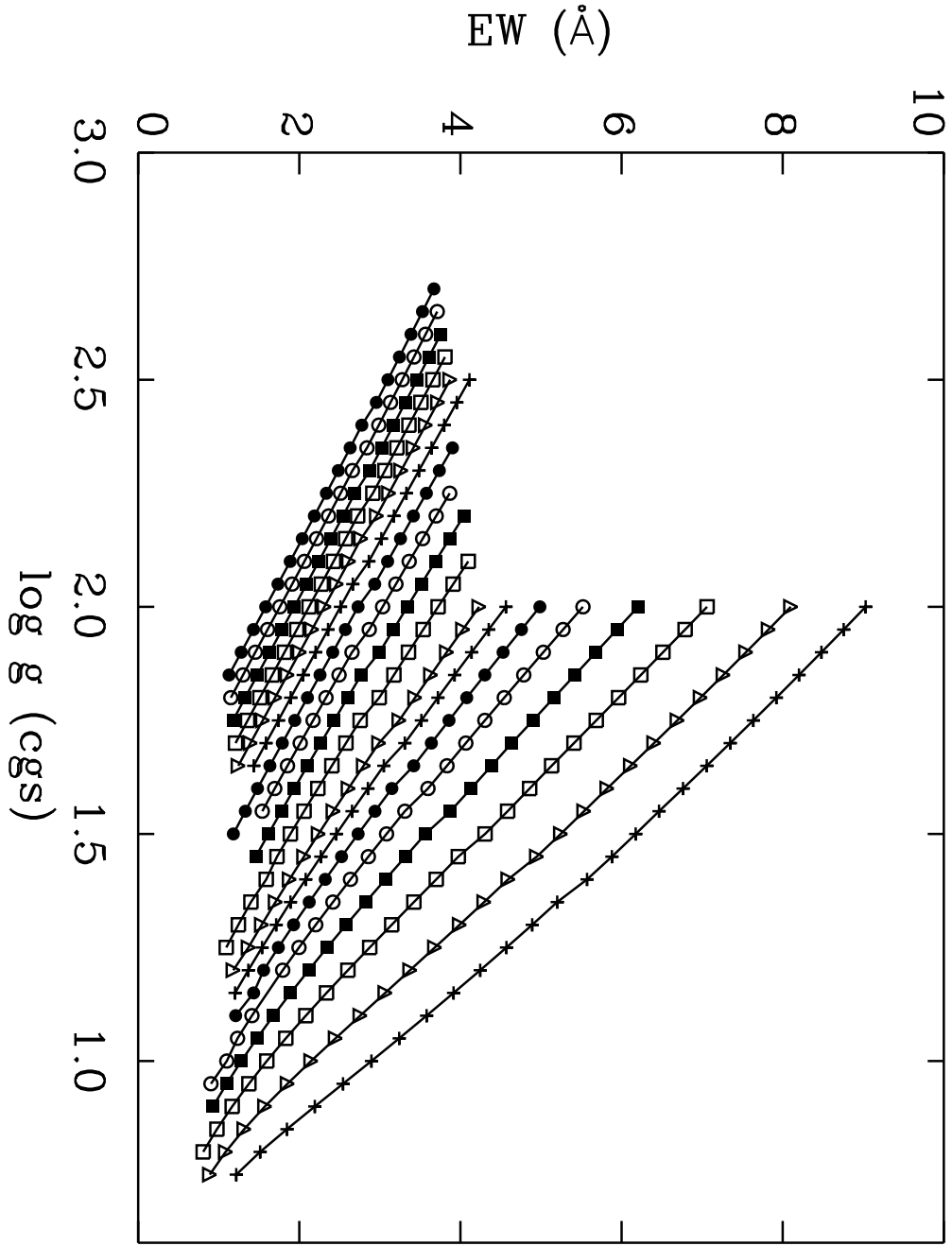}
\caption{Equivalent width of H${\delta}$ as a function of gravity $log~g$ for 
models of different effective temperature ranging between 8300~K (top curve, 
crosses) to 15000~K (bottom curve, filled circles). The \teff\/ steps can 
be inferred from Fig.~\ref{modelgrid}. The metallicity is [Z] = -0.5, but the 
curves at different metallicity are very similar.}
\label{h6gt}
\end{center}
\end{figure}

\begin{figure}
\begin{center}
\includegraphics[width=0.90\textwidth]{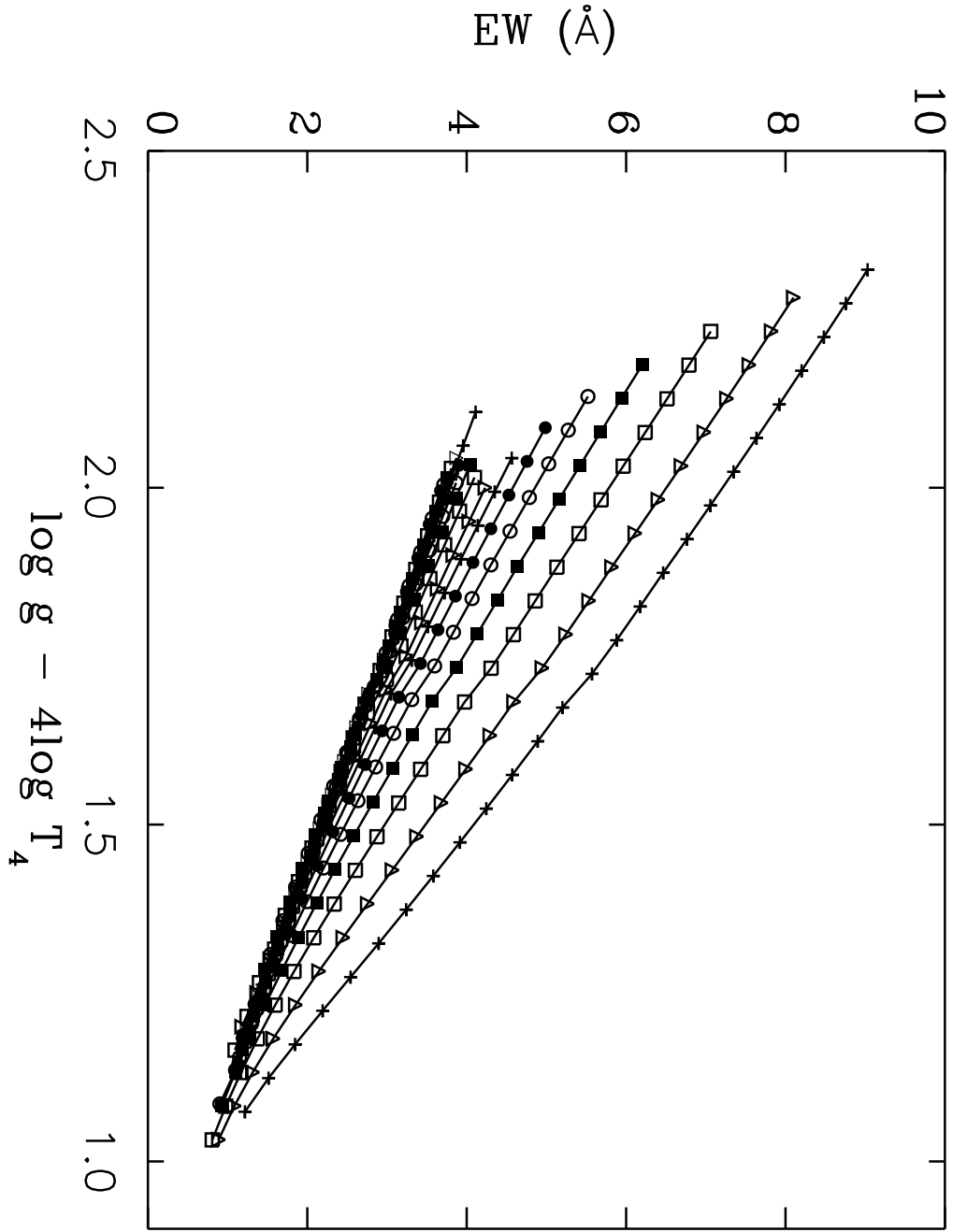}
\caption{Same as Fig.~\ref{h6gt}, but for the flux weighted gravity $log~g_{F}$.}
\label{h6gft}
\end{center}
\end{figure}

\begin{figure}
\begin{center}
\includegraphics[width=0.90\textwidth]{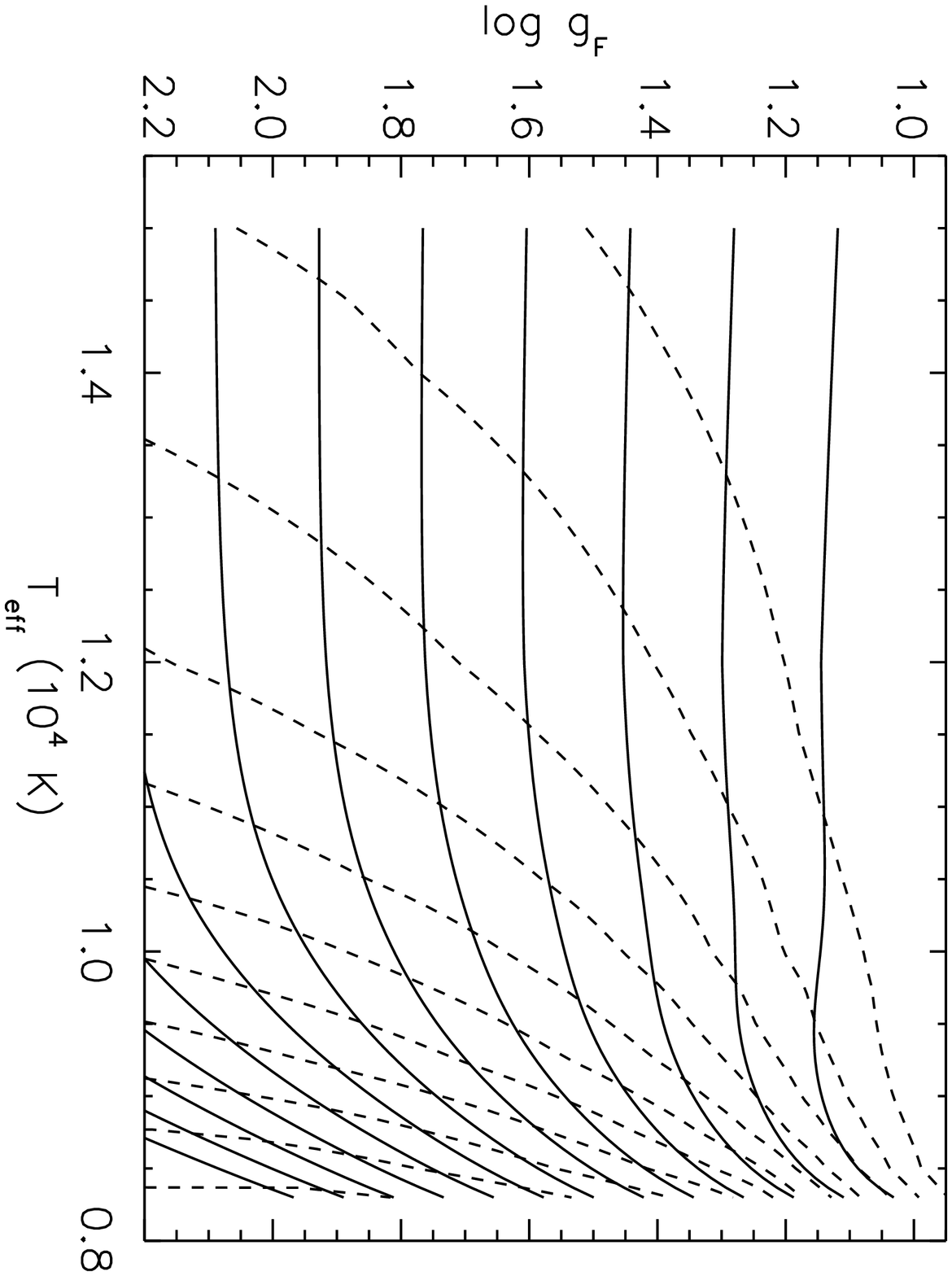}
\caption{Same as Fig.~\ref{isohdbgt} but for the flux weighted gravity $log~g_{F}$ 
instead of gravity $log~g$.}
\label{isohdbgft}
\end{center}
\end{figure}

In Fig.~\ref{h6gt} we show the dependence of the strength of the Balmer line 
H$_{\delta}$ characterized by its equivalent width W$_{\lambda}$(H${\delta})$ 
on gravity $log~g$ and effective temperature \teff\/. Obviously, 
W$_{\lambda}$(H${\delta})$ depends on both, gravity and temperature, and the 
non-horizontal isocontours displayed in Fig.~\ref{isohdbgt} are the result of 
this dependence. The situation changes dramatically, if one looks at the 
dependence of W$_{\lambda}$(H${\delta})$  on flux-weighted gravity $log~g_{F}$ 
as shown in Fig.~\ref{h6gft}. Now all the curves for higher effective temperatures are on 
top of each other and form a temperature independent relationship, which leads 
to practically horizontal isocontours in the $(log~g, log$~\teff) plane 
Fig.~\ref{isohdbgft}. All the other Balmer lines from H${\gamma}$ to H$_{10}$ 
show a very similar behaviour.

This has important consequences for the diagnostics of flux-weighted gravities. 
For objects hot enough the uncertainties of the effective temperature 
determination do not affect the determination of effective gravities. A simple 
measurement of the strengths of Balmer lines either through line profile fitting 
or by simply measuring the equivalent widths gives $log~g_{F}$ with rather 
high precision, as we have already pointed out in section 3.

In terms of distance determination using the FGLR this is very encouraging. It 
means that errors in \teff\/ stemming from the application of an inaccurate 
relationship between spectral types and effective temperatures caused by 
non-solar metallicities, as for instance in the work by \citet{kud03}, are 
less important as long as the flux-weighted gravities $log~g_{F}$ are well 
determined and the objects are hot enough. Errors are only introduced through 
erroneous bolometric corrections and errors in E(B-V), when the observed 
wavelength-dependent magnitudes have to be transformed into de-reddened 
bolometric magnitudes. While this may still be a serious error source and 
should be avoided through the application of our diagnostic technique described
in the previous chapters, the uncertainties of the primary quantity 
in the application of the method are significantly reduced increasing the 
reliability of the FGLR method.

Because of the importance of this effect for distance determinations we have 
investigated the physical reasons leading to Fig.~\ref{h6gft} and~\ref{isohdbgft}. 
We discuss them in the remaining part of this subsection. The strength of 
absorption lines depends on the line strength $\beta_{\nu}$

\begin{equation}
  \beta_{\nu} = \kappa^{L}_{\Delta \nu}/\kappa^{c}_{\nu},
\end{equation}

where $\kappa^{L}_{\Delta \nu}$ is the line absorption coefficient and 
$\kappa^{c}_{\nu}$ the continuum absorption coefficient at the frequency 
of the line. For H${\delta}$ and all other Balmer lines we have 

\begin{equation}
  \kappa^{L}_{\Delta \nu} \propto n_{2} \Phi(\Delta \nu)
\end{equation}

with $n_{2}$ the occupation number in the second level of hydrogen and 
$\Phi(\Delta \nu)$ the line broadening profile. Note that we have neglected the 
influence of stimulated emission, which is appropriate at the temperatures in 
question and for this qualitative discussion. For the supergiants the higher 
Balmer lines are always saturated in the Doppler line cores of the line profiles, 
where they simply reflect the source function in the outer atmosphere. However, 
the contribution to the equivalent widths of these Doppler cores is small. More 
important is the contribution from the line wings, where the line broadening is 
dominated by the Stark effect and the profile function is very well approximated 
by (see \citealt{unsoeld68}, p. 323 or \citealt{mihalas78}, p. 296) 

\begin{equation}
  \Phi(\Delta \nu) \propto n_{E} \Delta \lambda ^{-2.5}.
\end{equation}

$n_{E}$ is the local electron density. The continuum absorption coefficient at 
the wavelength of H$_{\delta}$ is well described by

\begin{equation}
  \kappa^{c}_{\nu} =  n_{E} \sigma _{E} (1 + n_{E} f(\lambda, T) \sigma/ \sigma _{E}).
\end{equation} 

$\sigma$ is a cross section of $10^{-18} cm^{2}$ and $\sigma _{E}$ is the cross 
section for Thompson scattering at free electrons. The function $f(\lambda, T)$ 
contains the contributions of all continuous true absorption processes, but is 
dominated by hydrogen free-free and bound-free absorption. In LTE a good 
approximation (better than 10\%) in the range between 8000~K and 15000~K and at 
the wavelength of H${\delta}$ is

\begin{equation}
 f(\lambda, T) \approx 1.06~10^{-19} T^{-3}_{4}
\end{equation} 

in units of cm$^{3}$. $T_{4}$ is the temperature in units of $10^{4}$~K. NLTE 
effects in the third level of hydrogen can have a small influence on
$f(\lambda, T)$, but will change mostly the value of the constant but not the 
temperature dependence. The occupation number of hydrogen in the second level is 
given by

\begin{equation}
  n_{2} \propto b_{2} n^{2}_{E} T^{-1.5} exp(E_{2}/kT)
\end{equation} 

with $b_{2}$ the non-LTE departure coefficient and $E_{2}$ the 
ionization energy from the second level. k is the Boltzmann constant. For this 
equation we have assumed that most of the free electrons in the atmospheres come 
from hydrogen, a valid approximation. In our temperature range we find that we 
can approximate (within 25\%)

\begin{equation}
  T^{-1.5}_{4} exp(E_{2}/kT) \approx 6.3 10^{1} T^{-6}_{4}
\end{equation} 

This leads to the following expression for the line strength $\beta_{\nu}$

\begin{equation}
  \beta_{\nu} \propto (n_{E} T^{-3}_{4})^{2}/(1 + 1.06~10^{-19} n_{E} T^{-3}_{4}\sigma /\sigma _{E}).
\end{equation} 

Thus, the line strength in the wings of the hydrogen Balmer lines is obviously a 
function of electron density devided by the third power of temperature in the 
layers of the formation of line wings, which is close to the continuum optical 
depth $\tau_{\nu}^{c}$ = 2/3. It is straightfoward to estimate  $n_{E} T^{-3}$ in 
these layers. Starting from the hydrostatic equation
\begin{equation}
  dP/d \tau^{c}_{\nu} = -g(1-\Gamma) \rho /\kappa^{c}_{\nu}
\end{equation} 

and with $\rho = m_{p}n_{p}(1 + 4Y),~n_{p} = n_{E},~P = n_{E}(2 + Y)kT$ and $\rho$, 
P, $n_{p}$, Y being mass density, gas pressure, proton number density and number 
ratio of helium to hydrogen, respectively, we obtain a differential equation for 
the electron density as a function of continuum optical depth (making the usual 
assumption that the logarithmic density gradient is much larger than the 
logarithmic temperature gradient) , which reads

\begin{equation}
  (1 + n_{E}f(\lambda,T) \sigma/\sigma_{E})dn_{E} = A(g,\Gamma,T)d\tau^{c}_{\nu},
\end{equation}  

where

\begin{equation}
  A(g,\Gamma,T) = g(1-\Gamma)m_{p}(1 + 4Y)/(\sigma_{E}(2 + Y)kT)
\end{equation} 

$\Gamma$ is the usual Eddington-ratio of Thompson scattering radiative acceleration 
devided by gravity, namely (assuming again n$_{E}$ = n$_{p}$)

\begin{equation}
  \Gamma = 7.49~T^{4}_{4}(1 + Y)/(g(1 + 4Y) ~ \propto g_{F}^{-1} 
\end{equation} 

The integration of this differential equation is simple (again regarding the 
temperature gradient as negligible). We obtain at every continuum optical 
depth$\tau_{c}$  

\begin{equation}
n_{E} + n_{E}^{2} \sigma f(\lambda,T)/(2\sigma_{E}) = A(g,\Gamma,T)\tau^{c}_{\nu}
\end{equation} 

or

\begin{equation}
n_{E}\sigma f(\lambda,T)/\sigma_{E} = (1 + 2A(g,\Gamma,T)\tau^{c}_{\nu}\sigma f(\lambda,T)/\sigma_{E})^{1/2}-1.
\end{equation} 

With $f(\lambda,T) \propto T^{-3}$ we see immediately that

\begin{equation}
n_{E}T^{-3} \propto (1 + g_{F}(1-\Gamma)*const.)^{1/2}.
\end{equation} 

which means that the line strength $\beta_{\nu}$ depends solely on the flux 
weighted gravity and not on effective temperature. We also note that 
depending on whether the second term under the square root is small or large 
compared with unity (i.e. $g_{F}(1-\Gamma)$ small or large)
we will have $n_{E}T^{-3} \propto g_{F}(1-\Gamma)$ or 
$\propto (g_{F}(1-\Gamma))^{0.5}$, respectively. Numerical solutions of the 
exact problem including full NLTE opacities and the temperature gradient 
show that this simple estimate is qualitatively correct and fairly accurate.

These simple calculations explain the physical reason why the the strengths 
of the Balmer lines depend mostly on the flux weighted gravity and why the 
isocontours of the equivalent widths in Fig.~\ref{isohdbgft} are horizontal. 
However, they do not explain, why the isocontours bend upwards at higher 
temperature. The reason is the recombination of hydrogen at lower temperatures. 
While it is  intuitively clear that recombination must lead to stronger hydrogen 
lines and thus to an upward turn of the isocontours, we verified this by simple 
numerical calculations solving the same problem as above but including the 
effects of recombination through the Saha formula. We also included the effects of atmospheric 
temperature stratifications by allowing for appropriate T($\tau^{c}_{\nu}$) 
relationships. At lower \teff\/ and higher $log~g$ hydrogen partly recombines. 
$\kappa^{c}_{\nu}$ becomes smaller ($\propto n_{E}^{2}$) and $\tau^{c}_{\nu}$ 
is reached in much larger geometrical depth of much higher pressure and total 
number density. As a result, the effect of recombination is overcompensated and $n_{E}$ 
at the optical depth of line wing formation is much larger than in the 
calculation without recombination at these temperatures and gravities.

\subsection{The FGLR of NGC 300}

With 24 A supergiants studied in this work and the 6 early type B supergiants 
investigated by \citet{urbaneja05} we have a large sample, which we can use for 
an observational test of the FGLR. For the A supergiants, all the data and their 
uncertainties have been discussed in the previous chapters. A similar 
discussion has been given by \citet{urbaneja05} for the B supergiants. Note, however, 
that \citet{urbaneja05} used the ground-based photometry by 
\citet{bresolin02}, which as discussed above has been superseded by the 
HST/ACS photometry presented and discussed by \citet{bresolin05}. With the effective 
temperatures, gravities and intrinsic energy distributions determined for 
each B supergiant by \citet{urbaneja05} we have used the new photometry 
to re-determine interstellar reddening and extinction, which then yields new 
stellar radii and absolute V and bolometric magnitudes. Note that 
\citet{urbaneja05} also used the old distance modulus by \citet{freedman01}, 
which is significantly different from the new distance modulus determined by 
\citet{gieren05} based on on optical and IR observations of a large sample 
of Cepheids minimizing the errors caused by reddening.

\begin{figure}
\begin{center}
\includegraphics[width=0.90\textwidth]{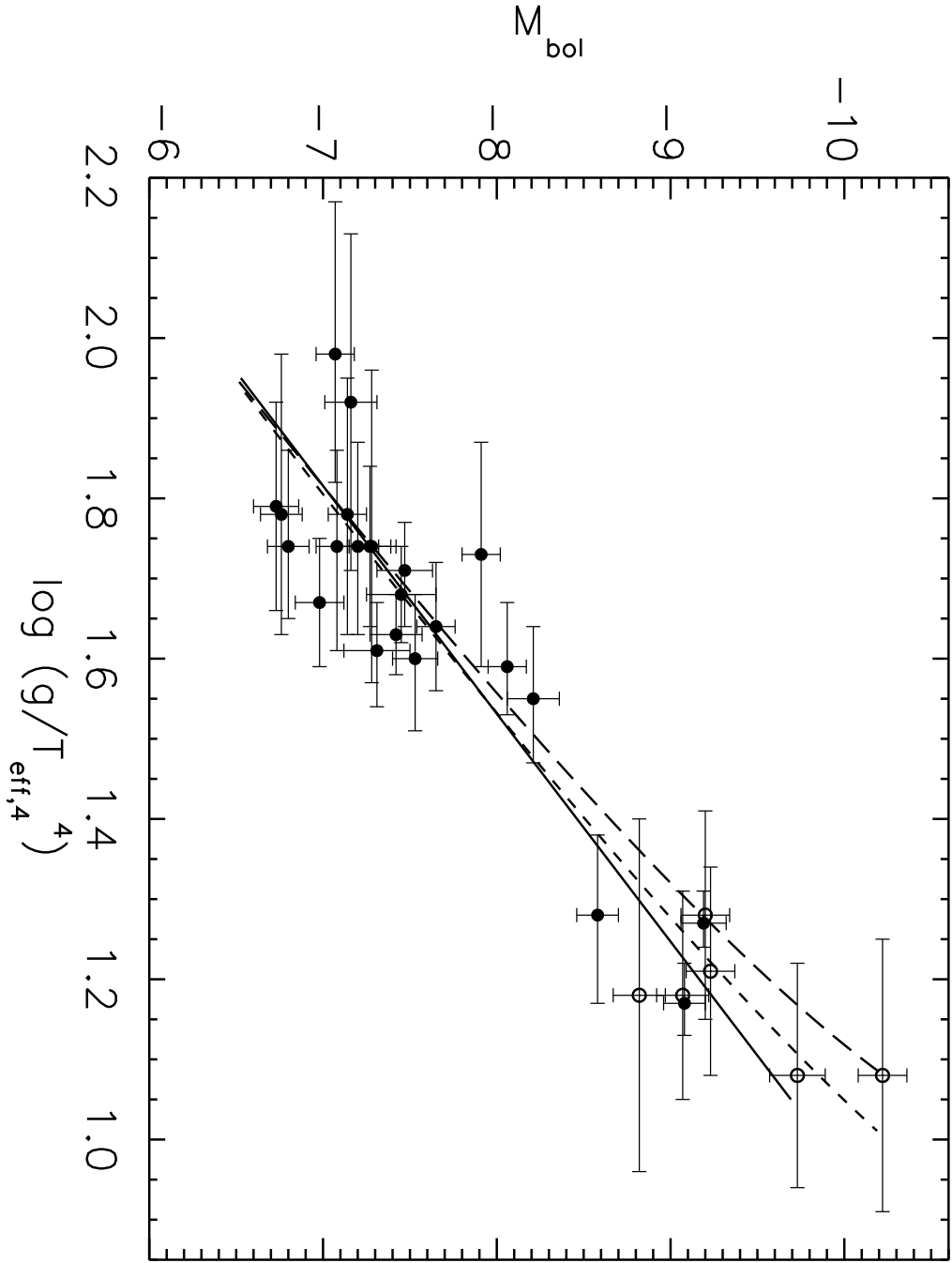}
\caption{The FGLR of A (solid circles) and B (open circles) supergiants in 
NGC 300 and the linear regression (solid). The stellar evolution 
FGLRs of Fig.~\ref{fglrev} for models with rotation are also overplotted 
(dashed: Milky Way metallicity, long-dashed: SMC metallicity).}
\label{fglrngc300}
\end{center}
\end{figure}

The result is shown in Fig.~\ref{fglrngc300}, which reveals a clear and rather 
tight relationship of flux weighted gravity $log~g_{F}$ with bolometric magnitude 
$M_{bol}$. A simple linear regression yields $b_{FGLR}$ = 8.11 for the zero point and 
$a_{FGLR}$ = -3.52 for the slope. The standard deviation from this relationship is 
$\sigma$ = 0.34 mag.

We also overplot the FGLRs predicted by stellar evolution theory (\citealt{meynet05}) as discussed 
in the previous paragraphs. At lower luminosities (or large $log~g_{F}$) they agree 
well with the observed linear regression. In particular, the FGLR for Milky Way
metallicity is very close. For higher luminosities (or small $log~g_{F}$) the linear 
regression and the stellar evolution FGLRs diverge because of the curvature of the 
latter, which is caused by the fact that the exponent of the mass-luminosity 
relationship becomes smaller. Looking at the scatter of the data points around 
either the linear regression or the FGLR for Milky Way metallicity it is hard 
to tell which of the two is the better fit. Approximating the stellar evolution
FGLRs by

\begin{equation}
\begin{array}{c}
\displaystyle M_{bol} = \alpha(log~g_{F})(log~g_{F}-x_{0}) + M_{bol}^{0} \\
\displaystyle \alpha(log~g_{F}) = \alpha_{0} + \alpha_{1}(x_{0} - log~g_{F}) + \alpha_{2}(x_{0} - log~g_{F})^{2} \\
\end{array} 
\end{equation}

%
%
(values for $\alpha_{0}, \alpha_{1}, \alpha_{2}$ and $x_{0}$ are given in Table 4) 
we can calculate standard deviations to characterize the scatter around those. For 
Milky Way metallicity we obtain $\sigma$ = 0.35 mag, practically identical to 
scatter around the linear regression. The FGLR for SMC metallicity has a much 
stronger curvature (as the result of less mass-loss in the O-star stage of stellar 
evolution) and is shifted towards higher luminosities and seems to be worse fit of 
the data. This is confirmed by the standard deviation of $\sigma$ = 0.41 mag for this 
relationship. Applying a vertical shift of $\Delta M_{b0l}^{0}$ = 0.16 mag to the SMC 
metallicity relationship minimizes its standard deviation to $\sigma$ = 0.38 mag, 
still worse than for the linear regression and the FGLR for Milky Way metallicity.
We note that most of our objects have metallicities between the Milky Way and the 
SMC comparable to the LMC. Unfortunatly, no complete set of models including the 
effects of rotation in the mass range needed is available at this metallicity at 
the moment.

The new linear regression coefficients for the FGLR in NGC 300 are different from 
those obtained by \citet{kud03} ($a_{FGLR}$ = -3.85 and $b_{FGLR}$ = 7.96) 
resulting now in a higher luminosity zero point and a somewhat shallower slope. This 
is caused by a combination of different factors, a new distance modulus to NGC 300, 
improved HST photometry leading to larger extinction and a systematic change of stellar 
parameters as the result of the new diagnostic technique of the low resolution spectra.

\subsection{The FGLR averaged over eight galaxies}

In their first investigation of the empirical FGLR \citet{kud03} have added A supergiants 
from six Local Group galaxies with stellar parameters obtained from quantitative studies 
of high resolution spectra (Milky Way, LMC, SMC, M31, M33, NGC 6822) to their results for NGC 300 to 
obtain a larger sample. They also added 4 objects from the spiral galaxy NGC 3621 
(at 6.7 Mpc) which were studied at low resolution. We will add exactly the same 
data set to our new enlarged NGC 300 sample, however, with a few minor modifications. 
For the Milky Way we include the latest results from \citet{przybilla06} and \citet{schiller07} and 
for the two objects in M31 we use the new stellar parameters obtained by 
\citet{przybilla06b}. For the objects in NGC 3621 we apply new HST photometry. 
(Note that for the 
objects \teff\/ and $log~g$ were obtained from a spectral fit assuming LMC 
metallicity, i.e. the detailed technique developed here for the NGC 300 objects has 
not yet been applied. However, eliminating those objects from the sample does not 
change the results significantly). We have also re-analyzed the LMC objects using ionization equilibria for the temperature determination.
The results are summarized in Table 5. The bolometric corrections for all objects are calculated using eq.(6).

\begin{figure}
\begin{center}
\includegraphics[width=0.90\textwidth]{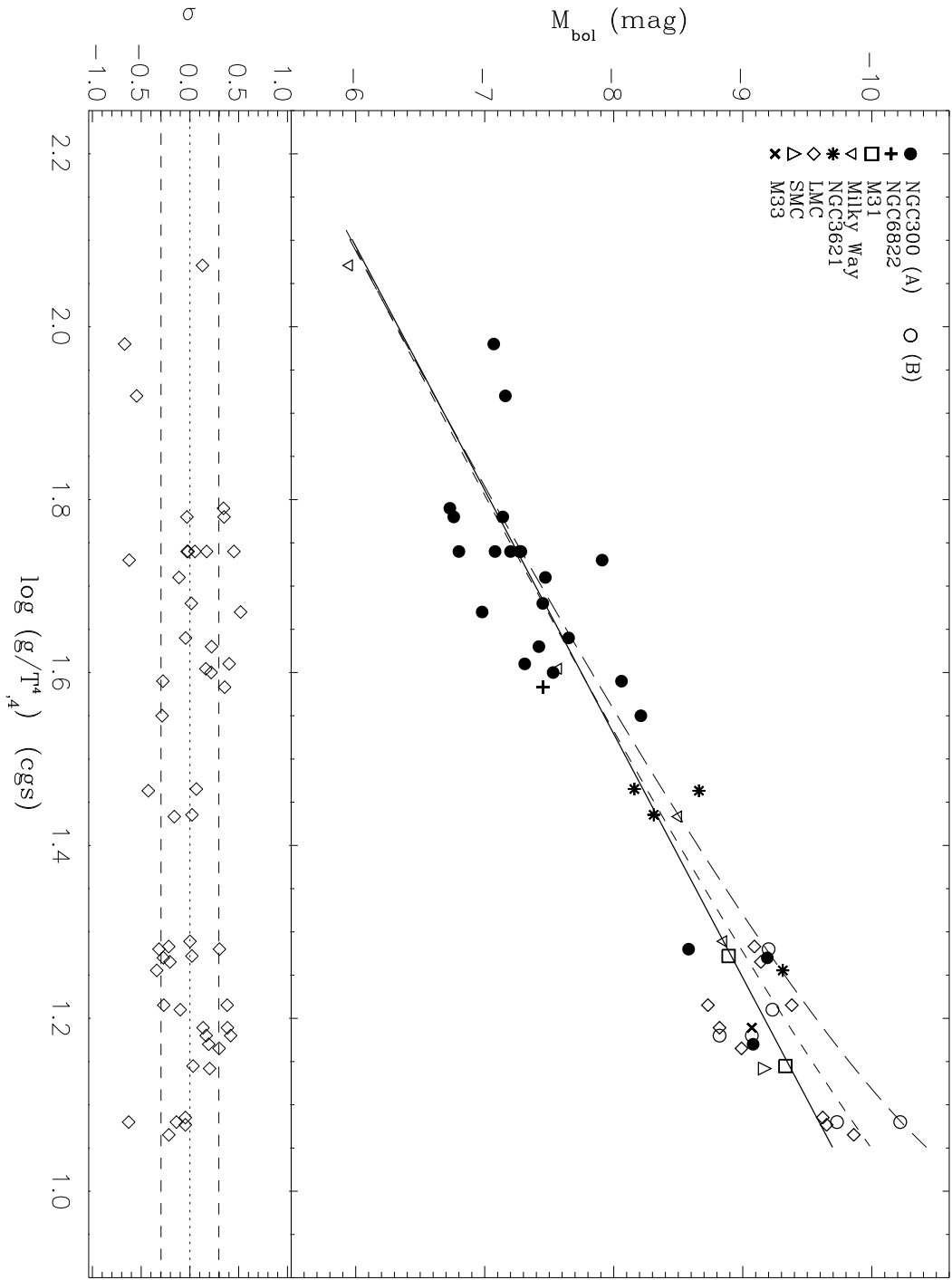}
\caption{Same as Fig.~\ref{fglrngc300}, but with additional objects from 7
other galaxies (see text).}
\label{fglrall}
\end{center}
\end{figure}

Fig.~\ref{fglrall} shows bolometric magnitudes and flux-weighted gravities for this 
full sample of eight galaxies revealing a tight relationship over one order of 
magnitude in flux-weighted gravity. The linear regression coefficients are 
$a_{FGLR} = -3.41\pm{0.16}$ and $b_{FGLR} = 8.02\pm{0.04}$, very simimilar to the NGC 300 sample 
alone. The standard deviation is 
$\sigma$ = 0.32 mag. The new coefficients are slightly different from the 
previous work by \citet{kud03} ($a_{FGLR}$ = -3.71 and $b_{FGLR}$ = 7.92) 
resulting in a FGLR brighter by 0.1 to 0.2 mag now in much closer agreement with 
stellar evolution. Compared with the new results the stellar evolution FGLR 
for Milky Way metallicity 
provides a fit of almost similar quality with a standard deviation of 
$\sigma$ = 0.31 mag. The stellar evolution FGLR fit for 
SMC metallicity is worse with $\sigma$ = 0.42 mag 
(applying $\Delta M_{b0l}^{0}$ = 0.20 mag reduces the standard deviation 
to $\sigma$ = 0.35 mag).

We conclude that the simple linear regression fit of this sample is the best way
to describe the empirical FGLR at this point. It is basically in agreement with 
stellar evolution theory, at least if mass-loss and rotational mixing affect the 
evolution as for the models with solar metallicity. We note that for the 
evolutionary models accounting for mass-loss and mixing based on SMC metallicity 
we have an offset relative to the observed FGLR by about 0.20 mag and a curvature 
stronger than observed. Again, we stress the need for a complete set of models 
including the effects of rotation at LMC metallicities. 

\section{Conclusions and Future Work}\label{sec_concl}

The goal of this work has been to demonstrate the astrophysical potential of low 
resolution spectroscopy of A supergiant stars in galaxies beyond the Local Group. 
By introducing a novel method for the quantitative spectral analysis we are 
able to determine accurate stellar parameters, which allow us to test stellar 
evolution models including the evolutionary time scales in the A supergiant stage. 
Through the spectroscopic determination of stellar parameters we can also 
constrain interstellar reddening and extinction by comparing the calculated SED 
with broad band photometry. We find a very patchy extintion pattern as to be 
expected for a star forming spiral galaxy. The average extinction is in agreement 
with multi-wavelength studies of Cepheids including K-band photometry.

The method also allows to determine stellar metallicities and to study stellar 
metallicity gradients. We find a metallicity close to solar in the center of 
NGC 300 and a gradient of $-$0.08 dex/kpc. To our knowledge this is the first 
systematic stellar metallicity study in galaxies beyond the Local Group 
focussing on iron group elements. In the future the method can be extended to 
not only determine metallicity but also the ratio of $\alpha$- to iron group 
elements as a function of galactocentric distance. The stellar metallicities 
obtained can be compared with oxygen abundance studies of HII regions using 
the strong line method. This allows us to discuss the various calibrations 
of the strong line method, which usually yield very different results.

The improved spectral diagnostic method presented here enables us to very 
accurately determine stellar flux weighted gravities $log~g_{F}$ = $log~g$/\teffq\/ 
and bolometric magnitudes, which we explain by a detailed discussion of the 
physical background of the spectroscopic diagnostics. We find that above a 
certain threshold in effective temperature a simple measurement of the strengths 
of the Balmer lines can be used to determine accurate values of $log~g_{F}$.

Absolute bolometric magnitudes $M_{bol}$ and flux-weighted gravities $log~g_{F}$ 
are tightly correlated. It is shown that such a correlation is expected for 
stars, which evolve at constant luminosity and mass. We discuss the observed 
``flux-weighted gravity - luminosity relationship (FGLR)'' in detail and 
compare with stellar evolution theory. We find resonable agreement with 
evolutionary tracks which account for mass-loss and rotational mixing assuming 
solar metallicity. The agreement is less good for similar tracks with SMC 
metallicity.

With a relatively small residual scatter of $\sigma$ = 0.3 mag the observed 
FGLR is an excellent tool to determine accurate spectroscopic distance to galaxies.
It requires multicolor photometry and low resolution ($5\AA$) spectroscopy to
determine effective temperature and gravity and, thus, flux-weighed gravity 
directly from the spectrum using the self-consistent method developed here.
With effective temperature, gravity and metallicity determined we also know the 
bolometric correction, which is small for A supergiants, which means that errors 
in the stellar parameters do not largely affect the determination of bolometric 
magnitudes. Moreover, we know the intrinsic stellar SED and, therefore, can 
determine interstellar reddening and extinction from the multicolor photometry, 
which will allow the accurate determination of the reddening-free apparent 
bolometric magnitude. The application of the FGLR will then yield absolute 
magnitudes and, thus, the distance modulus. With the intrinsic scatter of 
$\sigma$ = 0.3 mag and 30 targets per galaxy one can estimate an accuracy of 
0.05 mag in distance modulus (0.1 mag for 10 target stars).

The advantage of the FGLR method for distance determinations is its 
spectroscopic nature, which provides significantly more information about 
the physical status of the objects used for the distance determination than 
simple photometry methods. Most importantly, metallicity and interstellar 
extinction can be determined directly. The latter is crucial for spiral and 
irregular galaxies because of the intrinsic patchiness of reddening and 
extinction.

Since supergiant stars are known to show intrinsic photometric variability, 
the question arises whether the FGLR method is affected by such variability. 
For the our targets in NGC 300 this issue has been carefully investigated 
by \citet{bresolin04}, who studied their CCD photometry lightcurves 
obtained over many epochs in the parallel search for Cepheids in NGC 300. They 
concluded that amplitudes of photometric variability are very small and do 
not affect distance determinations using the FGLR method. The standard 
deviations from the mean V magnitude for subset of A supergiants identified 
as variable ranges between 0.03 to 0.05 magnitudes, with maximum amplitudes 
of the variability between 0.08 to 0.23 mag. This is clearly within the one 
$\sigma$ uncertainty found for the FGLR.

The effects of crowding and stellar multiplicity are also important. However,
in this regard A supergiants offer tremendous advantages relative to other 
stellar distance indicators. First of all, they are significantly brighter.
\citet{bresolin05} using HST ACS photometry compared to ground-based 
photometry have studied the effects of crowding on the Cepheid distance 
to NGC 300 and concluded that they are negligible. With A supergiants being 
3 to 6 magnitudes brighter than Cepheids it is clear that even with 
ground-based photometry only crowding is generally not an issue for these 
objects at the distance of NGC 300 and, of course, with HST photometry 
(and in the future JWST) one can reach much larger distances before crowding 
becomes important. In addition, any significant contribution by additional 
objects to the light of an A supergiant will become apparent in the spectrum, 
if the contaminators are not of a very similar spectral type, which is very 
unlikely because of the short evolutionary lifetime in the A supergiant stage. 
It is also important to note that A supergiants have evolutionary ages larger 
than 10 million years, which means that they have time to migrate into the 
field or that they are found in older clusters, which are usually less concentrated
than the very young OB associations.

It is evident that the type of work described in this paper can be 
in a straightforward way extended to the many spiral galaxies in the local volume 
at distances in the 4 to 7 Mpc range. Pushing the method we estimate 
that with present day 8m to 10m class telescopes and the existing very efficient 
multi-object spectrographs one can reach down with sufficient S/N to V = 22.5 mag in two nights of 
observing time under very good conditions. For objects brighter than 
$M_{V}$ = -8 mag this means metallicities and distances can be determined out 
to distances of 12 Mpc (m-M = 30.5 mag). This opens up a substantial volume 
of the local universe for metallicity and galactic evolution studies and 
independent distance determinations complementary to the existing methods. With 
the next generation of extremely large telescopes such as the TMT, GMT or the 
E-ELT the limiting magnitude can be pushed to V = 24.5 equivalent to 
distances of 30 Mpc (m-M = 32.5 mag).

Acknowledgements: WG and GP gratefully acknowledge financial support from the 
Chilean FONDAP Center of Astrophysics under grant 15010003, and from the Chilean 
Centro de Astrofisica y Tecnologias Afines (CATA). We like to thank 
Vivan U and John Hillier for the careful reading of the manuscript and 
stimulating discussion. We also thank the anonymous referee for his constructive
suggestions, which helped to improve the paper.

\clearpage

---------------------------------------------------------------------

\clearpage

\clearpage

\clearpage

\clearpage

\clearpage

\clearpage

\clearpage

\begin{deluxetable}{ccccccccccc}
\tabletypesize{\scriptsize}
\tablecolumns{11}
\tablewidth{0pt}
\tablecaption{NGC 300 - Spectroscopic targets and stellar parameters}

\tablehead{
\colhead{No.}     &
\colhead{name}            &
\colhead{$\rho/\rho_{0}$}           &
\colhead{spec.}     &
\colhead{$D_{B}$}     &
\colhead{$T_{eff}$}             &
\colhead{$log~g$}     &
\colhead{$log~g_{F}$} &
\colhead{[Z]}    &
\colhead{E(B-V)} & 
\colhead{}\\
\colhead{}   &
\colhead{}         &
\colhead{\tablenotemark{a}}         &
\colhead{type}     &
\colhead{}   &
\colhead{K}         &
\colhead{cgs}         &
\colhead{cgs}		&
\colhead{}			&
\colhead{mag}       &
\colhead{}\\[1mm]
\colhead{(1)}	&
\colhead{(2)}	&
\colhead{(3)}	&
\colhead{(4)}	&
\colhead{(5)}	&
\colhead{(6)}	&
\colhead{(7)}	&
\colhead{(8)}	&
\colhead{(9)}	&
\colhead{(10)}	&
\colhead{(11)}}
\startdata
\\[-1mm]
0  & A8  & 1.03 & B9-A0 &      & 10000$\rm^{+500}_{-500}$ & 1.60$\rm^{+0.17}_{-0.18}$ & 1.60$\rm^{+0.08}_{-0.09}$ & -0.44$\pm{0.15}$ & 0.09 & \tablenotemark{b}  \\[2pt]
1  & A10 & 0.81 &  A2   & 0.37 &  9250$\rm^{+270}_{-130}$ & 1.45$\rm^{+0.13}_{-0.08}$ & 1.59$\rm^{+0.08}_{-0.06}$ & -0.04$\pm{0.20}$ & 0.16 &     \\[2pt]
2  & A11 & 0.94 &  B8   & 0.08 & 10500$\rm^{+500}_{-500}$ & 1.25$\rm^{+0.14}_{-0.12}$ & 1.17$\rm^{+0.05}_{-0.04}$ & -0.65$\pm{0.15}$ & 0.16 & \tablenotemark{d}    \\[2pt]
3  & C6  & 0.85 &  A0   & 0.34 &  9750$\rm^{+400}_{-400}$ & 1.70$\rm^{+0.17}_{-0.18}$ & 1.74$\rm^{+0.10}_{-0.10}$ & -0.45$\pm{0.20}$ & 0.16 &     \\[2pt]
4  & C16 & 0.70 &  B9   & 0.13 & 10500$\rm^{+500}_{-500}$ & 1.35$\rm^{+0.12}_{-0.11}$ & 1.27$\rm^{+0.04}_{-0.03}$ & -0.43$\pm{0.15}$ & 0.14 & \tablenotemark{c,d}   \\[2pt]
5  & D12 & 0.64 &  A2   & 0.28 &  9000$\rm^{+400}_{-400}$ & 1.10$\rm^{+0.17}_{-0.19}$ & 1.28$\rm^{+0.10}_{-0.11}$ & -0.16$\pm{0.15}$ & 0.24 &     \\[2pt]
6  & A6  & 0.98 & A1-A2 & 0.56 &  8500$\rm^{+270}_{-200}$ & 1.50$\rm^{+0.22}_{-0.19}$ & 1.78$\rm^{+0.17}_{-0.15}$ & -0.48$\pm{0.30}$ & 0.17 &     \\[2pt]
7  & A13 & 0.83 &  B8   & 0.22 & 11500$\rm^{+730}_{-600}$ & 1.95$\rm^{+0.17}_{-0.16}$ & 1.71$\rm^{+0.06}_{-0.07}$ & -0.48$\pm{0.30}$ & 0.11 &     \\[2pt]
8  & A18 & 0.93 &  B8   & 0.19 & 12000$\rm^{+930}_{-670}$ & 2.00$\rm^{+0.19}_{-0.16}$ & 1.68$\rm^{+0.06}_{-0.06}$ & -0.36$\pm{0.30}$ & 0.13 & \tablenotemark{c} \\[2pt]
9  & B5  & 0.35 &  B8   & 0.23 & 11000$\rm^{+600}_{-530}$ & 1.80$\rm^{+0.14}_{-0.14}$ & 1.63$\rm^{+0.05}_{-0.05}$ & -0.29$\pm{0.20}$ & 0.11 & \tablenotemark{c}\\[2pt]
10 & B8  & 0.26 &  A0   &      &  9500$\rm^{+400}_{-400}$ & 1.65$\rm^{+0.19}_{-0.20}$ & 1.74$\rm^{+0.12}_{-0.13}$ & -0.14$\pm{0.25}$ & 0.12 & \tablenotemark{b,c} \\[2pt]
11 & B10 & 0.14 & A2-A3 & 0.40 &  9250$\rm^{+530}_{-330}$ & 1.60$\rm^{+0.32}_{-0.24}$ & 1.78$\rm^{+0.20}_{-0.15}$ & -0.07$\pm{0.30}$ & 0.15 &     \\[2pt]
12 & B11 & 0.13 &  A4   & 0.54 &  8500$\rm^{+330}_{-200}$ & 1.50$\rm^{+0.27}_{-0.19}$ & 1.74$\rm^{+0.22}_{-0.17}$ & -0.09$\pm{0.30}$ & 0.11 &     \\[2pt]
13 & B19 & 0.36 & B8-B9 & 0.26 & 10500$\rm^{+530}_{-530}$ & 1.75$\rm^{+0.16}_{-0.18}$ & 1.67$\rm^{+0.08}_{-0.08}$ & -0.30$\pm{0.20}$ & 0.15 &     \\[2pt]
14 & C1  & 1.18 &  A1   & 0.49 &  8750$\rm^{+270}_{-270}$ & 1.50$\rm^{+0.19}_{-0.20}$ & 1.73$\rm^{+0.14}_{-0.14}$ & -0.33$\pm{0.25}$ & 0.15 &     \\[2pt]
15 & C8  & 0.92 &  B9   & 0.19 & 11500$\rm^{+870}_{-670}$ & 1.85$\rm^{+0.19}_{-0.17}$ & 1.61$\rm^{+0.06}_{-0.07}$ & -0.45$\pm{0.30}$ & 0.12 &     \\[2pt]
16 & C9  & 0.93 &  B9   & 0.39 &  9500$\rm^{+470}_{-400}$ & 1.70$\rm^{+0.21}_{-0.20}$  & 1.79$\rm^{+0.13}_{-0.13}$ & -0.44$\pm{0.30}$ & 0.12 &     \\[2pt]
17 & C12 & 0.90 & B9-A0 & 0.49 &  8750$\rm^{+270}_{-330}$ & 1.45$\rm^{+0.18}_{-0.23}$ & 1.68$\rm^{+0.13}_{-0.16}$ & -0.54$\pm{0.25}$ & 0.09 &     \\[2pt]
18 & D2  & 0.57 &  A1   & 0.41 &  9250$\rm^{+400}_{-270}$ & 1.60$\rm^{+0.21}_{-0.16}$ & 1.74$\rm^{+0.13}_{-0.11}$ & -0.41$\pm{0.20}$ & 0.16 &     \\[2pt]
19 & D7  & 0.71 &  A4   & 0.56 &  8500$\rm^{+270}_{-200}$ & 1.70$\rm^{+0.24}_{-0.20}$ & 1.98$\rm^{+0.19}_{-0.16}$ & -0.31$\pm{0.20}$ & 0.23 & \tablenotemark{c}   \\[2pt]
20 & D10 & 0.57 &  A4   & 0.63 &  8300$\rm^{+270}_{-270}$ & 1.60$\rm^{+0.26}_{-0.26}$ & 1.92$\rm^{+0.21}_{-0.21}$ & -0.27$\pm{0.15}$ & 0.18 & \tablenotemark{c}   \\[2pt]
21 & D13 & 0.79 &  A0   & 0.27 & 10000$\rm^{+600}_{-470}$ & 1.55$\rm^{+0.19}_{-0.17}$ & 1.55$\rm^{+0.09}_{-0.08}$ & -0.39$\pm{0.15}$ & 0.13 &     \\[2pt]
22 & D17 & 0.82 & B9-A0 & 0.32 &  9750$\rm^{+270}_{-330}$ & 1.60$\rm^{+0.12}_{-0.15}$ & 1.64$\rm^{+0.08}_{-0.08}$ & -0.46$\pm{0.15}$ & 0.15 &     \\[2pt]
23 & D18 & 1.08 &  B9   & 0.41 &  9250$\rm^{+330}_{-200}$ & 1.60$\rm^{+0.18}_{-0.13}$ & 1.74$\rm^{+0.12}_{-0.09}$ & -0.66$\pm{0.20}$ & 0.07 &     \\[2pt]
\enddata
\tablenotetext{a}{Galactocentric distance, in units of the isophotal radius $\rho_{0}$ = 9.75 arcmin $\simeq$ 5.33 kpc}
\tablenotetext{b}{no Balmer jump measured; $T_{eff}$ from HeI lines}
\tablenotetext{c}{no HST photometry available; ground-based photometry used, see text}
\tablenotetext{d}{HeI lines also used for used $T_{eff}$, see text}
\end{deluxetable}

\clearpage

\begin{deluxetable}{ccccccccc}
\tabletypesize{\scriptsize}
\tablecolumns{9}
\tablewidth{0pt}
\tablecaption{Bolometric corrections and magnitudes, radii and masses}

\tablehead{
\colhead{No.}     &
\colhead{name}            &
\colhead{$m_{V}$} &
\colhead{$M_{V}$} &
\colhead{BC} &
\colhead{$M_{bol}$} &
\colhead{$R/R_{\odot}$} &
\colhead{$M/M_{\odot}$} &
\colhead{$M/M_{\odot}$}\\
\colhead{}   &
\colhead{}   &
\colhead{mag}	&
\colhead{\tablenotemark{a} mag}	&
\colhead{mag}	&
\colhead{\tablenotemark{a} mag}	&
\colhead{}      &
\colhead{spec}      &
\colhead{evol}\\[1mm]
\colhead{(1)}	&
\colhead{(2)}	&
\colhead{(3)}	&
\colhead{(4)}	&
\colhead{(5)}	&
\colhead{(6)}	&
\colhead{(7)}   &
\colhead{(8)}	&
\colhead{(9)}}
\startdata
\\[-1mm]
0  & A8  & 19.41 & -7.24 & -0.29 & -7.53$\pm{0.13}$ &  96 & 13.2 & 15.7 \\
1  & A10 & 18.93 & -7.94 & -0.12 & -8.06$\pm{0.05}$ & 142 & 20.8 & 18.8 \\
2  & A11 & 18.27 & -8.59 & -0.49 & -9.08$\pm{0.12}$ & 177 & 20.2 & 27.7 \\
3  & C6  & 19.84 & -7.01 & -0.26 & -7.27$\pm{0.12}$ &  89 & 14.4 & 14.4 \\
4  & C16 & 18.07 & -8.75 & -0.44 & -9.19$\pm{0.13}$ & 186 & 28.1 & 29.0 \\
5  & D12 & 18.66 & -8.45 & -0.13 & -8.58$\pm{0.04}$ & 191 & 16.7 & 22.7 \\
6  & A6  & 19.78 & -7.11 & -0.03 & -7.14$\pm{0.05}$ & 110 & 14.0 & 13.9 \\
7  & A13 & 19.85 & -6.86 & -0.61 & -7.47$\pm{0.13}$ &  70 & 16.0 & 15.4 \\
8  & A18 & 19.99 & -6.79 & -0.66 & -7.45$\pm{0.17}$ &  64 & 14.8 & 15.3 \\
9  & B5  & 19.77 & -6.94 & -0.48 & -7.42$\pm{0.11}$ &  75 & 12.9 & 15.2 \\
10 & B8  & 19.83 & -6.89 & -0.19 & -7.08$\pm{0.12}$ &  86 & 12.0 & 13.6 \\
11 & B10 & 19.69 & -7.14 & -0.14 & -7.28$\pm{0.10}$ &  99 & 14.3 & 14.5 \\
12 & B11 & 19.95 & -6.76 &  0.00 & -6.76$\pm{0.06}$ &  93 &  9.9 & 12.3 \\
13 & B19 & 20.23 & -6.60 & -0.38 & -6.98$\pm{0.06}$ &  67 &  9.2 & 13.2 \\
14 & C1  & 18.99 & -7.85 & -0.06 & -7.91$\pm{0.05}$ & 148 & 25.3 & 17.8 \\
15 & C8  & 20.01 & -6.74 & -0.57 & -7.31$\pm{0.16}$ &  65 & 10.9 & 14.6 \\
16 & C9  & 20.22 & -6.52 & -0.21 & -6.73$\pm{0.09}$ &  73 &  9.7 & 12.2 \\
17 & C12 & 20.34 & -6.31 & -0.06 & -6.37$\pm{0.04}$ &  73 &  5.5 & 11.0 \\
18 & D2  & 19.83 & -7.04 & -0.16 & -7.20$\pm{0.07}$ &  96 & 13.3 & 14.1 \\
19 & D7  & 20.02 & -7.06 & -0.01 & -7.07$\pm{0.05}$ & 107 & 20.8 & 13.6 \\
20 & D10 & 19.75 & -7.18 &  0.02 & -7.16$\pm{0.05}$ & 117 & 19.7 & 13.9 \\
21 & D13 & 18.86 & -7.92 & -0.29 & -8.21$\pm{0.11}$ & 131 & 22.0 & 19.8 \\
22 & D17 & 19.44 & -7.40 & -0.25 & -7.65$\pm{0.05}$ & 106 & 16.3 & 16.3 \\
23 & D18 & 19.96 & -6.62 & -0.18 & -6.80$\pm{0.06}$ &  80 &  9.2 & 12.5 \\
\enddata
\tablenotetext{a}{distance modulus = 26.37 mag, \citet{gieren05}}
\end{deluxetable}

\clearpage

\begin{deluxetable}{ccccc}
\tabletypesize{\scriptsize}
\tablecolumns{5}
\tablewidth{0pt}
\tablecaption{Central metallicity [Z] and metallicity gradient in NGC 300}

\tablehead{
\colhead{Source}     &
\colhead{Central abundance}            &
\colhead{Gradient} &
\colhead{Gradient} &
\colhead{Comments}\\
\colhead{}   &
\colhead{dex}   &
\colhead{$\rho/\rho_{0}$}	&
\colhead{dex/kpc}	&
\colhead{}\\[1mm]
\colhead{(1)}	&
\colhead{(2)}	&
\colhead{(3)}	&
\colhead{(4)}	&
\colhead{(5)}}
\startdata
\\[-1mm]
Dopita \& Evans(1986)     &  0.26$\pm{0.17}$ & -0.63$\pm{0.10}$ & -0.118$\pm{0.019}$ & HII, oxygen  \\
Zaritsky et al. (1994)    &  0.29$\pm{0.04}$ & -0.54$\pm{0.09}$ & -0.101$\pm{0.017}$ & HII, oxygen  \\
Kobulnicky et al. (1999)  &  0.06$\pm{0.04}$ & -0.27$\pm{0.09}$ & -0.051$\pm{0.017}$ & HII, oxygen  \\
Denicolo et al. (2002)    & -0.08$\pm{0.05}$ & -0.46$\pm{0.10}$ & -0.086$\pm{0.019}$ & HII, oxygen  \\
Pilyugin (2001)           & -0.17$\pm{0.06}$ & -0.28$\pm{0.12}$ & -0.053$\pm{0.023}$ & HII, oxygen  \\
Pettini \& Pagel (2004)   & -0.19$\pm{0.04}$ & -0.36$\pm{0.08}$ & -0.068$\pm{0.015}$ & HII, oxygen  \\
This work                 & -0.06$\pm{0.09}$ & -0.44$\pm{0.12}$ & -0.083$\pm{0.023}$ & stars, metals\\
\enddata
\end{deluxetable}

\begin{deluxetable}{ccccccccccc}
\tabletypesize{\scriptsize}
\tablecolumns{11}
\tablewidth{0pt}
\tablecaption{Fit coefficients for stellar evolution mass-luminosity relationship and FGLR}

\tablehead{
\colhead{Model}     &
\colhead{Metallicity}            &
\colhead{a} &
\colhead{b} &
\colhead{c} &
\colhead{$log(M^{min}/M_{\odot})$} &
\colhead{$\alpha_{0}$} &
\colhead{$\alpha_{1}$} &
\colhead{$\alpha_{2}$} &
\colhead{$x_{0}$} &
\colhead{$M_{bol}^{0}$}\\[1mm]
\colhead{(1)}	&
\colhead{(2)}	&
\colhead{(3)}	&
\colhead{(4)}	&
\colhead{(5)}	&
\colhead{(6)}	&
\colhead{(7)}	&
\colhead{(8)}	&
\colhead{(9)}	&
\colhead{(10)}	&
\colhead{(11)}}
\startdata
\\[-1mm]
with rotation &  Milky Way & 3.80 & 1.40 & 4.095 & 0.943 & 3.505 & 0.15 & 0.30 & 1.957 & -6.464 \\
with rotation &     SMC    & 3.10 & 1.20 & 4.560 & 1.077 & 3.659 & 0.32 & 0.32 & 1.967 & -6.439 \\
no rotation   &  Milky Way & 4.25 & 2.00 & 3.825 & 0.943 &   --  &  --  &  --  &   --  &   --   \\
no rotation   &     SMC    & 4.10 & 2.05 & 4.030 & 0.954 &   --  &  --  &  --  &   --  &   --   \\
\enddata
\end{deluxetable}

\clearpage

\begin{deluxetable}{cccccccc}
\tabletypesize{\scriptsize}
\tablecolumns{8}
\tablewidth{0pt}
\tablecaption{Stellar parameters of A supergiants in 7 galaxies}

\tablehead{
\colhead{No.}     &
\colhead{name}            &
\colhead{galaxy} &
\colhead{m-M} &
\colhead{$T_{eff}$} &
\colhead{$log~g$} &
\colhead{$M_{bol}$} &
\colhead{comments}\\
\colhead{}   &
\colhead{}   &
\colhead{}	&
\colhead{mag}	&
\colhead{K}	&
\colhead{cgs}      &
\colhead{mag}   &
\colhead{ }\\[1mm]
\colhead{(1)}	&
\colhead{(2)}	&
\colhead{(3)}	&
\colhead{(4)}	&
\colhead{(5)}	&
\colhead{(6)}	&
\colhead{(7)}	&
\colhead{(8)}}
\startdata
\\[-1mm]
0  & $\eta$ Leo  & Milky Way &  9.00 &  9600 & 2.00 &  -5.73 & \tablenotemark{a} \\
1  & HD111613    & Milky Way & 11.80 &  9150 & 1.45 &  -7.43 & \tablenotemark{a} \\
2  & HD92207     & Milky Way & 12.40 &  9500 & 1.20 &  -9.04 & \tablenotemark{a} \\
3  & Rigel       & Milky Way &  7.80 & 12000 & 1.75 &  -8.49 & \tablenotemark{a} \\
4  & Deneb       & Milky Way &  9.50 &  8525 & 1.10 &  -8.48 & \tablenotemark{b} \\
5  & HD12953     & Milky Way & 11.80 &  9250 & 1.20 &  -8.08 & \tablenotemark{c,d} \\
5  & HD14489     & Milky Way & 11.80 &  9000 & 1.35 &  -7.81 & \tablenotemark{c,d} \\
5  & HD223385    & Milky Way & 12.00 &  8500 & 0.95 &  -8.43 & \tablenotemark{c,d} \\
6  & Sk-69 211   & LMC       & 18.50 & 11000 & 1.30 &  -9.00 & \tablenotemark{e} \\
7  & Sk-67 17    & LMC       & 18.50 & 10500 & 1.20 &  -9.53 & \tablenotemark{e} \\
8  & Sk-67 207   & LMC       & 18.50 & 10000 & 1.20 &  -8.50 & \tablenotemark{e} \\
9  & Sk-66 58    & LMC       & 18.50 &  9500 & 1.10 &  -8.74 & \tablenotemark{e} \\
10 & Sk-67 201   & LMC       & 18.50 & 10000 & 1.20 &  -9.14 & \tablenotemark{e} \\
11 & Sk-69 239   & LMC       & 18.50 & 10000 & 1.05 &  -9.68 & \tablenotemark{e} \\
12 & Sk-69 170   & LMC       & 18.50 & 10000 & 1.20 &  -8.89 & \tablenotemark{e} \\
13 & Sk-69 299   & LMC       & 18.50 &  8750 & 1.00 &  -8.93 & \tablenotemark{e} \\
14 & Sk-67 44    & LMC       & 18.50 &  8000 & 0.65 &  -9.54 & \tablenotemark{e} \\
15 & AV457       & SMC       & 19.00 &  8500 & 0.90 &  -9.10 & \tablenotemark{e} \\
16 & 41-3654     & M31       & 24.47 &  9200 & 1.00 &  -9.08 & \tablenotemark{f,k} \\
16 & 41-3712     & M31       & 24.47 &  8850 & 1.00 &  -8.79 & \tablenotemark{f,k} \\
17 & 117-A       & M33       & 24.47 &  9500 & 1.10 &  -8.29 & \tablenotemark{g,d} \\
18 & NGC6822m    & NGC 6822  & 23.31 &  9000 & 1.40 &  -7.23 & \tablenotemark{h,i} \\
19 & No. 1       & NGC 3621  & 29.08 &  9250 & 1.30 &  -8.21 & \tablenotemark{d,j} \\
20 & No. 9       & NGC 3621  & 29.08 &  9250 & 1.12 &  -9.20 & \tablenotemark{d,j} \\
21 & No. 16      & NGC 3621  & 29.08 & 10500 & 1.55 &  -8.05 & \tablenotemark{d,j} \\
22 & No. 17      & NGC 3621  & 29.08 &  8350 & 1.15 &  -8.58 & \tablenotemark{d,j} \\
\enddata
\tablenotetext{a}{\citet{przybilla06}}
\tablenotetext{b}{\citet{schiller07}}
\tablenotetext{c}{\citet{kud99}}
\tablenotetext{d}{\citet{kud03}}
\tablenotetext{e}{this work}
\tablenotetext{f}{\citet{przybilla06b}}
\tablenotetext{g}{distance modulus from \citet{bonanos06}}
\tablenotetext{h}{distance modulus from \citet{gieren06}}
\tablenotetext{i}{\citet{przybilla02}}
\tablenotetext{j}{\citet{bresolin01}}
\tablenotetext{k}{distance modulus from \citet{holland98}}
\end{deluxetable}

\clearpage

\end{document}